\documentclass[twocolumn,preprintnumbers,amsmath,amssymb]{revtex4}
\usepackage{amsmath,amsfonts,latexsym,amssymb,graphics,epsfig,subfigure,color,makeidx,inputenc}
\usepackage{xcolor}
\usepackage{graphicx}
\usepackage{hyperref}
\usepackage{diagbox}
\usepackage{multirow}

\def\gw#1{gravitational wave#1 (GW#1)\gdef\gw{GW}}
\def\bbh#1{binary black hole#1 (BBH#1)\gdef\bbh{BBH}}
\def\bh#1{black hole#1 (BH#1)\gdef\bh{BH}}

\graphicspath{{./Figs/}}

\begin{document}

\title{Scalar wave dark matter imprints on binary black hole mergers: dependence on field mass and density}

\author{Yu-Peng Zhang \footnote{zhangyupeng@lzu.edu.cn},
        Shao-Wen Wei\footnote{weishw@lzu.edu.cn},
        Yu-Xiao Liu\footnote{liuyx@lzu.edu.cn}
}
\affiliation{
        Lanzhou Center for Theoretical Physics, Key Laboratory of Theoretical Physics of Gansu Province, Key Laboratory of Quantum Theory and Applications of MoE, Gansu Provincial Research Center for Basic Disciplines of Quantum Physics,\\
Institute of Theoretical Physics \& Research Center of Gravitation, School of Physical Science and Technology, Lanzhou University, Lanzhou 730000, China
	}

\begin{abstract}

Dark matter environments can modify binary black hole mergers and imprint their gravitational-wave signals. We perform numerical-relativity simulations of nonspinning binary black holes immersed in complex scalar-field clouds, considering mass ratios $q=1,2,3,$ and $4$ and varying both the dimensionless scalar-field mass $\hat m_{\Phi}=m_{\Phi}M_0$ and initial amplitude $\Pi_0$, which controls the density. Our initial data keep the scalar energy-density profile and binary geometry unchanged when varying $\hat m_{\Phi}$ at fixed $\Pi_0$, allowing the dynamical effect of the field mass to be isolated. We find that the scalar environment generally accelerates the merger and significantly modifies the gravitational-wave signal. For unequal-mass binaries, differential scalar-field accretion changes the horizon-mass ratio during inspiral, weakly at low density but substantially at high density, thereby increasing the binary asymmetry before merger. Dense scalar environments suppress the energy and angular momentum carried by gravitational waves. Lighter scalar fields propagate more efficiently to large radii, whereas heavier fields remain more localized and show stronger inward energy transport. Scalar angular-momentum transport is more complex, showing nonmonotonic behavior and even direction reversal near the binary. For unequal-mass systems, gravitational and scalar radiation both carry net linear momentum, and their relative orientation determines whether the recoil is enhanced or suppressed. The largest kick is $2063.6~{\rm km,s^{-1}}$ for $q=2$, $\hat m_{\Phi}=0$, and $\Pi_0=0.0100$, compared with $137.8~{\rm km,s^{-1}}$ in vacuum. These results show that the scalar-field mass and density jointly control the binary asymmetry, dynamics, radiative channels, and momentum balance.

\end{abstract}

\maketitle

\section{Introduction}
\label{sec:intro}

The direct detection of gravitational waves (GWs) from the binary black hole (BBH) merger GW150914 opened a new window for probing gravity in the strong-field and highly dynamical regime \cite{Abbott:2016blz}. Since then, the LIGO--Virgo--KAGRA network has established compact-binary coalescences as a major class of GW sources \cite{Abbott:2021dja,LIGOScientific:2025slb}. Besides constraining the intrinsic masses and spins of the binary, GW observations can also probe the environments in which compact binaries evolve. Astrophysical binaries need not be isolated: gas accretion, dynamical drag, and disk-driven migration can modify the inspiral rate and accumulate phase shifts \cite{Kocsis:2011environment,Yunes:2011migration,Barausse:2014environment,Caputo:2020accreting}, while dark-matter structures and other extended media can alter the orbital dynamics and waveform \cite{Cardoso:2020constraints,Coogan:2022dm,Becker:2023env}. The environment can further affect the tidal response and the generation or propagation of GWs \cite{Bonvin:2017matter,Cardoso:2020tidal,Cardoso:2022galaxies}. These effects have motivated dedicated studies of their detectability and distinguishability in both stellar-mass and massive-BH binaries \cite{Sberna:2022qbn,Cole:2022yzw,Leong:2023nuk,Zwick:2024yzh}, including possible parameter biases and environmental dephasing in future observations \cite{Yuan:2024duo,Zwick:2025wkt,Zwick:2025qzv}. Environmental effects are therefore both a potential systematic in precision GW inference and a possible probe of matter in the strong-gravity regime.

Dark matter is a particularly important environment because its existence is supported by several independent classes of observations. Galactic rotation curves provide classic evidence for a substantial nonluminous mass component \cite{Rubin:1980rotation,Begeman:1991rotation,Persic:1995ru,Corbelli:1999af}; gravitational lensing in colliding galaxy clusters gives complementary evidence that the dominant gravitating component is spatially separated from the baryonic plasma \cite{Clowe:2006bullet}; and measurements of the cosmic microwave background and the cosmological matter budget require a large nonbaryonic dark-matter component \cite{WMAP:2010qai,WMAP:2010sfg,Aghanim:2018eyx}. Its microscopic nature, however, remains unknown. Among the many candidates, ultralight bosonic fields are especially interesting because their large occupation numbers and macroscopic coherence scales allow them to behave as classical waves \cite{Hu:2000ke,Hui:2016ltb,Marsh:2016axion,Ferreira:2021uldm,Niemeyer:2020fuzzy}. Cosmological simulations of fuzzy or wave dark matter have revealed interference patterns, suppressed small-scale structure, and self-gravitating solitonic cores in halos \cite{Schive:2014nature,Schive:2014core,Schwabe:2016cores,Mocz:2017becdm}, together with a rich nonlinear halo and filamentary structure \cite{Veltmaat:2018halos,Mocz:2019filaments,May:2021fuzzy}. Black hole superradiance and related searches provide complementary probes of ultralight bosons in a very different mass regime \cite{Brito:2015oca,Cardoso:2019rvt,Chen:2023ryb}. These wave-like properties make enhanced scalar concentrations around compact objects natural targets for strong-field studies.

Numerical relativity is particularly well suited to following the nonlinear interaction between scalar fields and compact objects \cite{Healy:2011ef,Cardoso:2014uka}. For isolated black holes and collapsing systems, simulations have investigated scalar accretion, gravitational radiation, and long-lived quasibound configurations \cite{Barranco:2011dmBH,Nunez:2011scalar,SanchisGual:2015collapse,SanchisGual:2016accreting}, as well as nonlinear massive-scalar dynamics and collapse in the vicinity of black holes \cite{Okawa:2014nda,Okawa:2014nea}. Fully relativistic calculations have also studied axion-star collapse and collisions with black holes or neutron stars \cite{Helfer:2017axion,Clough:2018axion}. Self-gravitating bosonic compact objects provide another closely related arena: boson-star binaries exhibit characteristic orbital dynamics and merger remnants \cite{Palenzuela:2008boson,Bezares:2017boson,Siemonsen:2023boson,Atteneder:2024boson,Ge:2024itl,Ge:2025btw}, together with distinctive GW emission and recoil phenomenology \cite{Palenzuela:2017bosonGW,Bezares:2022bosonKick,Evstafyeva:2024boson}. Scalar environments have also been explored in binary-neutron-star mergers and mixed boson-star--black-hole systems \cite{Bezares:2019dmBNS,Srikanth:2026scalarBNS,Ning:2026BSBH}.

For BBHs embedded directly in scalar environments, fully nonlinear simulations have shown that the surrounding field can modify the orbital evolution, accretion, merger waveform, radiative losses, and properties of the remnant \cite{Healy:2011ef,Yang:2017xlf,Choudhary:2020pxy,Bamber:2022pbs,Zhang:2022rms}. In our previous work \cite{Zhang:2022rms}, we focused on the recoil of BBH merger remnants at a fixed scalar field mass while varying the initial scalar density, and found that differential scalar accretion can change the individual black hole properties and substantially modify the final kick. More recent studies have shown that the scalar field mass can strongly affect the binary evolution, particularly when the characteristic scalar wavelength is comparable to the orbital scale \cite{Aurrekoetxea:2023jwk}, while self-interactions can further modify the accumulation and response of wave dark matter \cite{Aurrekoetxea:2024zsu}. Simulations covering different scalar masses have also found substantial changes in the merger dynamics and in both gravitational and scalar radiation \cite{Cheng:2025xxx}, and the corresponding analyses have begun to assess the observability of dense scalar environments in GW data \cite{Roy:2025xxx}.

Despite this progress, the combined dependence on the scalar field mass and density has not yet been systematically characterized in the fully nonlinear BBH regime. The field mass sets a characteristic length and time scale, whereas the density controls the amount of scalar matter available for accretion and backreaction. Their interplay can therefore affect the inspiral, the evolving binary asymmetry, the radiative energy and angular-momentum budget, and the linear-momentum balance that determines the remnant recoil.

In this work, we perform a systematic numerical-relativity study of nonspinning BBH mergers in scalar field clouds, considering mass ratios $q=1,2,3,$ and $4$ while varying both the dimensionless scalar field mass and the initial scalar density. Our initial-data construction keeps the scalar energy-density profile and binary geometry unchanged when the field mass is varied at fixed initial amplitude, allowing the dynamical mass dependence to be isolated. We follow the scalar distribution, orbital dynamics, and individual horizon masses, including the evolution of the effective mass ratio caused by differential accretion. We then analyze the GW signal and the energy, angular momentum, and linear momentum transported through both gravitational and scalar channels, and determine the recoil of unequal-mass remnants. This two-dimensional mass--density survey provides a unified characterization of how the characteristic scalar scale and the amount of surrounding scalar matter jointly control BBH dynamics, radiation, and momentum transfer.

\section{Theoretical framework and numerical setup}
\label{sec:method}
We consider nonspinning BBHs immersed in scalar field environments with mass ratios
\begin{equation}
    q \equiv \frac{m_1}{m_2}
    =\left(1,\;2,\;3,\;4\right),
    \qquad m_1\geq m_2 ,
\label{eq:mass_ratio}
\end{equation}
where $m_1$ and $m_2$ denote the masses of the two black holes. The initial total mass of the binary,
\begin{equation}
    M_0=m_1+m_2
\end{equation}
is used as the characteristic mass scale throughout this work and we set $M_0=1$. All dimensionful quantities are expressed in units of $M_0$.

\subsection{Einstein--Klein--Gordon system} \label{sec:theory}

The system is described by general relativity minimally coupled to a massive complex scalar field $\Phi$, with action
\begin{equation}
S =
\int d^4x \sqrt{-g}
\left[
\frac{R}{16\pi}
-\frac{1}{2}
\left(
\partial^\nu \Phi^* \partial_\nu \Phi
+m_{\Phi}^2 |\Phi|^2
\right)
\right],
\label{eq:action}
\end{equation}
where $R$ is the Ricci scalar, $g$ is the determinant of the spacetime metric $g_{\mu\nu}$, and $m_{\Phi}$ denotes the mass parameter of the scalar field. The corresponding characteristic length scale of the scalar field is set by $m_{\Phi}^{-1}$. In the following, the scalar field mass parameter is characterized by the dimensionless combination
\begin{equation}
    \hat{m}_{\Phi} \equiv m_{\Phi} M_0 .
\label{eq:mudimless}
\end{equation}
Unlike our previous study \cite{Zhang:2022rms}, in which $\hat{m}_{\Phi}$ was fixed, here we vary $\hat{m}_{\Phi}$ together with the initial scalar field density in order to explore their combined effects on the BBH dynamics.

Variation of the action \eqref{eq:action} with respect to the metric $g_{\mu\nu}$ yields the Einstein equations
\begin{equation}
G_{\mu\nu}=8\pi T_{\mu\nu},
\label{eq:einstein}
\end{equation}
where the stress-energy tensor of the complex scalar field is
\begin{align}
T_{\mu\nu}
=&\,
\frac{1}{2}
\left(
\partial_\mu\Phi^* \partial_\nu\Phi
+
\partial_\nu\Phi^* \partial_\mu\Phi
\right)
\nonumber\\
&-\frac{1}{2}g_{\mu\nu}
\left(
\partial_\alpha\Phi^*\partial^\alpha\Phi
+m_{\Phi}^2|\Phi|^2
\right).
\label{eq:stressenergy}
\end{align}
Variation with respect to the scalar field $\Phi$ gives the massive Klein--Gordon equation,
\begin{equation}
\nabla_\mu\nabla^{\mu}\Phi-m_{\Phi}^2\Phi=0,
\label{eq:KG}
\end{equation}
together with the complex-conjugate equation for $\Phi^*$.

\subsection{$3+1$ decomposition and evolution equations} \label{sec:evolution}

We perform a standard $3+1$ decomposition of the spacetime metric,
\begin{equation}
g_{\mu\nu}=\gamma_{\mu\nu}-n_\mu n_\nu ,
\end{equation}
where $\gamma_{\mu\nu}$ is the spatial metric induced on the $t={\rm const.}$ hypersurfaces and $n^\mu$ is the future-directed timelike unit normal,
\begin{equation}
n^\mu=
\frac{1}{\alpha}
\left(1,-\beta^i\right).
\end{equation}
Here $\alpha$ and $\beta^i$ denote the lapse function and shift vector, respectively.

We introduce the conjugate momentum of the scalar field as
\begin{equation}
\Pi=-n^\mu\nabla_\mu\Phi
   =-\frac{1}{\alpha}
     \left(\partial_t-\beta^i\partial_i\right)\Phi .
\label{eq:Pi}
\end{equation}
Defining
\begin{equation}
    \partial_0\equiv\partial_t-\beta^i\partial_i ,
\end{equation}
the Klein--Gordon equation can be written as the first-order-in-time system
\begin{equation}
\partial_0\Phi=-\alpha\Pi ,
\label{eq:evol_phi}
\end{equation}
and
\begin{equation}
\partial_0\Pi
=
-\alpha D_iD^i\Phi
-D^i\alpha D_i\Phi
+\alpha K\Pi
+\alpha m_{\Phi}^2\Phi ,
\label{eq:evol_pi}
\end{equation}
where $D_i$ is the covariant derivative associated with $\gamma_{ij}$ and $K=\gamma^{ij}K_{ij}$ is the trace of the extrinsic curvature.

The scalar field enters the evolution of the geometry through the projections of its stress-energy tensor. The energy density measured by normal observers is
\begin{equation}
\rho =
\frac{1}{2}
\left(
|\Pi|^2
+D_i\Phi^*D^i\Phi
+m_{\Phi}^2|\Phi|^2
\right),
\label{eq:rho}
\end{equation}
while the spatial stress tensor is
\begin{align}
S_{ij}
=&\,
\frac{1}{2}\left(D_i\Phi^*D_j\Phi+D_i\Phi D_j\Phi^*\right)
\nonumber\\
&+\frac{1}{2}\gamma_{ij}
\left(
\Pi^* \Pi
-D_k\Phi^*D^k\Phi
-m_{\Phi}^2|\Phi|^2
\right).
\label{eq:Sij}
\end{align}

\subsection{Initial data and system evolution}
\label{sec:initial}
The initial data on a spacelike hypersurface must satisfy the Hamiltonian constraint
\begin{equation}
R+K^2-K_{ij}K^{ij}=16\pi\rho ,
\label{eq:Hamiltonian}
\end{equation}
and the momentum constraints
\begin{equation}
D_jK^j_{\ i}-D_iK=8\pi S_i,
\label{eq:momentum_constraint}
\end{equation}
respectively. Here $S_i$ is given by
\begin{equation}
S_i = \frac{1}{2}(\Pi^* D_i \Phi + \Pi~D_i \Phi^*).
\label{eq:momentum_density}
\end{equation}

Following our previous work \cite{Zhang:2022rms}, we construct the initial data using the York--Lichnerowicz conformal decomposition. We assume conformal flatness and maximal slicing
\begin{equation}
\gamma_{ij}=\psi^4\eta_{ij},
\qquad
K=0 ,
\label{eq:conformalmetric}
\end{equation}
and write the traceless part of the extrinsic curvature as
\begin{equation}
K_{ij}=A_{ij}=\psi^{-2}\widetilde{A}_{ij},
\qquad
\widetilde{A}^i_{\ i}=0 .
\label{eq:Aij}
\end{equation}
The Bowen--York solution is used for $\widetilde{A}_{ij}$ to describe the orbital linear momenta and spins of the two punctures
\cite{Bowen:1980yu}.

For the scalar field, we adopt the conformal rescaling \cite{Laguna:1991zs,Balakrishna:2006ru}
\begin{equation}
\Phi=\widetilde{\Phi},
\qquad
\Pi=\psi^{-6}\widetilde{\Pi}.
\label{eq:scalarconformal}
\end{equation}

We choose
\begin{equation}
    \Phi|_{t=0}=0 .
\label{eq:phiinitial}
\end{equation}
This choice is particularly convenient for puncture initial data, since terms proportional to positive powers of the conformal factor that would otherwise become singular at the punctures vanish identically. Moreover, the scalar momentum density vanishes initially, $S_i=0$, and the Hamiltonian and momentum constraints reduce to
\begin{equation}
\widetilde{\Delta}\psi
+
\left(
\frac{1}{8}
\widetilde{A}_{ij}\widetilde{A}^{ij}
+\pi|\widetilde{\Pi}|^2
\right)
\psi^{-7}
=0 ,
\label{eq:Hamiltonian_reduced}
\end{equation}
and
\begin{equation}
\partial_j\widetilde{A}^{ij}=0 ,
\label{eq:Momentum_reduced}
\end{equation}
respectively, where
$\widetilde{\Delta}=\eta^{ij}\partial_i\partial_j$.

An important consequence of the choice $\Phi|_{t=0}=0$ is that the scalar field mass term $m_{\Phi}^2|\Phi|^2$ in Eq. \eqref{eq:rho} does not contribute to the initial Hamiltonian constraint. Therefore, for fixed binary parameters and fixed $\Pi_0$, changing $m_{\Phi}$ leaves the initial scalar energy-density profile and the constraint equation for the conformal factor unchanged. The dependence on $m_{\Phi}$ enters only through the subsequent Klein--Gordon evolution, Eq.~(\ref{eq:evol_pi}). This construction provides a controlled comparison in which the dynamical effect of the scalar field mass can be isolated. We consider
\begin{equation}
   \hat{m}_{\Phi} =   \left(0,\;0.25,\;0.50,\;0.75,\;1.00\right) .
\label{eq:scalar_mass_values}
\end{equation}
The case $\hat m_{\Phi}=0$ is a massless reference model, while the remaining cases describe massive scalar fields with progressively shorter characteristic Compton wavelengths,
\begin{equation}
    \lambda_{\Phi}\sim m_{\Phi}^{-1}.
\end{equation}
The range considered here therefore allows us to probe different relations between the characteristic length scale of the scalar field and the dynamical scale of the binary.

For each value of $\hat{m}_{\Phi}$, we further vary the initial amplitude of the scalar cloud. Following Sec.~\ref{sec:initial}, the scalar field is initialized through its conjugate momentum,
\begin{equation}
\widetilde{\Pi}(t=0,r)
=
\Pi_0
\exp\left[
-\frac{1}{2}
\left(\frac{r-r_0}{\sigma}\right)^2
\right],
\label{eq:Pi_parameter}
\end{equation}
with a set of different values of $\Pi_0$. The parameters $r_0$ and $\sigma$, which determine the location and width of the initial scalar cloud, are kept fixed within each sequence. Increasing $\Pi_0$ increases the initial energy density of the surrounding scalar field and therefore allows us to continuously vary the strength of its interaction with the binary.

Consequently, for each binary mass ratio we explore a two-dimensional scalar field parameter space characterized by
\begin{equation}
    \left(\hat{m}_{\Phi},\,\Pi_0\right),
\end{equation}
or, equivalently, by the scalar field mass and its initial energy
density. This parameter survey enables us to determine separately and
jointly the effects of the characteristic scalar field scale and the
amount of scalar matter surrounding the binary.

We solve Eq.~(\ref{eq:Hamiltonian_reduced}) using our modified version of the \texttt{TwoPunctures} solver \cite{Ansorg:2004ds}, in which the scalar contribution $\pi|\widetilde{\Pi}|^2\psi^{-7}$ is included explicitly.

The spacetime geometry is evolved with the Baumgarte--Shapiro--Shibata--Nakamura (BSSN) formulation \cite{Baumgarte:1998te,Shibata:1995we}, while Eqs.~(\ref{eq:evol_phi}) and (\ref{eq:evol_pi}) are evolved simultaneously with the geometry, including the matter source terms from Eq.~(\ref{eq:stressenergy}). The simulations are carried out with the \textsc{Maya} numerical-relativity infrastructure \cite{Haas:2012bk,Evans:2015fha,Clark:2016ppe,Jani:2016wkt} within the \textsc{Einstein Toolkit} \cite{Loffler:2011ay}; the spacetime evolution uses the \textsc{McLachlan} thorn \cite{Brown:2008sb}. We employ the moving-puncture gauge \cite{Campanelli:2005dd,Baker:2005vv} for the lapse and shift throughout the simulations.

Compared with our previous study \cite{Zhang:2022rms}, we additionally employ the \texttt{Llama} multipatch infrastructure \cite{Pollney:2009yz,Reisswig:2012nc}, which provides an extended wave zone and improves the extraction of gravitational and scalar radiation. Mesh refinement is used to resolve the individual black holes. Fine refinement levels follow the black holes during the inspiral, whereas progressively coarser levels cover the wave-extraction region and the outer computational domain.

The mesh-refinement hierarchy is adjusted according to the binary mass ratio. For the equal-mass case ($q=1$), both black holes are covered by identical refinement structures with six refinement levels, and the finest grid spacing is $0.8/2^5\,M_0$. For unequal-mass binaries, the larger black hole is covered by the same refinement hierarchy as in the $q=1$ case, while additional refinement levels are introduced around the smaller black hole to properly resolve its smaller length scale. Specifically, seven refinement levels are used around the smaller black hole for $q=2$, while eight levels are used for $q=3$ and $q=4$. For the \texttt{Llama} multipatch region, we use a radial resolution of $\Delta r=0.8\,M_0$, which matches the largest spatial grid spacing of the central Cartesian mesh, and an angular resolution of $28\times28$ grid points on each patch. In all simulations, the outer boundary is placed at $1200\,M_0$.

The initial configuration for all BBH systems has a coordinate separation with $d=8\,M_0$. The scalar field momentum shell has a radius $r_0=12\,M_0$ and a thickness $\sigma=1\,M_0$.

\subsection{Black hole and radiation diagnostics}
\label{sec:diagnostics}

We monitor the individual and common apparent horizons throughout the
evolution. The black hole masses and angular momenta are calculated
from the apparent-horizon geometry using the isolated/dynamical
horizon formalism \cite{Ashtekar:2004cn,Thornburg:2003sf}. The irreducible mass is
defined by
\begin{equation}
M_{\rm irr}
=
\sqrt{\frac{A_{\rm AH}}{16\pi}},
\end{equation}
where $A_{\rm AH}$ is the apparent-horizon area. The horizon mass is then obtained from the Christodoulou relation \cite{Christodoulou:1970wf} as follows
\begin{equation}
M_{\rm H}
=
\sqrt{
M_{\rm irr}^2
+
\frac{J_{\rm BH}^2}{4M_{\rm irr}^2}
},
\label{eq:christodoulou}
\end{equation}
where $J_{\rm BH}$ is the angular momentum of the black hole.

Gravitational radiation is extracted through the Newman--Penrose Weyl scalar $\Psi_4$ \cite{Ruiz:2007yx}. Then the corresponding gravitational-wave energy flux is calculated as
\begin{equation}
\frac{dE_{\rm GW}}{dt}
=
\lim_{r\rightarrow\infty}
\frac{r^2}{16\pi}
\oint
\left|
\int_{-\infty}^{t}\Psi_4(r, t')dt'
\right|^2
d\Omega ,
\label{eq:EGW}
\end{equation}
and the linear-momentum flux is
\begin{equation}
\frac{dP_i^{\rm GW}}{dt}
=
\lim_{r\rightarrow\infty}
\frac{r^2}{16\pi}
\oint
\hat l_i
\left|
\int_{-\infty}^{t}\Psi_4(r, t')dt'
\right|^2
d\Omega ,
\label{eq:PGW}
\end{equation}
where
\begin{equation}
\hat l_i=
(\sin\theta\cos\varphi,
 \sin\theta\sin\varphi,
 \cos\theta)
\end{equation}
and
\begin{equation}
d\Omega = \sin\theta d\theta d\varphi.
\end{equation}
The angular-momentum flux is computed from
\begin{align}
\frac{dJ_i^{\rm GW}}{dt}
=&
-\lim_{r\rightarrow\infty}
\frac{r^2}{16\pi}
{\rm Re}
\Bigg[
\oint
\left(
\int_{-\infty}^{t}\bar{\Psi}_4(r, t') dt'
\right)
\nonumber\\
&\times
\hat{J}_i
\left(
\int_{-\infty}^{t}
\int_{-\infty}^{t'}
\Psi_4(r, t'') dt''dt'
\right)
d\Omega
\Bigg],
\label{eq:JGW}
\end{align}
where $\hat{J}_i$ denotes the angular-momentum operator.

In addition to gravitational radiation, the scalar field carries energy, linear momentum, and angular momentum away from the system. These fluxes are computed \cite{Witek:2018dmd} directly from the scalar field stress-energy tensor \eqref{eq:stressenergy},
\begin{equation}
\frac{dE_{\Phi}}{dt}
=
\lim_{r\rightarrow\infty}
r^2
\oint T_{tr}\,d\Omega ,
\label{eq:Ephi_flux}
\end{equation}
\begin{equation}
\frac{dP_i^{\Phi}}{dt}
=
\lim_{r\rightarrow\infty}
r^2
\oint T_{ir}\,d\Omega ,
\label{eq:Pphi_flux}
\end{equation}
and, for the $z$ component of angular momentum,
\begin{equation}
\frac{dJ_z^{\Phi}}{dt}
=
\lim_{r\rightarrow\infty}
r^2
\oint T_{\varphi r}\,d\Omega .
\label{eq:Jphi_flux}
\end{equation}
Because the computational domain has a finite spatial extent, gravitational and scalar radiation can only be evaluated on extraction spheres at finite coordinate radii. For the gravitational channel, $\Psi_4$ is measured at each extraction radius and used to construct the corresponding energy, linear-momentum, and angular-momentum fluxes. The scalar fluxes are evaluated directly from the stress-energy tensor on the corresponding extraction surfaces.

Here the Newman--Penrose scalar $\Psi_4$ itself is kept at the finite extraction radii, which allows us to reduce the systematic uncertainty associated with finite-radius wave extraction while avoiding an additional extrapolation of the gravitational waveform. Instead, when asymptotic integrated radiative quantities are required, we extrapolate the finite-radius results to spatial infinity in order to reduce finite-radius effects \cite{Scheel:2008rj,Boyle:2009vi}. We use a second-order polynomial in the inverse extraction radius,
\begin{equation}
Q(r)
=
Q_{\infty}
+
\frac{a_1}{r}
+
\frac{a_2}{r^2},
\label{eq:radial_extrapolation}
\end{equation}
where $Q_{\infty}$ represents the corresponding asymptotic value in the limit $r\rightarrow\infty$. Equivalently, $Q_{\infty}$ is the intercept of the polynomial fit at $1/r=0$. The extrapolation is performed using results obtained at several extraction radii $r/M_0=(200, 300, 400)$.

For the radiated linear momentum, each Cartesian component is
extrapolated independently for both radiative channels,
\begin{equation}
P_i^{X}(r)
=
P_{i,\infty}^{X}
+
\frac{b_{i,1}^{X}}{r}
+
\frac{b_{i,2}^{X}}{r^2},
\label{eq:momentum_extrapolation}
\end{equation}
where $i=x,y,z,\quad X\in\{{\rm GW},\Phi\}.$ The magnitude is constructed only after extrapolation,
\begin{equation}
P_{X}^{\infty}
=
\sqrt{
\left(P_{x,\infty}^{X}\right)^2+
\left(P_{y,\infty}^{X}\right)^2+
\left(P_{z,\infty}^{X}\right)^2
}.
\label{eq:Pgw_inf}
\end{equation}

For configurations with vanishing initial total linear momentum, the
recoil velocity of the merger remnant follows from momentum
conservation. Including both gravitational and scalar field channels,
we calculate \cite{Gonzalez:2006md}
\begin{equation}
\mathbf{v}_{\rm kick}
=
-\frac{
\mathbf{P}_{\rm rad}^{\rm GW}
+
\mathbf{P}_{\rm rad}^{\Phi}}
{M_f},
\label{eq:vkick}
\end{equation}
where $M_f$ denotes the final black hole mass. This allows us to
separate the gravitational-wave and scalar field contributions to the
total recoil and to study their dependence on both the scalar field
mass and its initial density.

\section{Effects of scalar field mass and density}
\label{sec:results}

We now present the numerical results and examine how the scalar field
mass and initial density jointly affect BBH dynamics and radiation. For
each nonspinning binary with
\begin{equation}
q=\left(1,2,3,4\right),
\end{equation}
we consider
\begin{equation}
\hat m_{\Phi}=\left(0,0.25,0.50,0.75,1.00\right)
\end{equation}
and
\begin{equation}
\Pi_0=\left(0.0025,0.0050,0.0100,0.0150,0.0200\right).
\end{equation}
Because $\Phi=0$ initially, the scalar contribution to the initial
energy density is purely kinetic, $\rho_{\Phi}=\frac12|\Pi|^2$.
Accordingly, $\Pi_0$ provides a direct control parameter for the
initial scalar density, while $\hat m_{\Phi}$ changes the subsequent
dynamics without altering the initial scalar energy-density profile at
fixed $\Pi_0$.

The total ADM masses of the initial configurations are listed in
Table~\ref{table:amdmasstotal}. For fixed $q$ and $\Pi_0$, the ADM mass
is independent of $\hat m_{\Phi}$ by construction. The strong increase
of $M_{\rm ADM}$ with $\Pi_0$ also illustrates the broad range of scalar
environments covered by our parameter survey.

\begin{table}
\centering
\caption{
Initial ADM mass $M_{\rm ADM}/M_0$ for the simulated binaries.
The columns correspond to the initial scalar field amplitude $\Pi_0$.
For fixed $q$ and $\Pi_0$, the initial data are independent of
$\hat m_{\Phi}$.
}
\label{table:amdmasstotal}
\begin{tabular}{c|ccccc}
\hline\hline
\multirow{2}{*}{~~$q$~~} &
\multicolumn{5}{c}{$\Pi_0$}
\\
\cline{2-6}
&~$0.0025$~&~$0.0050$~&~$0.0100$~&~$0.0150$~&~$0.0200$~\\
\hline
1 & 1.0178 & 1.1049 & 1.4187 & 1.8529 & 2.3431 \\
2 & 1.0192 & 1.1064 & 1.4202 & 1.8545 & 2.3447 \\
3 & 1.0210 & 1.1082 & 1.4221 & 1.8563 & 2.3466 \\
4 & 1.0224 & 1.1096 & 1.4235 & 1.8578 & 2.3481 \\
\hline\hline
\end{tabular}
\end{table}

An important feature of the initial-data construction in
Sec.~\ref{sec:initial} is that simulations with different
$\hat m_{\Phi}$ but the same $q$ and $\Pi_0$ start from the same binary
geometry and scalar energy-density profile. Differences among those
simulations therefore arise dynamically from the scalar mass term in
the Klein--Gordon equation. In contrast, varying $\Pi_0$ changes the
amount of scalar matter available to interact with the binary.

We organize the discussion as follows. We first examine the scalar field
evolution, differential accretion onto the two black holes, the resulting
evolution of the horizon-mass ratio, and the orbital dynamics and merger
time. We then analyze the gravitational-wave signal and the energy and
angular momentum transported through the gravitational and scalar channels.
Finally, we study the linear-momentum balance and the recoil of the merger
remnant.

\subsection{Binary dynamics and scalar field evolution}
\label{sec:dynamics}

We begin by examining the dynamical evolution of the scalar field and its interaction with the BBH. This provides the physical basis for understanding the changes in the gravitational-wave emission and the properties of the merger remnant discussed in the following subsections. In particular, we study how the scalar field mass and initial density affect the redistribution and accretion of the scalar cloud, the orbital evolution of the binary, and the time of merger.

The scalar field mass introduces an intrinsic length scale, $\lambda_{\Phi}\sim m_{\Phi}^{-1}$, which determines how the scalar field responds to the characteristic length and time scales of the binary. On the other hand, the initial amplitude $\Pi_0$ controls the amount of scalar matter surrounding the black holes and therefore the strength of its backreaction on the binary. To disentangle these two effects, we perform two complementary comparisons. First, for fixed $q$ and $\Pi_0$, we compare simulations with different values of $\hat{m}_{\Phi}$. Second, for fixed $q$ and $\hat{m}_{\Phi}$, we vary $\Pi_0$ to investigate the dependence on the initial scalar field density.

\begin{figure*}
	\includegraphics[width=0.95\linewidth]{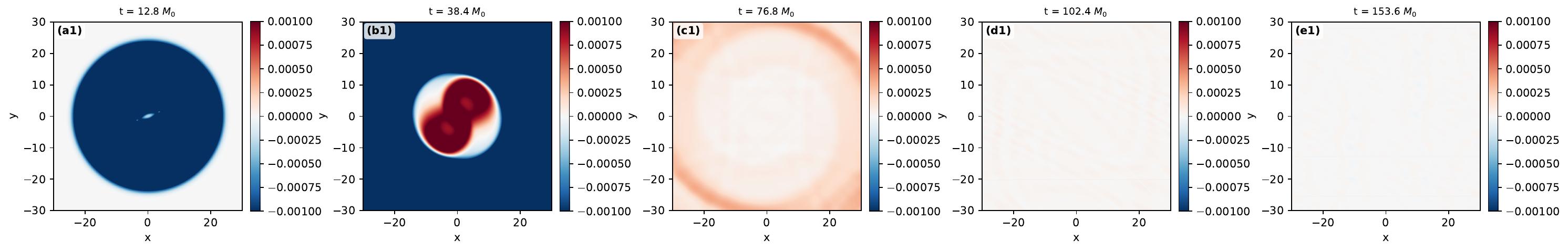}
    \includegraphics[width=0.95\linewidth]{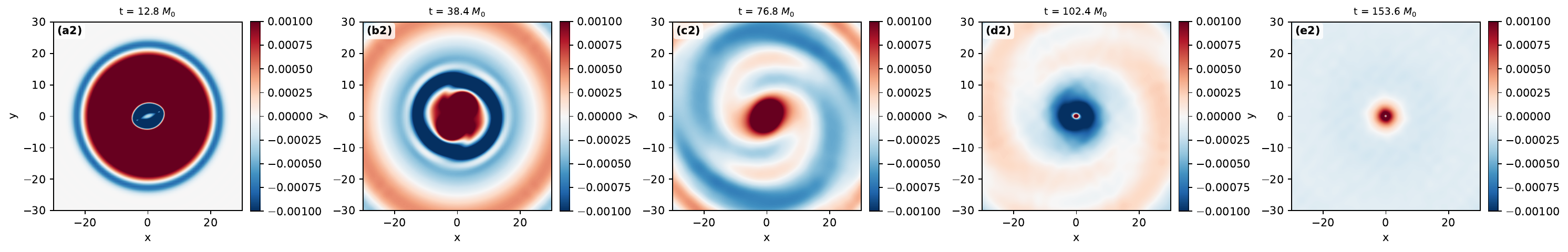}
    \includegraphics[width=0.95\linewidth]{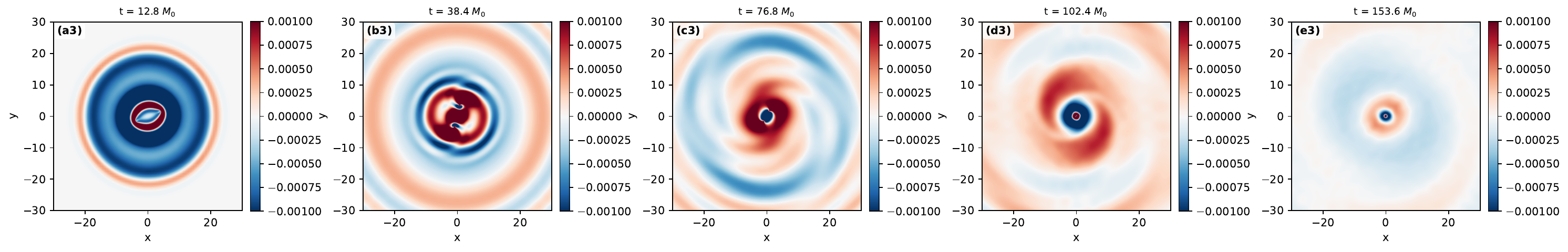}
    \includegraphics[width=0.95\linewidth]{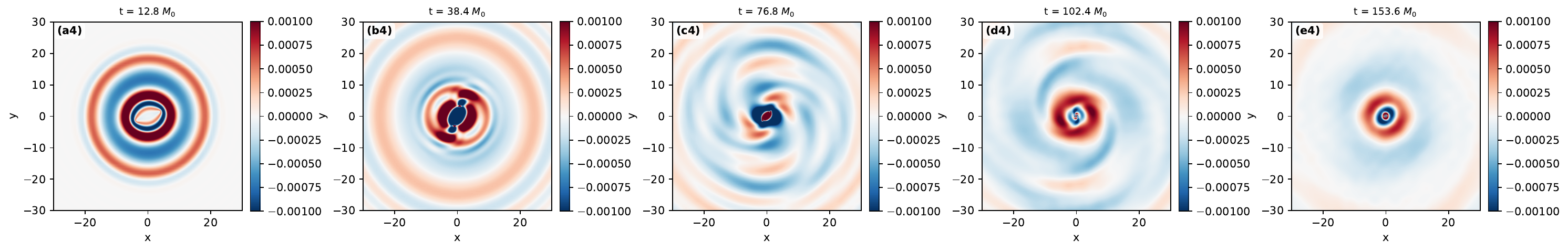}
    \includegraphics[width=0.95\linewidth]{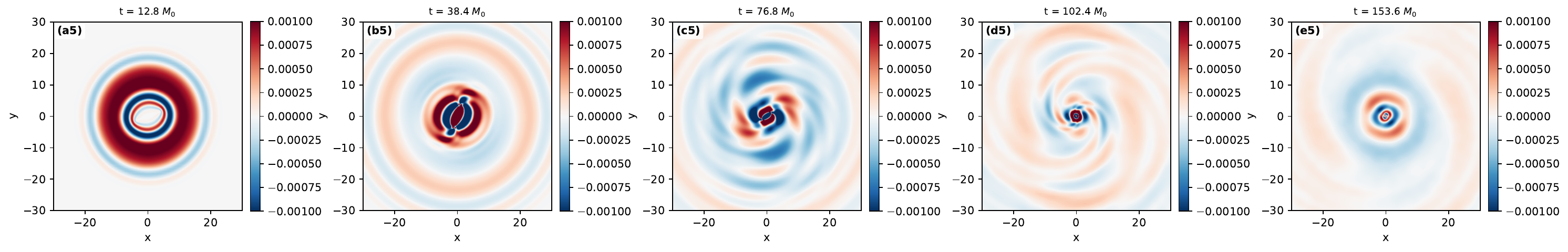}
	\caption{Evolution of the real part of the scalar field, $\mathrm{Re}(\Phi)$, on the equatorial ($x$--$y$) plane for the equal-mass binary, $q=1$, with fixed initial amplitude $\Pi_0=0.005$. From top to bottom, the rows correspond to $\hat m_{\Phi}=0$, $0.25$, $0.50$, $0.75$, and $1.00$, while from left to right the columns show $t/M_0=12.8$, $38.4$, $76.8$, $102.4$, and $153.6$. The same color scale is used in all panels. The massless field propagates rapidly outward, whereas increasing $\hat m_{\Phi}$ leads to more persistent and localized structures around the binary and the remnant.}
	\label{fig:scalar_evolution}
\end{figure*}

Figure~\ref{fig:scalar_evolution} shows representative snapshots of $\mathrm{Re}(\Phi)$ for the equal-mass binary with a fixed initial scalar field amplitude $\Pi_0=0.005$. Thus, the only scalar field parameter varied among the different rows is its mass $\hat m_{\Phi}$. For a fixed $\Pi_0$, all these models start from the same scalar field energy-density profile and the same solution of the Hamiltonian constraint, as discussed in Sec.~\ref{sec:initial}. The different propagation and localization patterns that subsequently develop therefore arise dynamically from the scalar field mass term in the Klein--Gordon equation, directly illustrating the important role of $m_{\Phi}$ in the interaction between the scalar field and the binary.

The comparison between different rows clearly demonstrates the
importance of the scalar field mass even for the same equal-mass BBH
configuration. For the massless case, $\hat m_{\Phi}=0$, the scalar
field propagates efficiently outward, and the field amplitude in the
central region becomes very small at late times. In contrast, for
nonzero scalar field masses, a larger fraction of the field remains
localized in the vicinity of the binary and the final black hole.
The massive cases also exhibit more persistent oscillatory structures
around the central region at late times. Since all these $q=1$ models
with the same initial scalar field amplitude start from the same initial
geometry and scalar field energy-density profile, the differences
between the rows are generated dynamically by the scalar field mass
term in the Klein--Gordon equation. Figure~\ref{fig:scalar_evolution}
therefore provides a direct illustration of how $m_{\Phi}$ controls the
propagation, redistribution, and localization of the scalar field
during the inspiral, merger, and post-merger evolution.

\begin{figure*}
    \centering
    \includegraphics[width=0.245\linewidth]{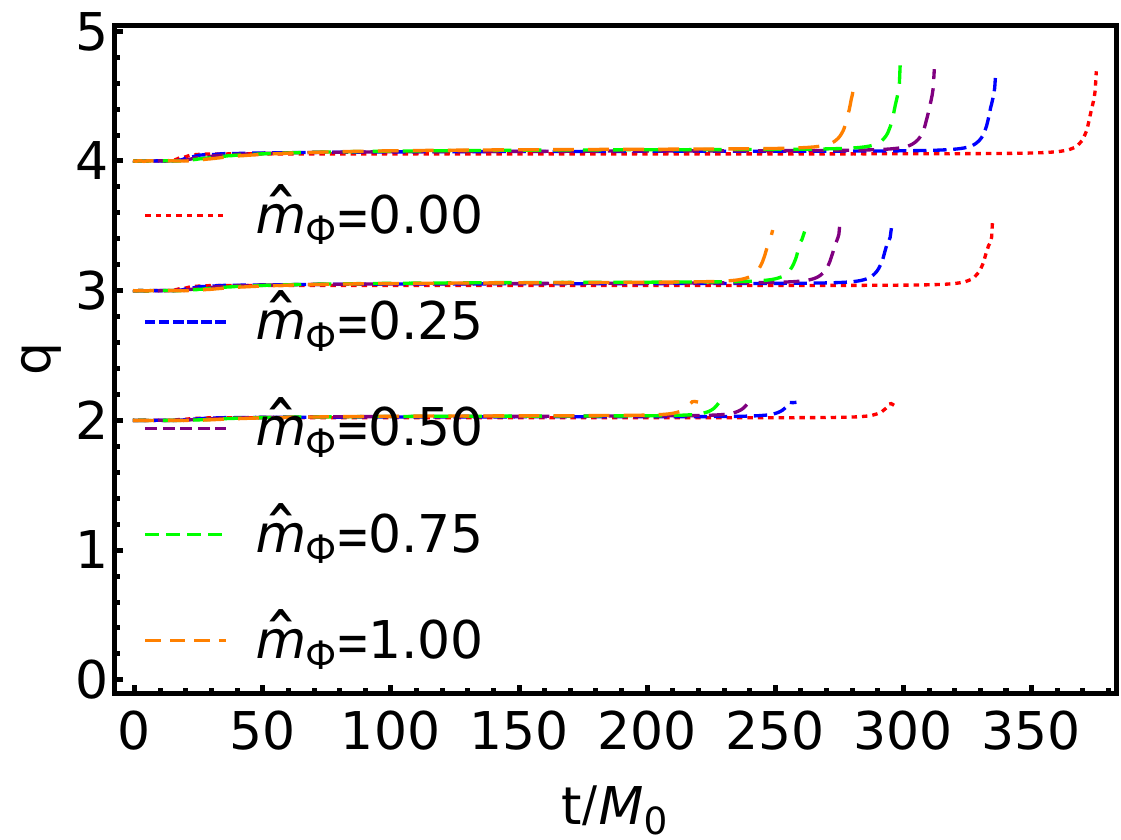}
    \includegraphics[width=0.245\linewidth]{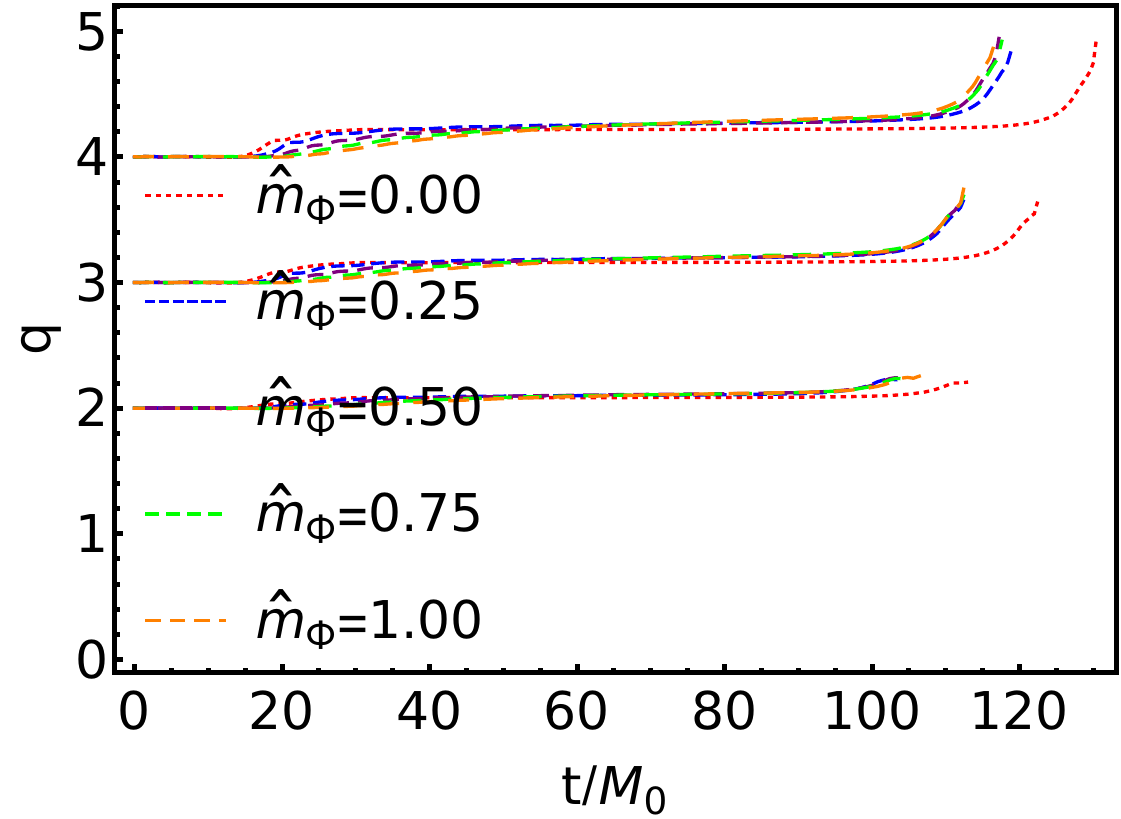}
    \includegraphics[width=0.245\linewidth]{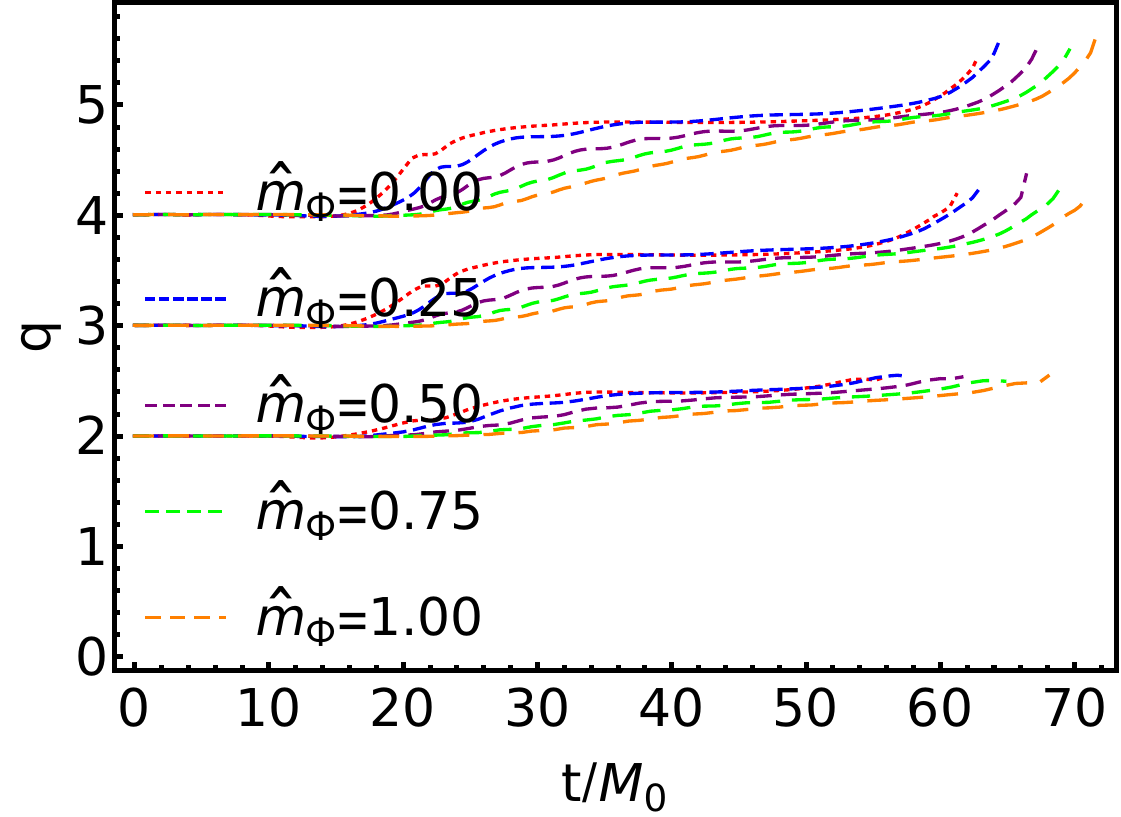}
    \includegraphics[width=0.245\linewidth]{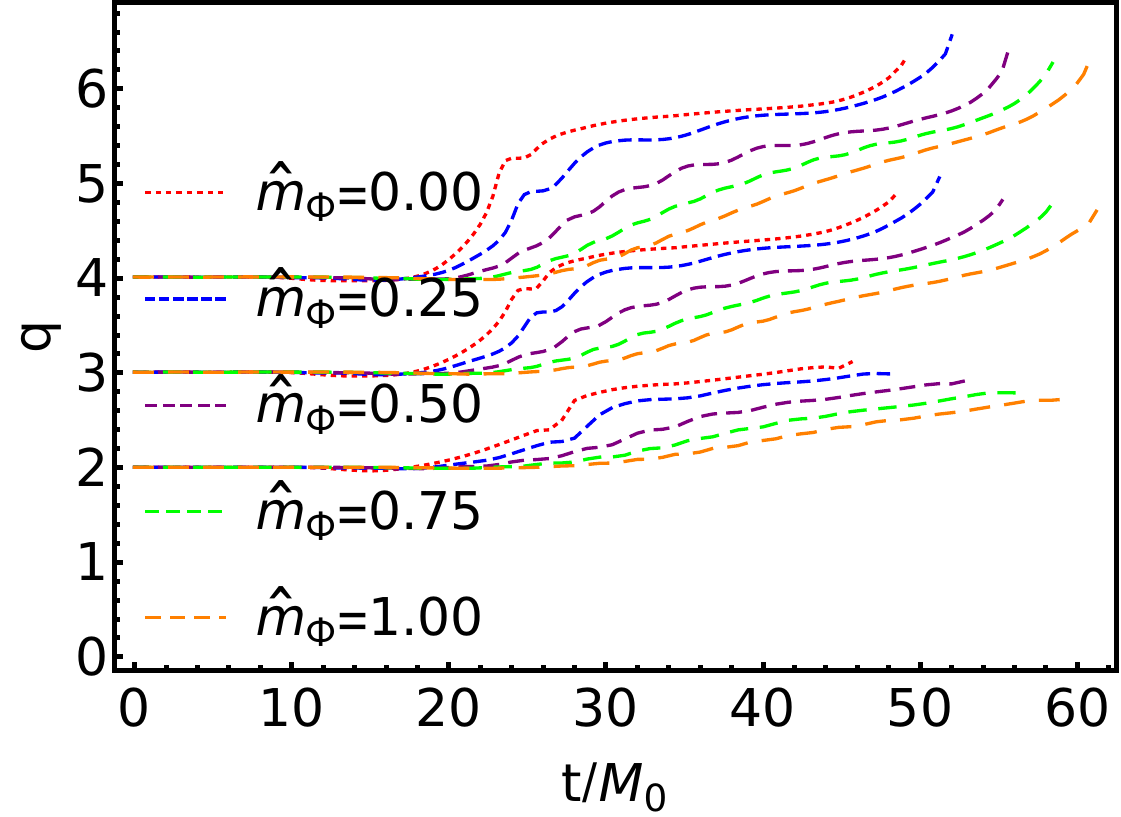}
    \caption{
    Evolution of the horizon-mass ratio
    $q_{\rm H}(t)=m_{{\rm H},1}(t)/m_{{\rm H},2}(t)$ for unequal-mass
    BBHs. From left to right, the panels correspond to
    $\Pi_0=0.0025$, $0.0050$, $0.0100$, and $0.0150$, respectively.
    In each panel, the three groups of curves start from
    $q=2$, $3$, and $4$, while the different curves denote
    $\hat m_{\Phi}=0$, $0.25$, $0.50$, $0.75$, and $1.00$.
    The curves terminate near the formation of a common apparent horizon.
    }
    \label{fig:massratiochange}
\end{figure*}

For unequal-mass binaries, scalar field accretion provides an additional
dynamical effect: the two black holes do not accrete the surrounding field
at the same fractional rate. To quantify this differential accretion, we
introduce the instantaneous horizon-mass ratio
\begin{equation}
    q(t)
    =
    \frac{m_{{\rm BH},1}(t)}{m_{{\rm BH},2}(t)},
    \qquad m_{{\rm BH},1}\geq m_{{\rm BH},2},
\label{eq:qH}
\end{equation}
where the individual horizon masses are calculated from
Eq.~\eqref{eq:christodoulou}. For $q=1$, symmetry keeps the two horizon masses equal to numerical
accuracy and hence $q_{\rm H}\simeq1$. We therefore focus in
Fig.~\ref{fig:massratiochange} on the unequal-mass binaries with initial
mass ratios $q=2$, $3$, and $4$.

Figure~\ref{fig:massratiochange} shows that the scalar environment can
change the binary mass ratio appreciably during the inspiral. At the
lowest density, $\Pi_0=0.0025$, $q_{\rm H}$ remains close to its initial
value through most of the evolution and increases mainly during the late
inspiral. As the scalar field amplitude is increased, the departure from
the initial mass ratio begins progressively earlier and becomes much
larger. For $\Pi_0=0.0100$ and $0.0150$, a clear secular growth of
$q_{\rm H}$ is already visible well before merger, and the change can
become of order unity in the densest configurations shown.

The systematic increase of $q_{\rm H}$ implies that the larger black hole gains
mass more efficiently in fractional terms than the smaller one. The
strength of this differential accretion depends on both $\Pi_0$ and
$\hat m_{\Phi}$. At low density, the scalar mass mainly changes the time
at which the late growth sets in, consistently with the mass-dependent
merger time. At higher densities, the different scalar masses lead to
clearly separated mass-ratio evolutions, with the massless and lighter
fields generally producing the largest departures from the initial
$q$ at comparable stages of the inspiral. Thus, scalar accretion changes
not only the total masses of the components but also the intrinsic
asymmetry of the binary. This provides an important indirect mechanism
for modifying the gravitational-wave linear-momentum flux and the final
recoil, in addition to the linear momentum carried directly by the scalar
field, as discussed in Sec.~\ref{sec:recoil}.

\begin{figure}
	\includegraphics[width=0.7\linewidth]{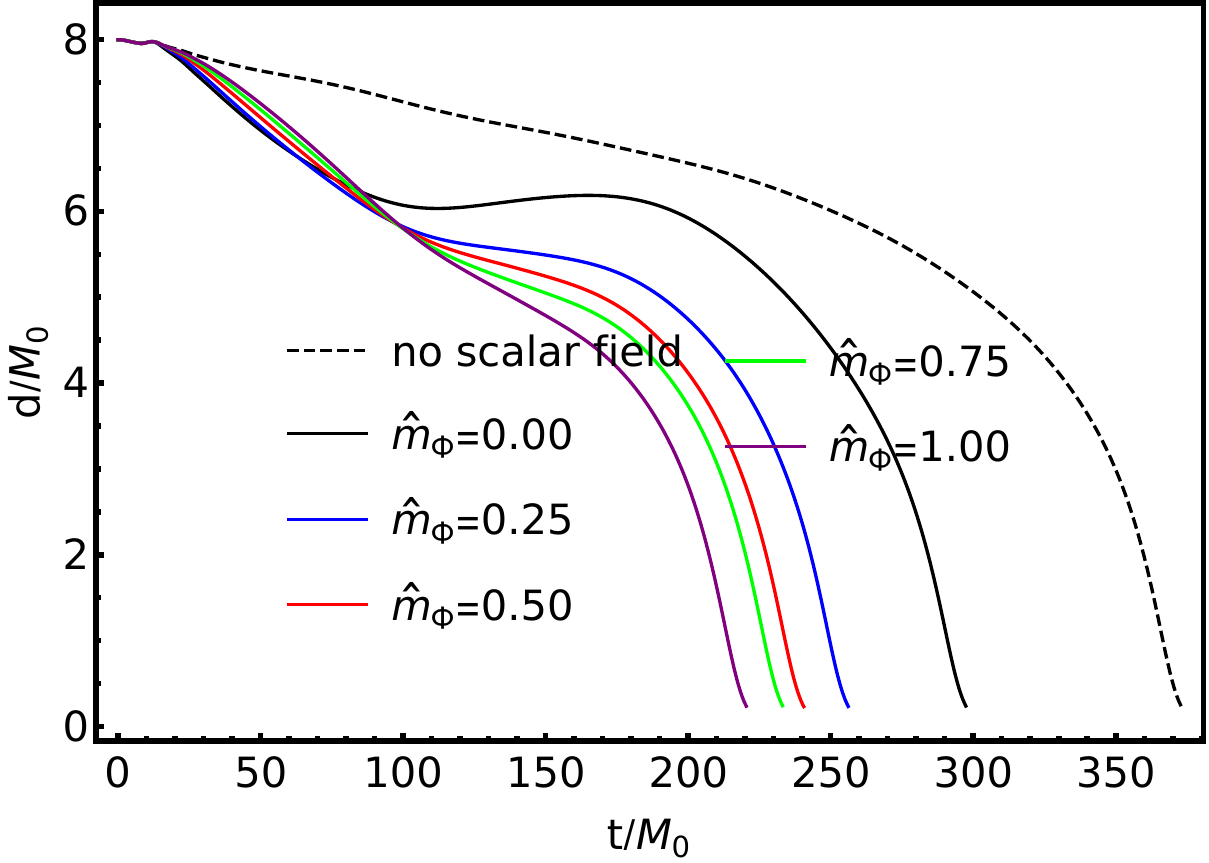}
	\caption{Coordinate separation $d/M_0$ between the two black holes as a function of time for $q=1$ and $\Pi_0=0.005$. The colored curves correspond to $\hat m_{\Phi}=0$, $0.25$, $0.50$, $0.75$, and $1.00$, while the dashed black curve shows the vacuum binary.}
	\label{fig:separation}
\end{figure}

The modification of the orbital dynamics is illustrated in
Fig.~\ref{fig:separation}, where we show the coordinate separation
between the two black holes for a representative equal-mass binary,
$q=1$, with the representative low-density amplitude
$\Pi_0=0.005$. To avoid overcrowding the figure, we only display this
representative set while varying the scalar field mass
$\hat m_{\Phi}=0,\,0.25,\,0.50,\,0.75,$ and $1.00$.
At early times, the binaries follow very similar trajectories, while
clear differences gradually develop during the subsequent evolution.
Within the parameter range considered here, the merger occurs
progressively earlier as $\hat m_{\Phi}$ increases: the massless case
has the longest inspiral, whereas the $\hat m_{\Phi}=1.00$
configuration merges first.

This behavior demonstrates that the scalar field mass can strongly
affect the orbital evolution even when the initial scalar field
amplitude and binary parameters are kept fixed. Qualitatively, this is
consistent with previous numerical-relativity studies showing that
wave-like scalar environments can produce significant dephasing and
shifts in the merger time, with the magnitude of the effect depending
sensitively on the scalar field mass
\cite{Aurrekoetxea:2023jwk,Cheng:2025xxx}.
In particular, Ref.~\cite{Aurrekoetxea:2023jwk} found that the
interaction becomes especially important when the characteristic
Compton wavelength of the scalar field is comparable to the binary
orbital scale. Together with the scalar field distributions shown in
Fig.~\ref{fig:scalar_evolution}, our results suggest that the different
propagation and localization properties associated with $m_{\Phi}$
modify the exchange of energy and angular momentum between the scalar
field and the binary, thereby changing the inspiral rate and merger
time. The relative importance of accretion and radiative losses will
be quantified below.

A convenient quantity for characterizing this effect is the merger
time $t_{\rm m}$, which we define as the time at which a common
apparent horizon is first detected. We measure the shift relative to
the corresponding vacuum binary through
\begin{equation}
    \Delta t_{\rm m}
    =
    t_{\rm m}
    -
    t_{\rm m}^{\rm vac}.
\label{eq:delta_tm}
\end{equation}
Thus, $\Delta t_{\rm m}<0$ indicates an accelerated merger, whereas
$\Delta t_{\rm m}>0$ corresponds to a delayed merger.

\begin{figure*}
    \centering
    \includegraphics[width=0.245\linewidth]{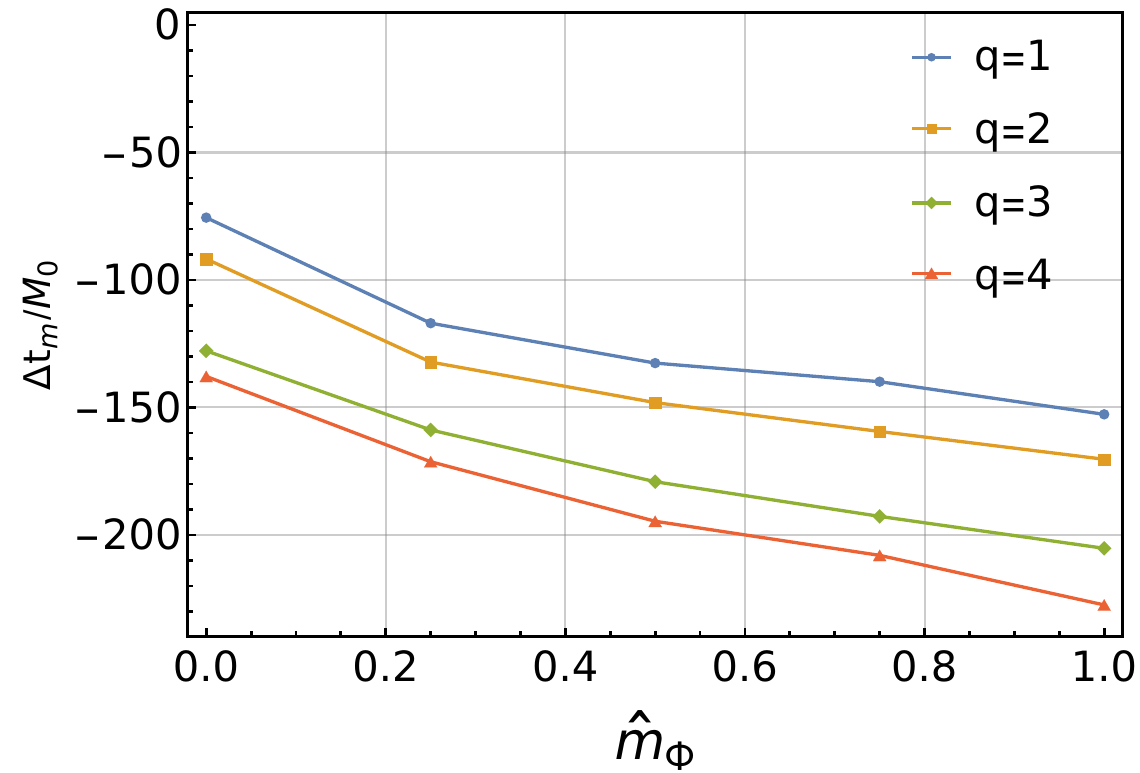}
    \includegraphics[width=0.245\linewidth]{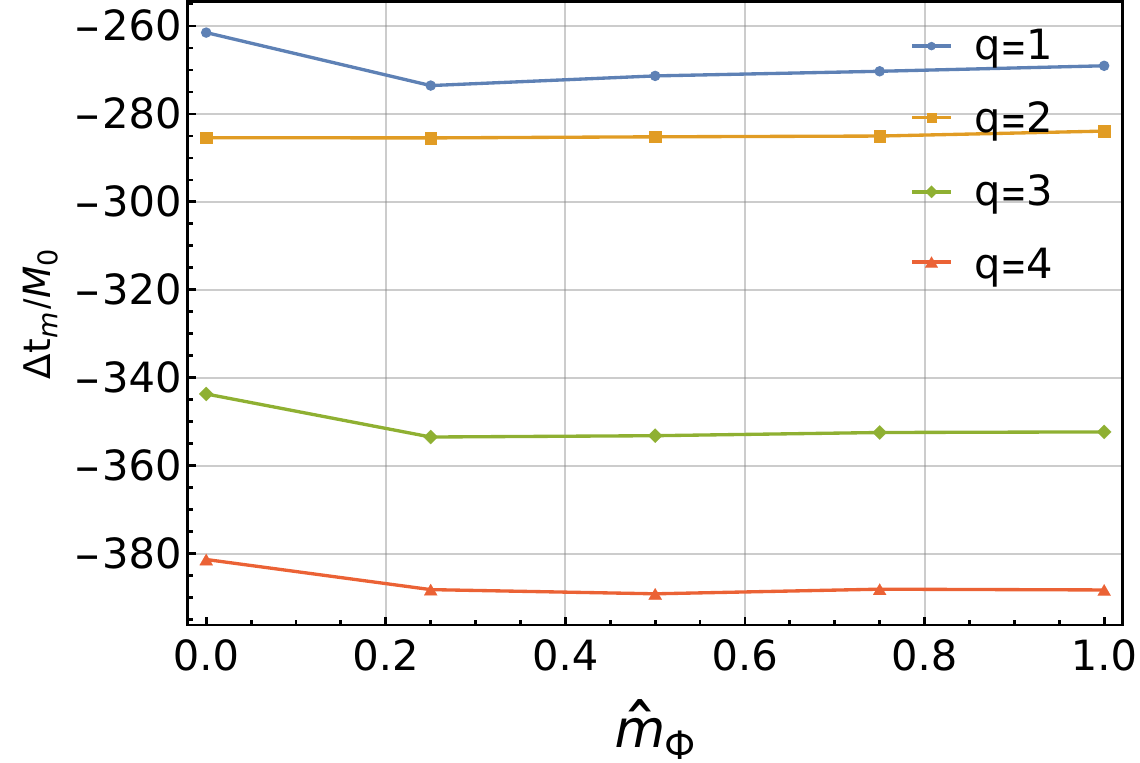}
    \includegraphics[width=0.245\linewidth]{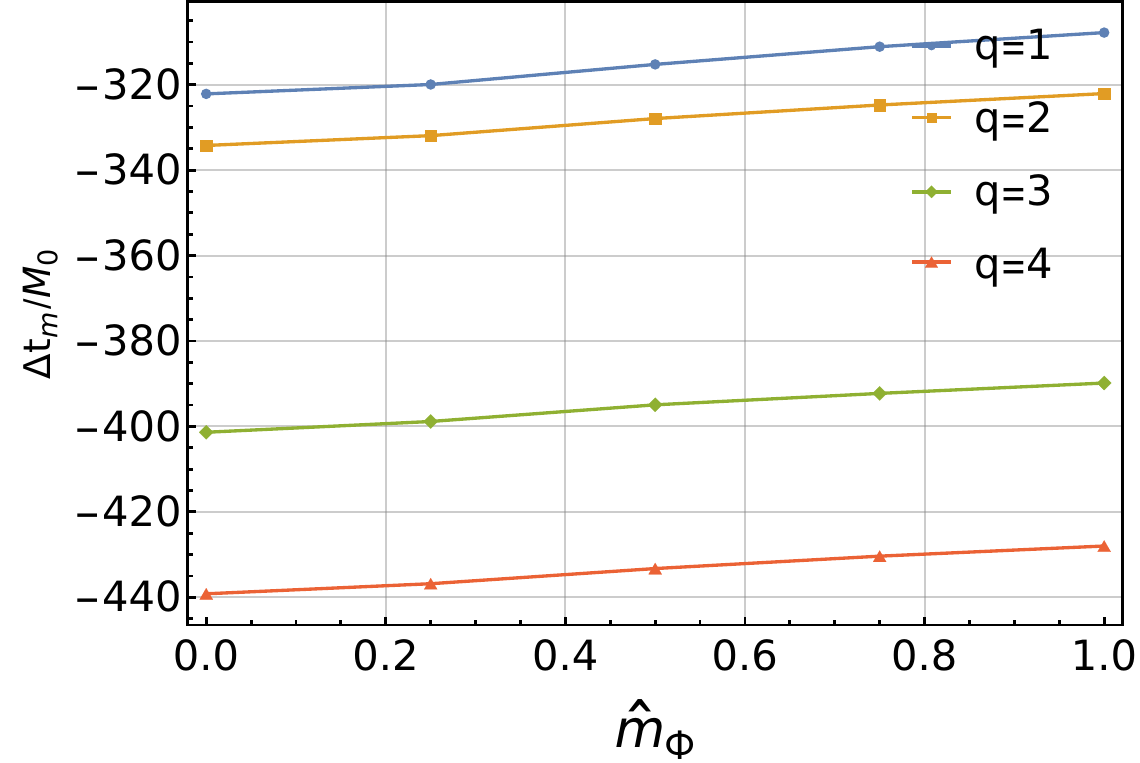}
    \includegraphics[width=0.245\linewidth]{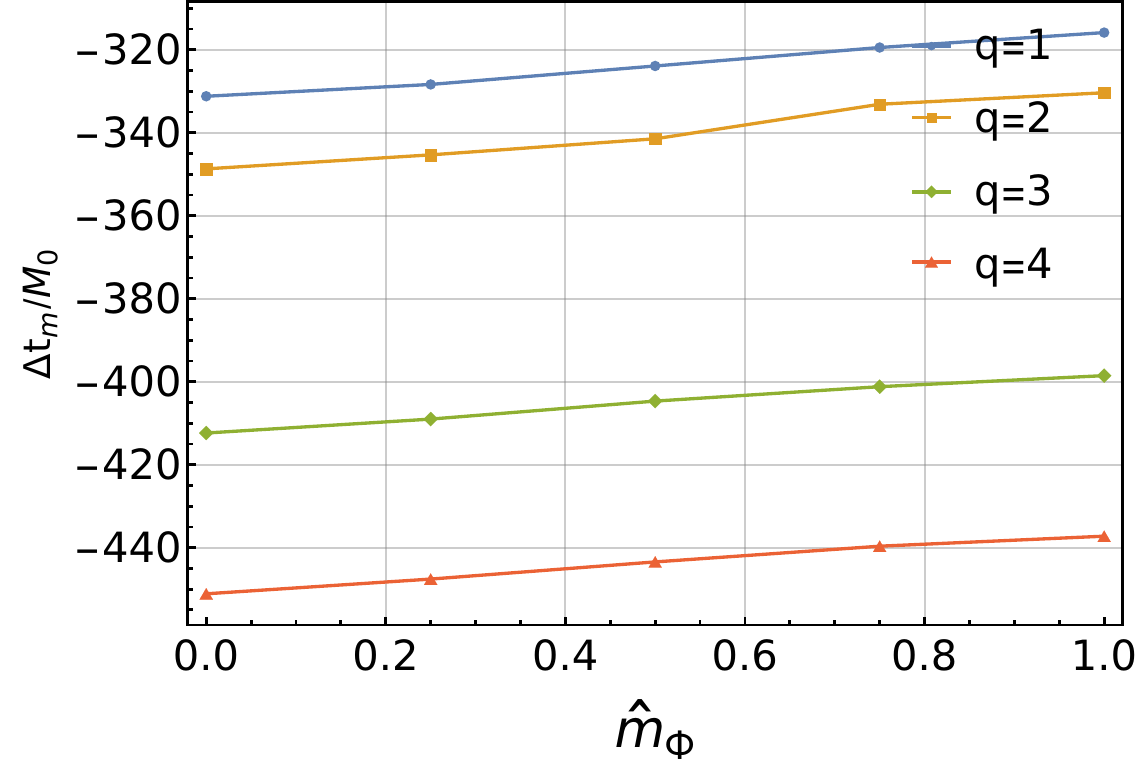}
    \caption{
    Shift of the merger time relative to the corresponding vacuum
    binary, $\Delta t_{\rm m}/M_0=(t_{\rm m}-t_{\rm m}^{\rm vac})/M_0$,
    as a function of the scalar field mass $\hat m_{\Phi}$.
    From left to right, the panels correspond to
    $\Pi_0=0.0025$, $0.0050$, $0.0100$, and $0.0150$, respectively.
    In each panel, the curves show $q=1$, $2$, $3$, and $4$.
    Negative values indicate an earlier merger than in vacuum.
    }
    \label{fig:merger_time}
\end{figure*}

Figure~\ref{fig:merger_time} summarizes $\Delta t_{\rm m}$ for the full
set of mass ratios and for four representative scalar amplitudes. All
configurations shown have $\Delta t_{\rm m}<0$, demonstrating that the
scalar environment accelerates the merger throughout this part of the
parameter space. At the lowest amplitude, $\Pi_0=0.0025$, the mass
dependence is especially clear: increasing $\hat m_{\Phi}$ makes
$\Delta t_{\rm m}$ progressively more negative for all four mass
ratios. This trend is consistent with Fig.~\ref{fig:scalar_evolution},
where more massive fields remain concentrated in the strong-field
region for longer times.

As $\Pi_0$ is increased, the density-driven acceleration becomes much
stronger and the additional dependence on $\hat m_{\Phi}$ becomes
weaker and, for some mass ratios, nonmonotonic. The comparison therefore
shows that the field mass and density do not act as independent
perturbations. Instead, the characteristic scalar length scale and the
amount of scalar matter jointly determine how efficiently the scalar
environment exchanges energy and angular momentum with the BBH.

The scalar field evolution, differential accretion, mass-ratio evolution,
orbital separation, and merger-time shifts therefore give a consistent
dynamical picture. The surrounding scalar field can substantially shorten
the inspiral while simultaneously changing the component masses and, for
unequal-mass binaries, the effective binary asymmetry. The detailed
response depends jointly on $\Pi_0$ and $\hat m_{\Phi}$. We now examine
how these changes appear in the gravitational-wave signal.

\subsection{Gravitational-wave signals}
\label{sec:waveforms}

We now examine how the surrounding scalar field modifies the
gravitational-wave emission from the binary. As shown in the previous
subsection, the scalar environment can significantly alter the orbital
evolution and merger time. These changes are directly reflected in the
gravitational-wave signal through shifts in the merger time, waveform
phase, and amplitude.

\begin{figure*}
    \centering
    \includegraphics[width=0.245\linewidth]{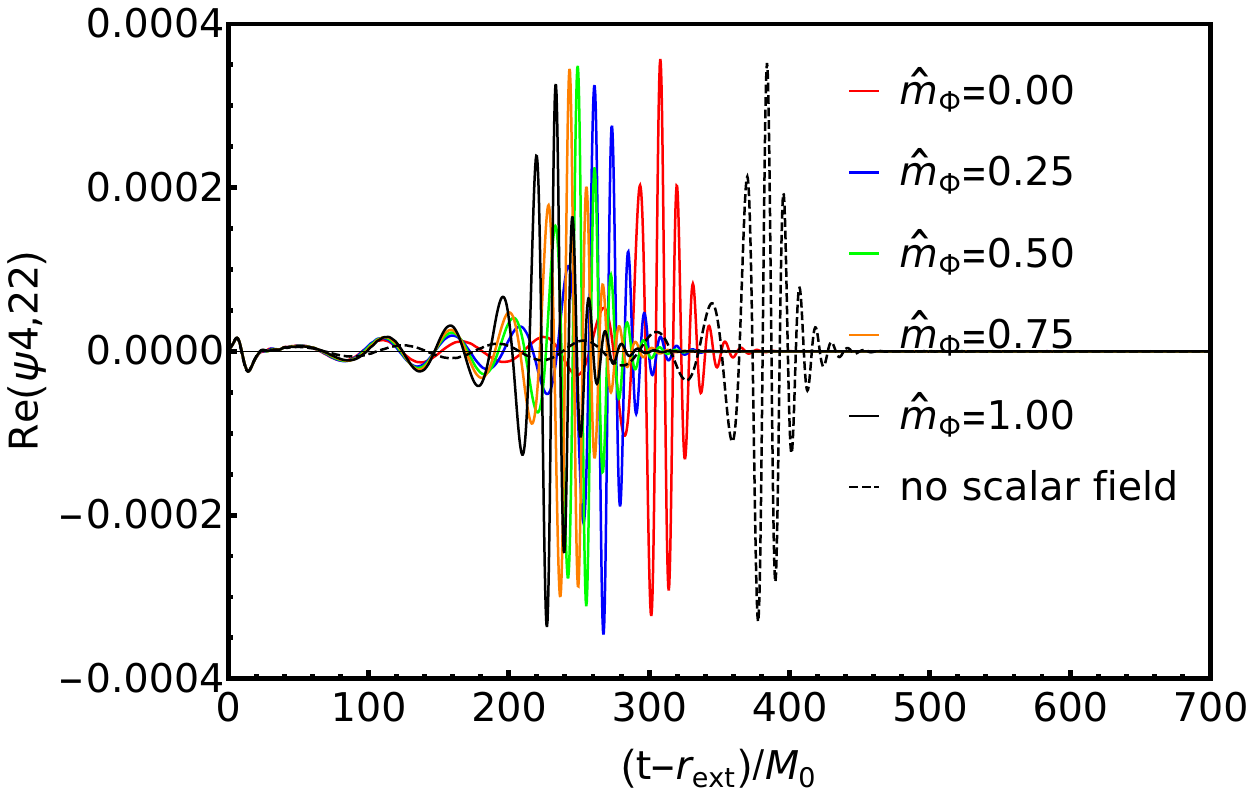}
    \includegraphics[width=0.245\linewidth]{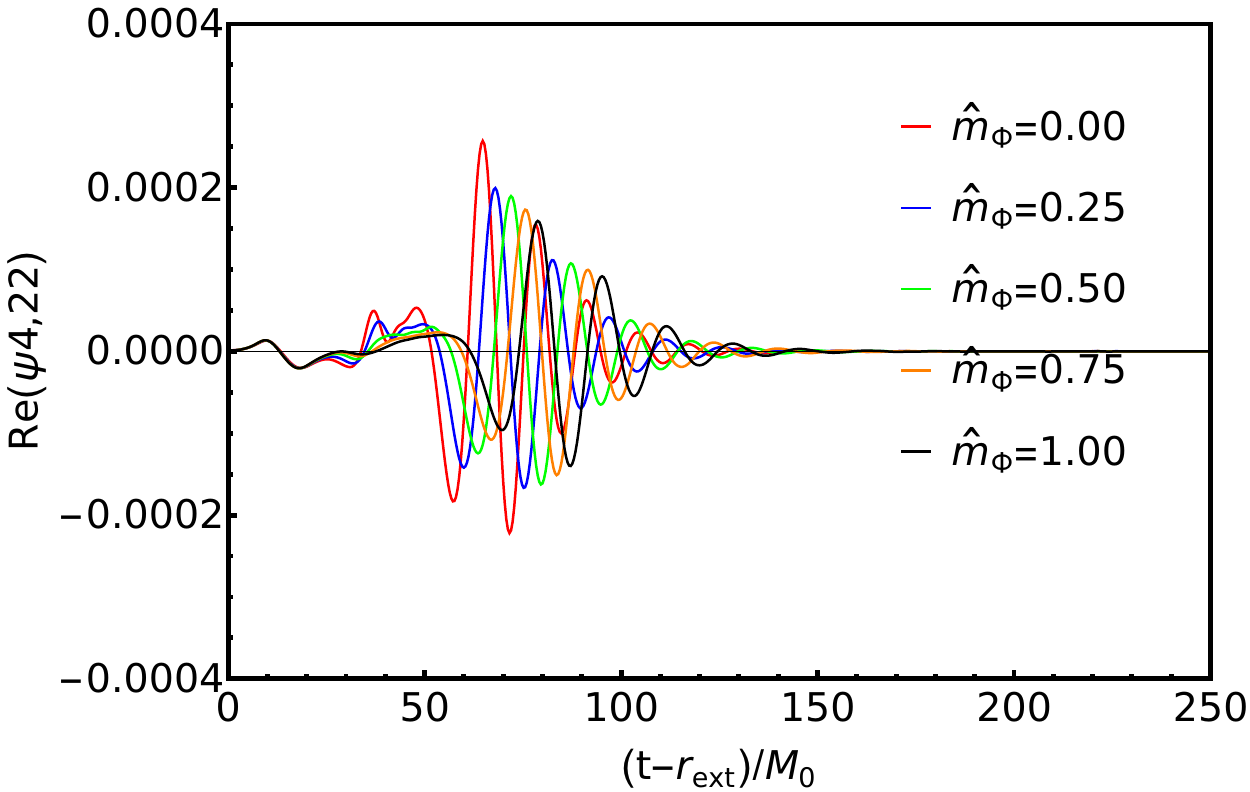}
    \includegraphics[width=0.245\linewidth]{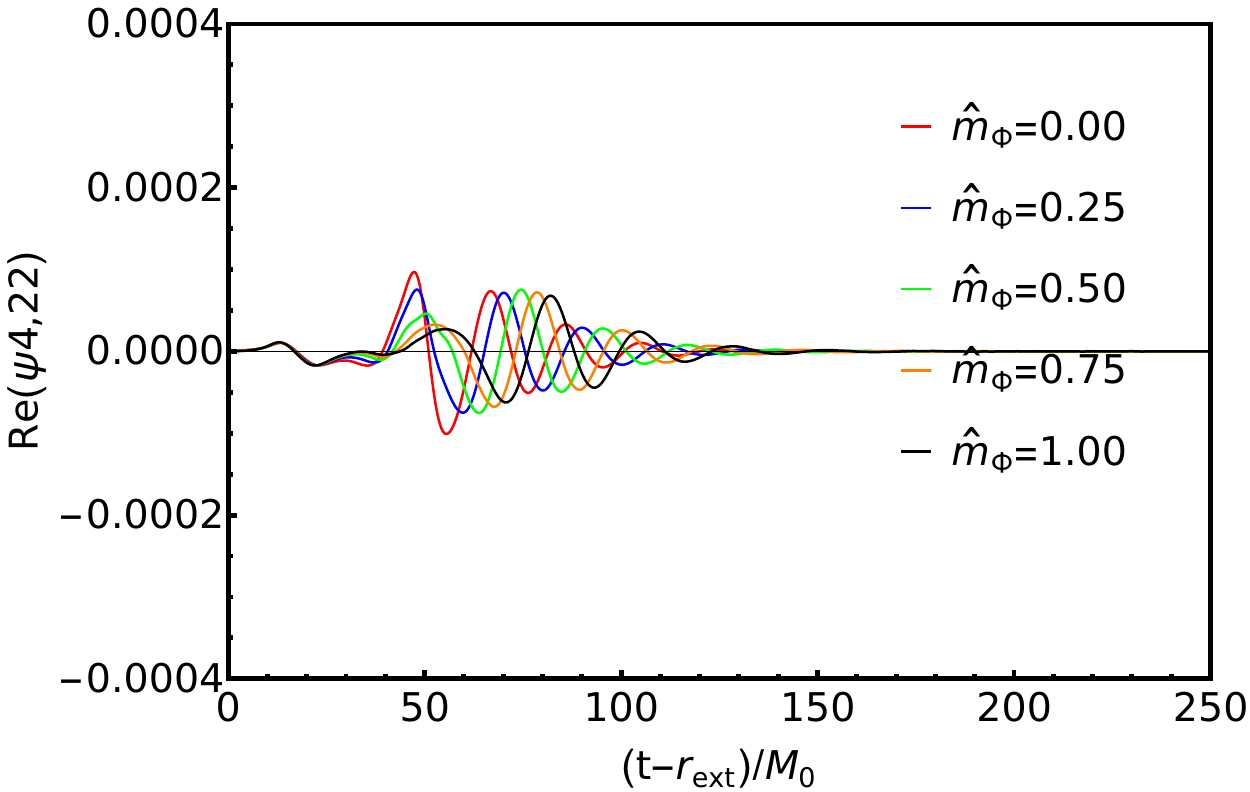}
    \includegraphics[width=0.245\linewidth]{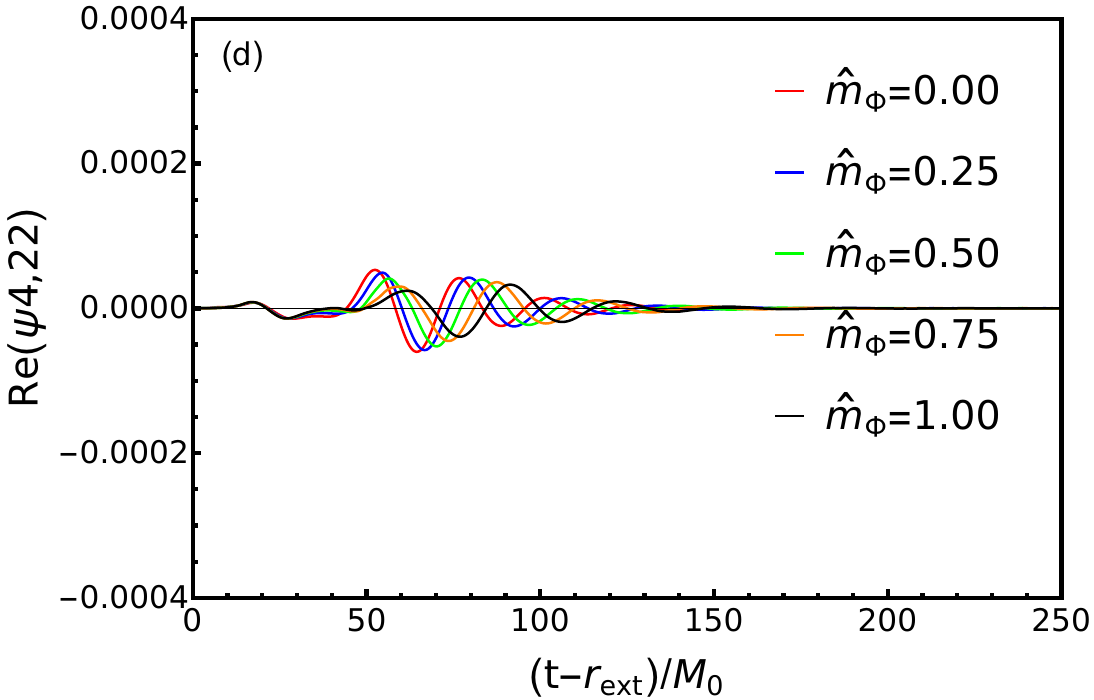}
    \caption{
    Dominant $(\ell,m)=(2,2)$ mode of $\Psi_4$ for equal-mass BBH
    mergers, $q=1$, with different scalar field masses
    $\hat m_{\Phi}=0$, $0.25$, $0.50$, $0.75$, and $1.00$.
    From left to right, the initial scalar field amplitudes are
    $\Pi_0=0.0025$, $0.0050$, $0.0100$, and $0.0150$,
    respectively. The vacuum waveform is shown in black when it lies
    within the displayed time window. The waveforms are extracted at
    $r_{\rm ext}=200\,M_0$ and plotted against the retarded time
    $(t-r_{\rm ext})/M_0$.
    }
    \label{fig:gw_mass}
\end{figure*}

The gravitational radiation is characterized by the Newman--Penrose
scalar $\Psi_4$, decomposed into spin-weighted spherical harmonics as
described in Sec.~\ref{sec:diagnostics}. Here we focus on the dominant
$(\ell,m)=(2,2)$ mode. Since the full parameter survey contains a large
number of configurations, displaying all waveforms would make the
comparison unnecessarily complicated. We therefore show in
Fig.~\ref{fig:gw_mass} representative results for the equal-mass
binary, $q=1$, while varying both the scalar field mass and its initial
amplitude.

The waveforms are extracted at a fixed coordinate radius
$r_{\rm ext}=200\,M_0$ and are plotted as functions of the retarded
time $(t-r_{\rm ext})/M_0$. From left to right,
Fig.~\ref{fig:gw_mass} corresponds to
$\Pi_0=0.0025$, $0.0050$, $0.0100$, and $0.0150$,
respectively. Within each panel, the five curves correspond to
$\hat m_{\Phi}=0$, $0.25$, $0.50$, $0.75$, and $1.00$.
The corresponding vacuum waveform is also included as a reference
where it lies within the displayed time range. As discussed in
Sec.~\ref{sec:diagnostics}, the waveform itself is not extrapolated
to infinity.

Figure~\ref{fig:gw_mass} shows that both the scalar field mass and
density have a pronounced influence on the gravitational-wave signal.
For the two lower-density configurations,
$\Pi_0=0.0025$ and $0.0050$, increasing $\hat m_{\Phi}$ shifts the
merger signal to earlier times, consistent with the accelerated
orbital evolution discussed in Sec.~\ref{sec:dynamics}. Even for the
smallest amplitude, the presence of the scalar field produces a
substantial shift relative to the vacuum waveform, demonstrating that
the merger time is sensitive to the scalar environment.

As the initial scalar field amplitude is increased further, the binary
merges much earlier and the waveform changes more significantly.
Interestingly, the dependence on $\hat m_{\Phi}$ is no longer described
by a simple monotonic trend at the larger amplitudes. For
$\Pi_0=0.0100$ and $0.0150$, the relative ordering of the merger signals
with scalar field mass differs from that found at lower density. This
indicates that the effects of the scalar field mass and density are
not independent: changing the amount of scalar matter can alter how
the characteristic scale $m_{\Phi}^{-1}$ couples to the binary
dynamics.

The waveform amplitude also changes noticeably as the scalar field
density increases. In the higher-density cases, the peak amplitude of
the $(2,2)$ mode is generally reduced compared with the lower-density
configurations. This behavior is consistent with the strong
modification of the orbital evolution: a faster merger leaves less time
for the binary to radiate through the standard inspiral channel, while
part of the energy and angular momentum can also be exchanged with the
surrounding scalar field. A quantitative assessment of this
redistribution requires the corresponding gravitational and scalar
energy fluxes, which will be discussed in the next subsection.

These results demonstrate that the gravitational-wave modification
cannot be characterized by either $\hat m_{\Phi}$ or $\Pi_0$ alone.
Instead, the waveform reflects their combined influence on the binary
dynamics. The full dependence on the binary mass ratio and scalar field
parameters will be quantified below using the merger time and the
radiated energy, angular momentum, and linear momentum.

\subsection{Energy and angular-momentum radiation}
\label{sec:energy_angular}

We next quantify how the scalar environment modifies the energy and
angular-momentum budget of the binary. While the gravitational-wave
signals discussed in the previous subsection provide a direct view of
the changes in the binary dynamics, the accumulated radiative
quantities allow us to further determine how the energy and angular
momentum of the system are redistributed among gravitational radiation,
scalar field propagation, and the central black hole region during the
inspiral and merger.

For the gravitational-wave channel, we calculate the accumulated
radiated energy and angular momentum at a fixed finite extraction
radius $r=200\,M_0$. Since the same extraction radius is used for all
configurations, these quantities provide a consistent comparison of
their temporal evolution and relative dependence on the scalar field
parameters. The finite-radius results shown in this subsection are
mainly used for such comparisons. When accurate asymptotic quantities
are required, especially for the final linear-momentum balance and
recoil calculation, results obtained at several extraction radii are
extrapolated to $r\rightarrow\infty$ following the procedure described
in Sec.~\ref{sec:diagnostics}.

For the scalar field, we additionally evaluate the accumulated energy
and angular-momentum fluxes through two representative spheres at
$r=5\,M_0$ and $r=100\,M_0$. The outer surface characterizes the
scalar quantities transported away from the binary, whereas the inner
surface probes the transport of scalar energy and angular momentum in
the strong-field region close to the black holes. The quantities
evaluated at $r=5\,M_0$ should therefore be interpreted as local
transport through the inner surface rather than radiation measured at
infinity.

\subsubsection{Energy transport}

\begin{figure*}
    \centering
    \includegraphics[width=0.245\linewidth]{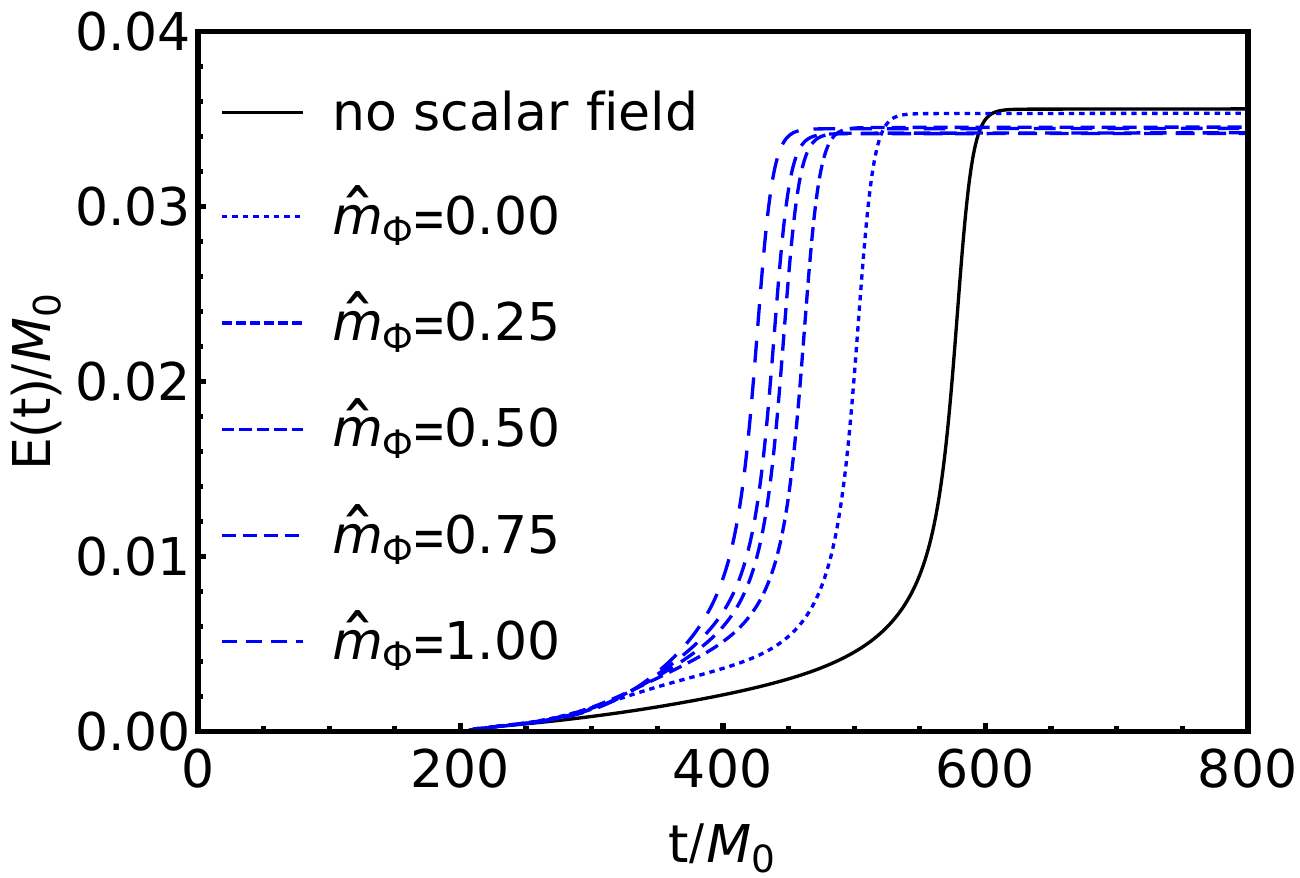}
    \includegraphics[width=0.245\linewidth]{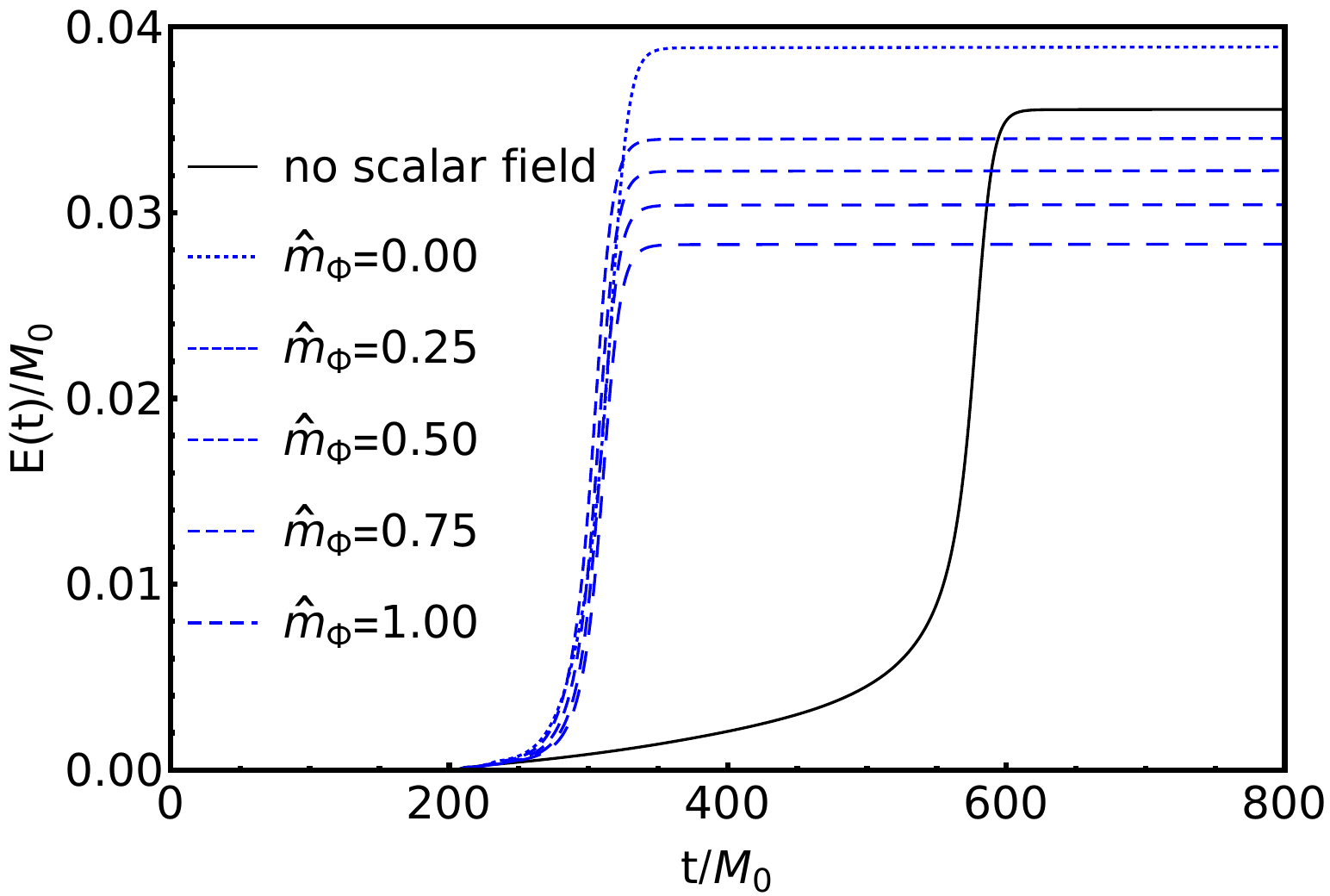}
    \includegraphics[width=0.245\linewidth]{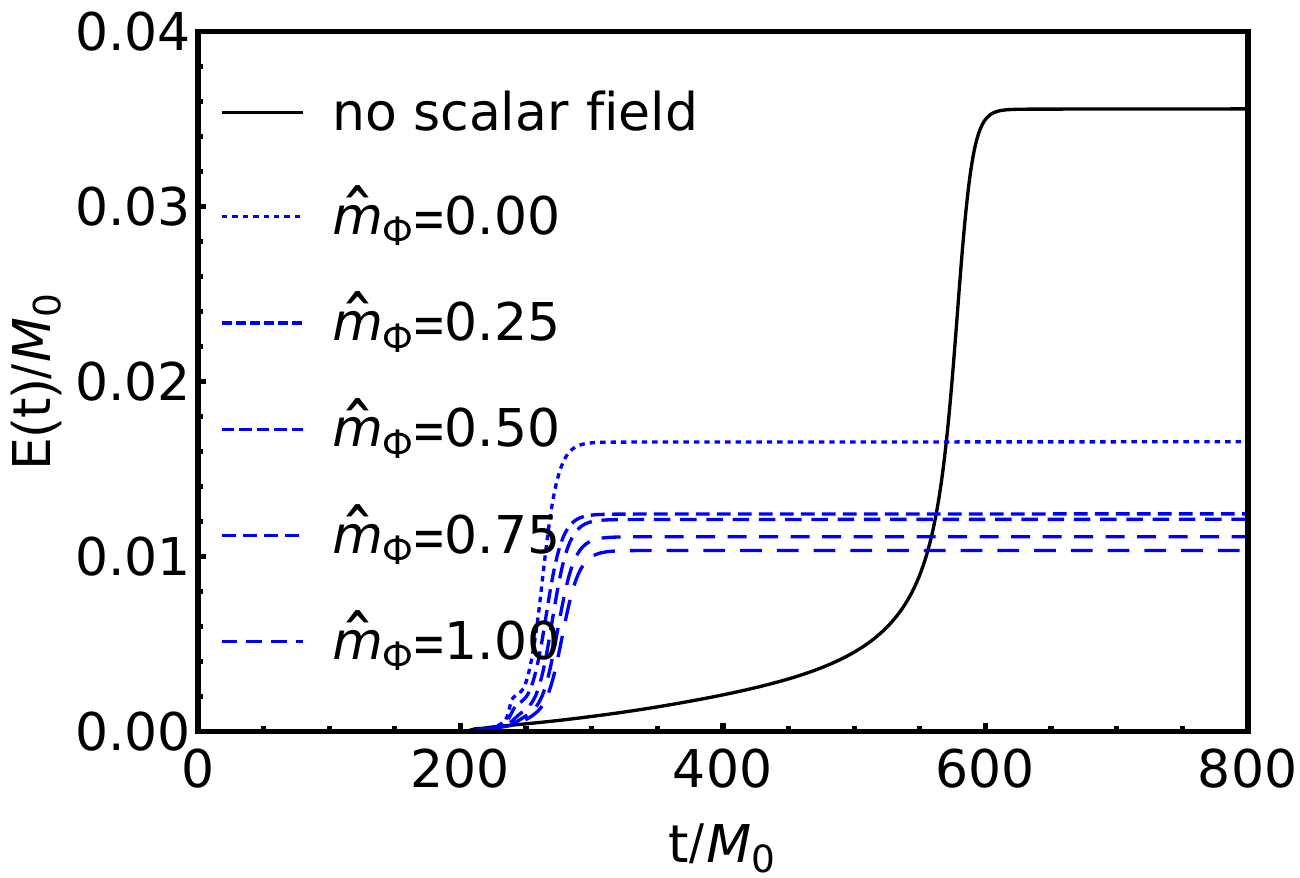}
    \includegraphics[width=0.245\linewidth]{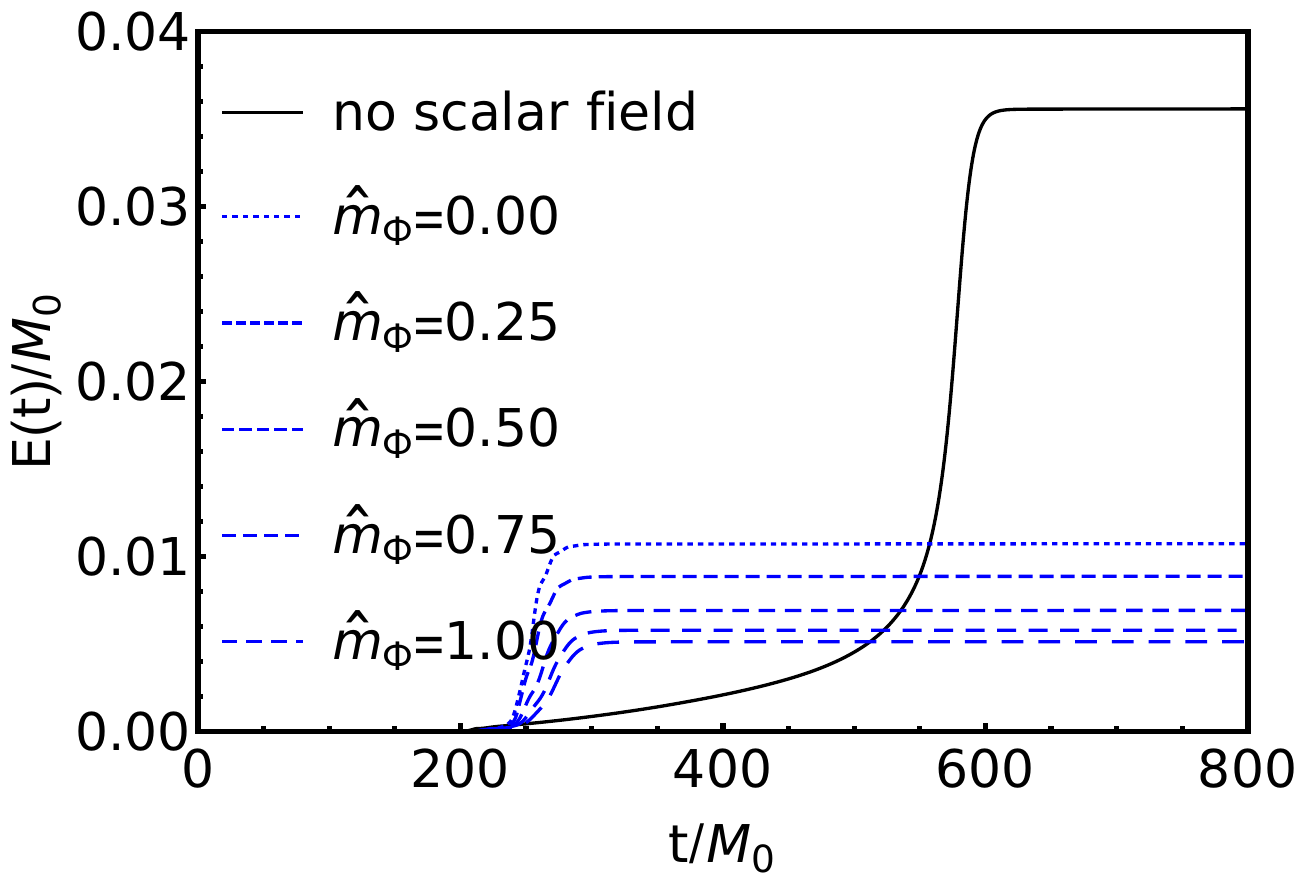}
    \caption{
    Accumulated gravitational-wave energy $E_{\rm GW}(t)/M_0$
    for equal-mass BBH mergers, $q=1$.
    From left to right, the panels correspond to
    $\Pi_0=0.0025$, $0.0050$, $0.0100$, and $0.0150$,
    respectively. Different line styles denote
    $\hat m_{\Phi}=0$, $0.25$, $0.50$, $0.75$, and $1.00$.
    The corresponding vacuum result is shown in black.
    All quantities are evaluated at the fixed extraction radius
    $r=200\,M_0$.
    }
    \label{fig:energy_time}
\end{figure*}

We first consider the accumulated gravitational-wave energy,
\begin{equation}
E_{\rm GW}(t)
=
\int_{t_0}^{t}
\frac{dE_{\rm GW}}{dt'}\,dt' .
\label{eq:EGW_acc}
\end{equation}
Since the complete parameter survey contains a large number of
configurations, we again use the equal-mass binaries, $q=1$, as
representative examples. Figure~\ref{fig:energy_time} shows
$E_{\rm GW}(t)/M_0$ for four different initial scalar field
amplitudes and scalar field mass parameter. Here the
corresponding vacuum evolution is also included for comparison.

Two main effects can be identified in Fig.~\ref{fig:energy_time}.
First, the rapid increase of $E_{\rm GW}(t)$ occurs substantially
earlier in the presence of the scalar field than in the corresponding
vacuum binary. The shift becomes more pronounced as the initial
scalar field amplitude increases, in agreement with the accelerated
inspiral and earlier merger found from the orbital separation and the
gravitational-wave waveform. The scalar field mass further changes the
time at which the main gravitational-wave energy is emitted,
confirming that $m_{\Phi}$ plays an important role in controlling the
binary dynamics.

Second, the final gravitational-wave energy depends strongly on both
$\Pi_0$ and $\hat m_{\Phi}$. At the lowest amplitude,
$\Pi_0=0.0025$, the final values remain relatively close to the vacuum
result despite the considerable shift in merger time. At
$\Pi_0=0.0050$, the dependence on the scalar field mass becomes more
pronounced, while for $\Pi_0=0.0100$ and $0.0150$ the
gravitational-wave energy is strongly suppressed relative to the
vacuum binary for all scalar field masses.

The suppression of $E_{\rm GW}$ at large scalar field amplitudes is
consistent with the substantially shortened inspiral. The binary has
less time to lose energy through the usual gravitational-wave inspiral
channel before merger. In addition, the scalar field provides
alternative channels through which energy can be redistributed,
including outward scalar radiation and inward transport toward the
black holes.

\begin{figure*}
    \centering
    \includegraphics[width=0.24\linewidth]{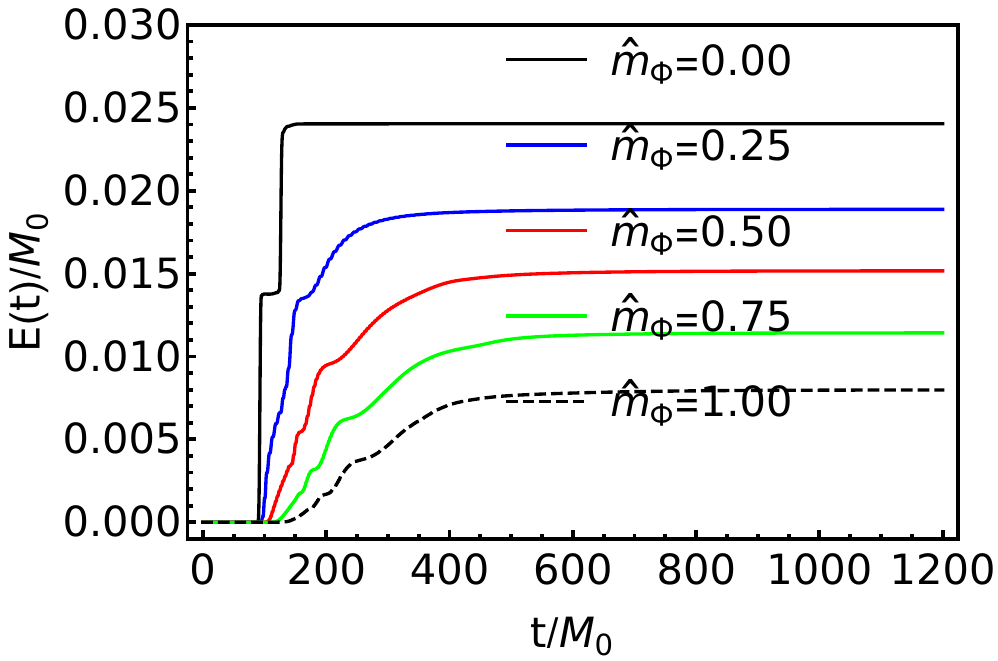}
    \includegraphics[width=0.24\linewidth]{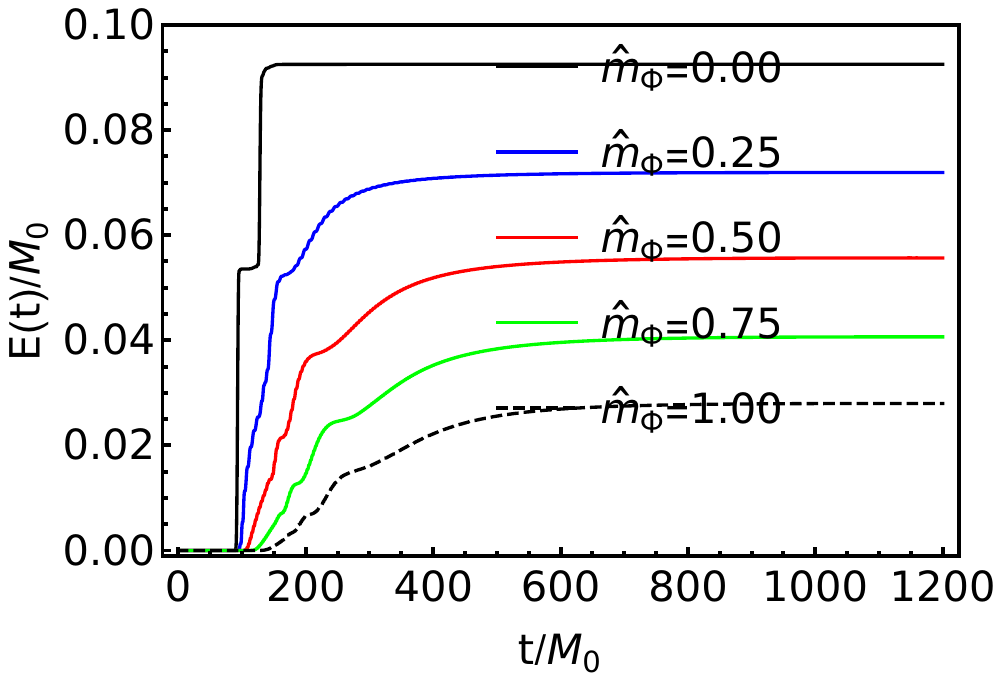}
    \includegraphics[width=0.24\linewidth]{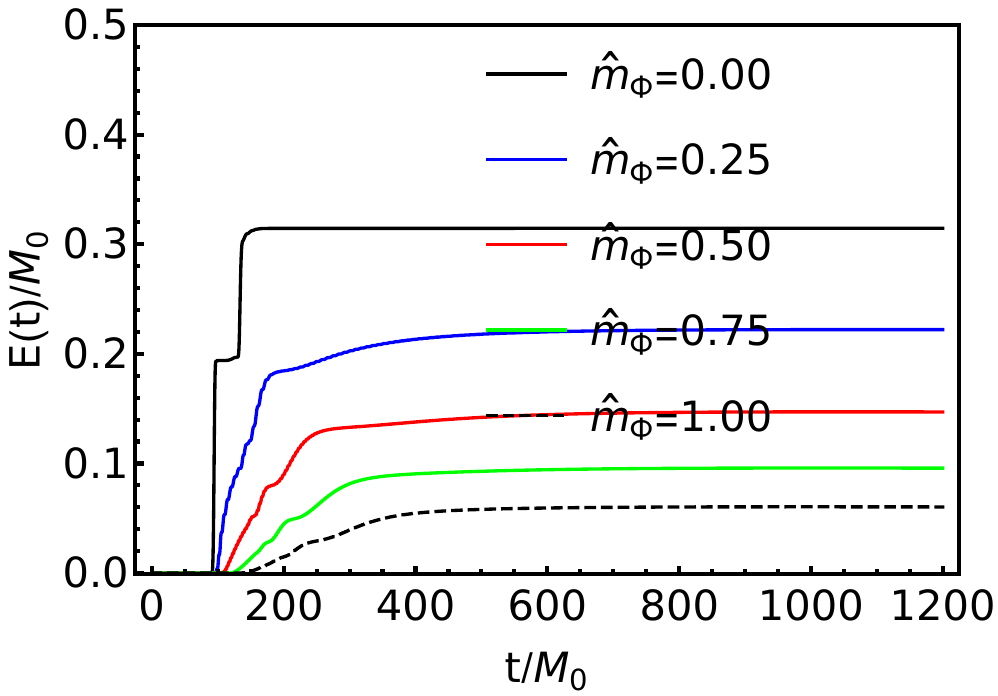}
    \includegraphics[width=0.24\linewidth]{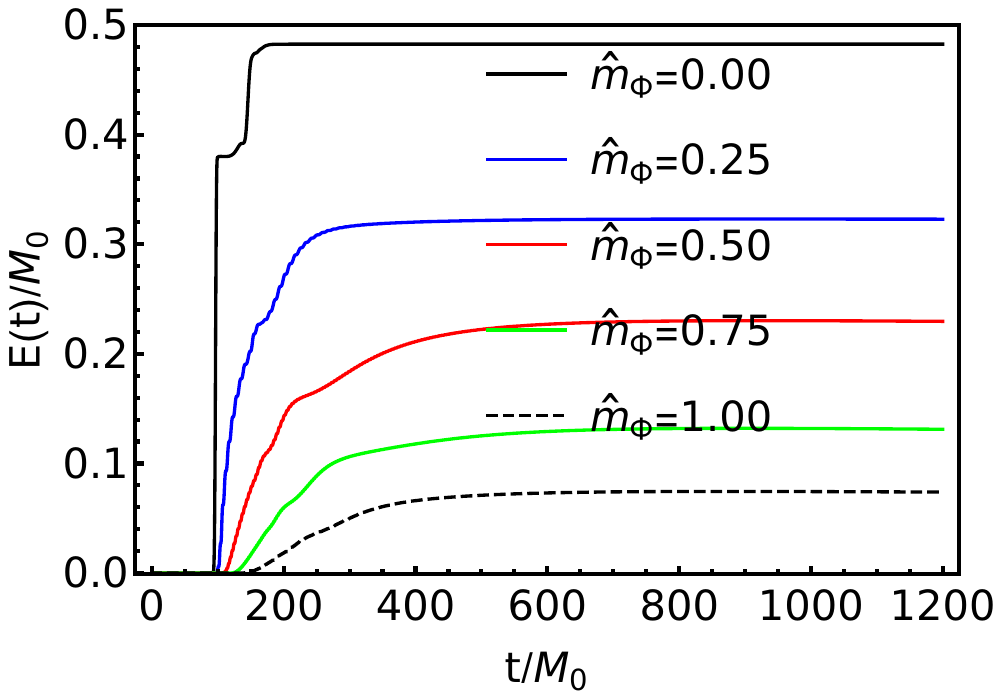}
    \includegraphics[width=0.24\linewidth]{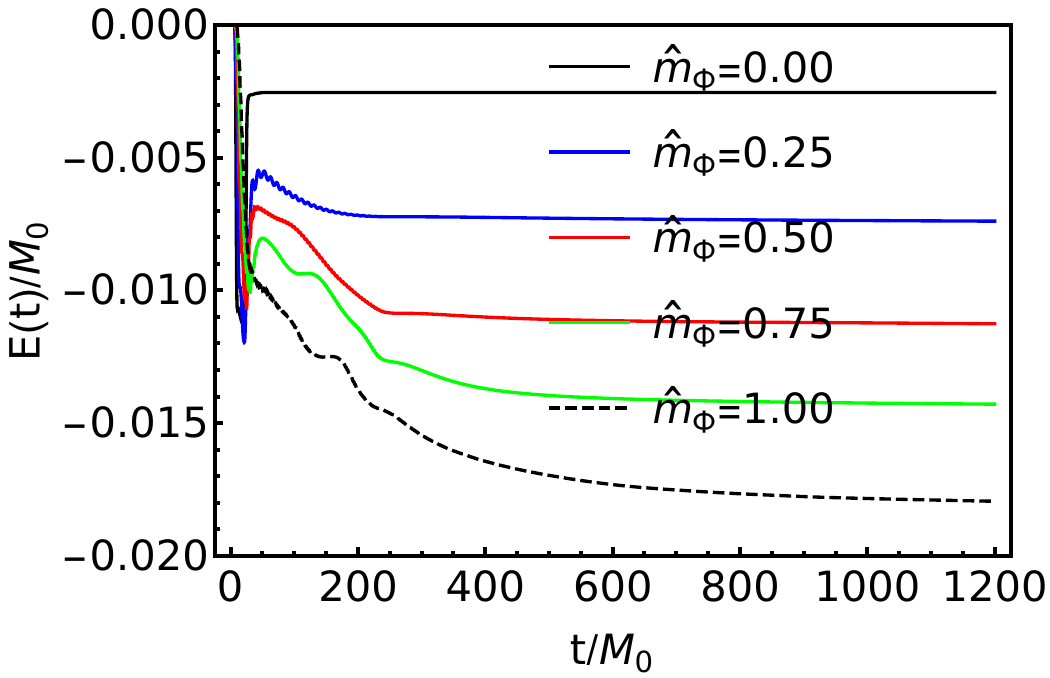}
    \includegraphics[width=0.24\linewidth]{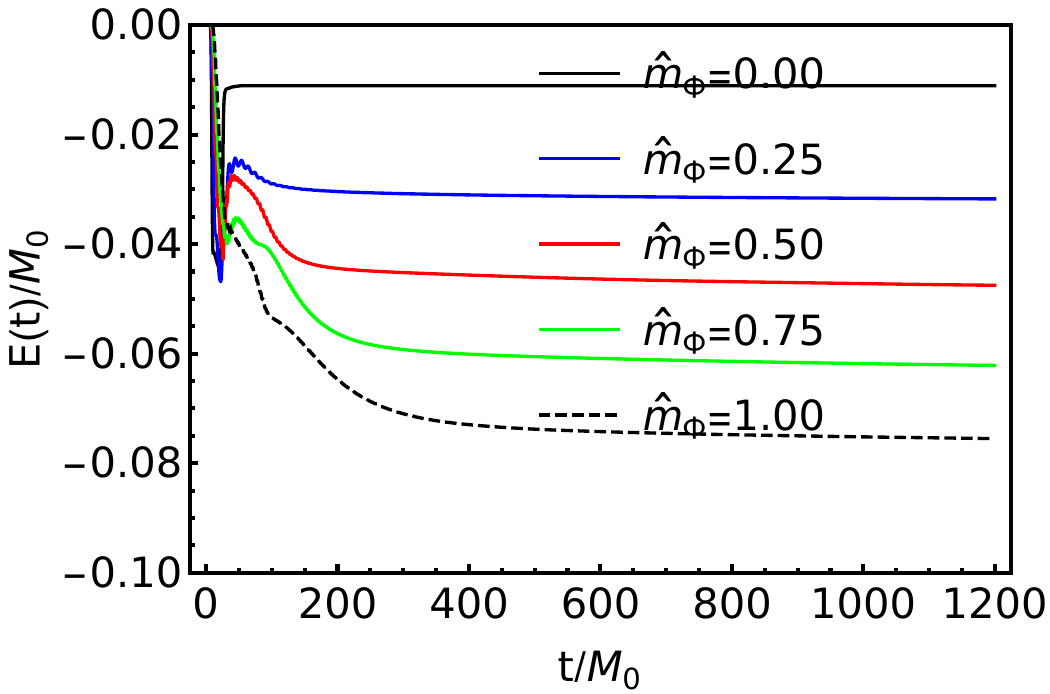}
    \includegraphics[width=0.24\linewidth]{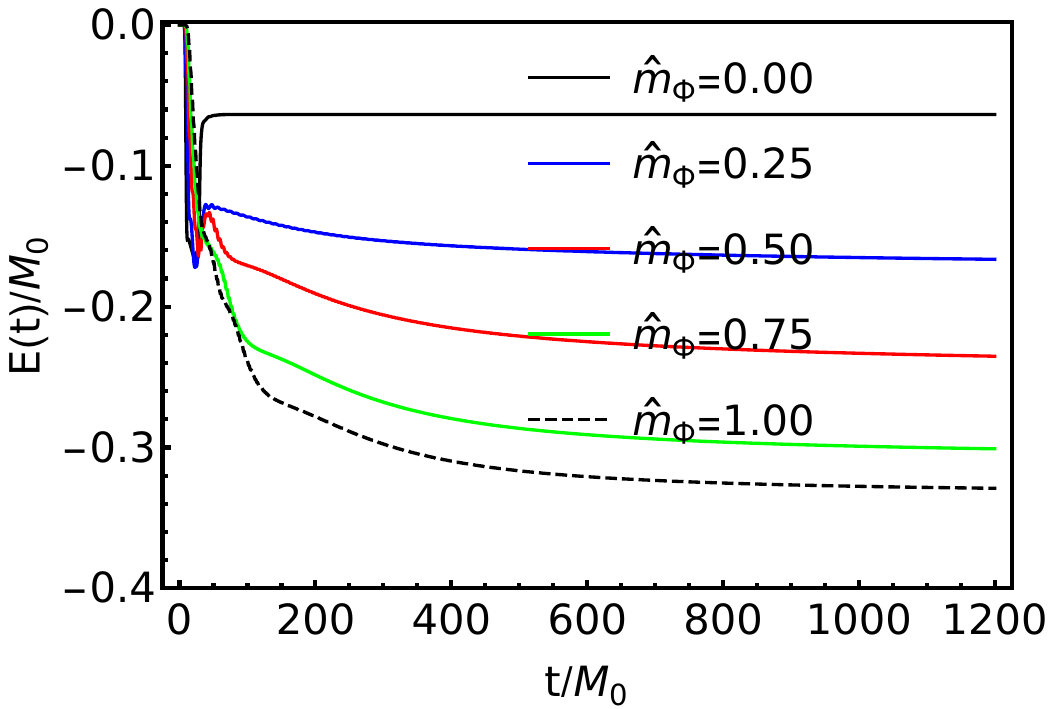}
    \includegraphics[width=0.24\linewidth]{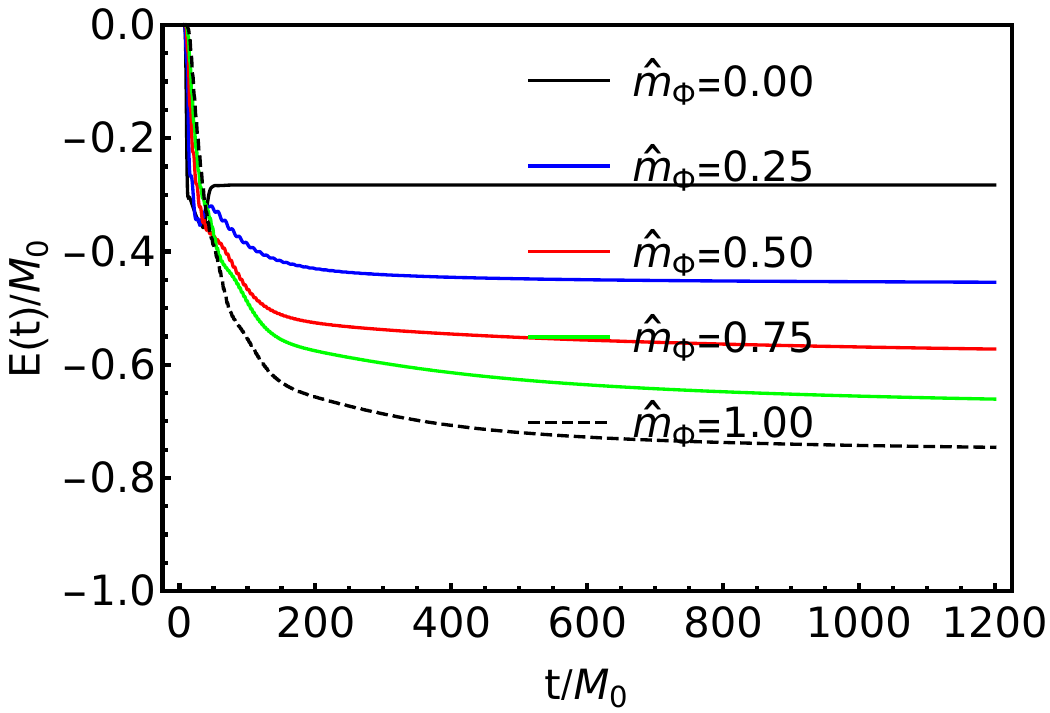}
    \caption{
    Accumulated scalar field energy for equal-mass BBH mergers,
    $q=1$, evaluated at two representative radii.
    The upper and lower rows show the quantities evaluated at
    $r=100\,M_0$ and $r=5\,M_0$, respectively.
    From left to right, the columns correspond to
    $\Pi_0=0.0025$, $0.0050$, $0.0100$, and $0.0150$.
    Different curves denote
    $\hat m_{\Phi}=0$, $0.25$, $0.50$, $0.75$, and $1.00$.
    Positive values at the outer radius represent net outward
    transport of scalar field energy, whereas negative values at the
    inner radius indicate net inward transport toward the central
    strong-field region.
    }
    \label{fig:sf_energy_time}
\end{figure*}

To examine these channels directly, we calculate the accumulated
scalar field energy crossing a sphere of radius $r$,
\begin{equation}
E_{\Phi}(t;r)
=
\int_{t_0}^{t}
\frac{dE_{\Phi}(r,t')}{dt'}\,dt' .
\label{eq:Ephi_acc}
\end{equation}
Figure~\ref{fig:sf_energy_time} shows the results at $r=100\,M_0$ in
the upper row and at $r=5\,M_0$ in the lower row. From left to right,
the columns correspond to $\Pi_0=0.0025$, $0.0050$, $0.0100$, and
$0.0150$.

At $r=100\,M_0$, the accumulated scalar field energy is positive and
approaches a plateau at late times, indicating a net outward transport
of scalar energy. The amount of outward scalar radiation increases
strongly with the initial scalar field amplitude. At fixed $\Pi_0$,
however, the final outward energy decreases systematically with
increasing $\hat m_{\Phi}$. The massless field carries the largest
amount of energy to large radii, whereas the largest scalar field mass
considered here produces the smallest outward contribution.

The opposite behavior is found at $r=5\,M_0$. The accumulated energy
there is negative, showing that the net scalar-energy flux through the
inner sphere is directed toward the central region. Its magnitude
increases markedly with the initial scalar field amplitude and, in
general, also with the scalar field mass. Thus, increasing
$\hat m_{\Phi}$ suppresses long-range scalar radiation while enhancing
the inward transport of scalar field energy.

This behavior is consistent with the scalar field distributions shown
in Fig.~\ref{fig:scalar_evolution}. The massless field propagates
efficiently to large distances, whereas massive fields remain more
strongly localized around the binary and the merger remnant. The
scalar field mass therefore controls not only the dynamical
interaction with the binary but also the partition of scalar energy
between outward propagation and the strong-field region.

\subsubsection{Angular-momentum transport}

\begin{figure*}
    \centering
    \includegraphics[width=0.24\linewidth]{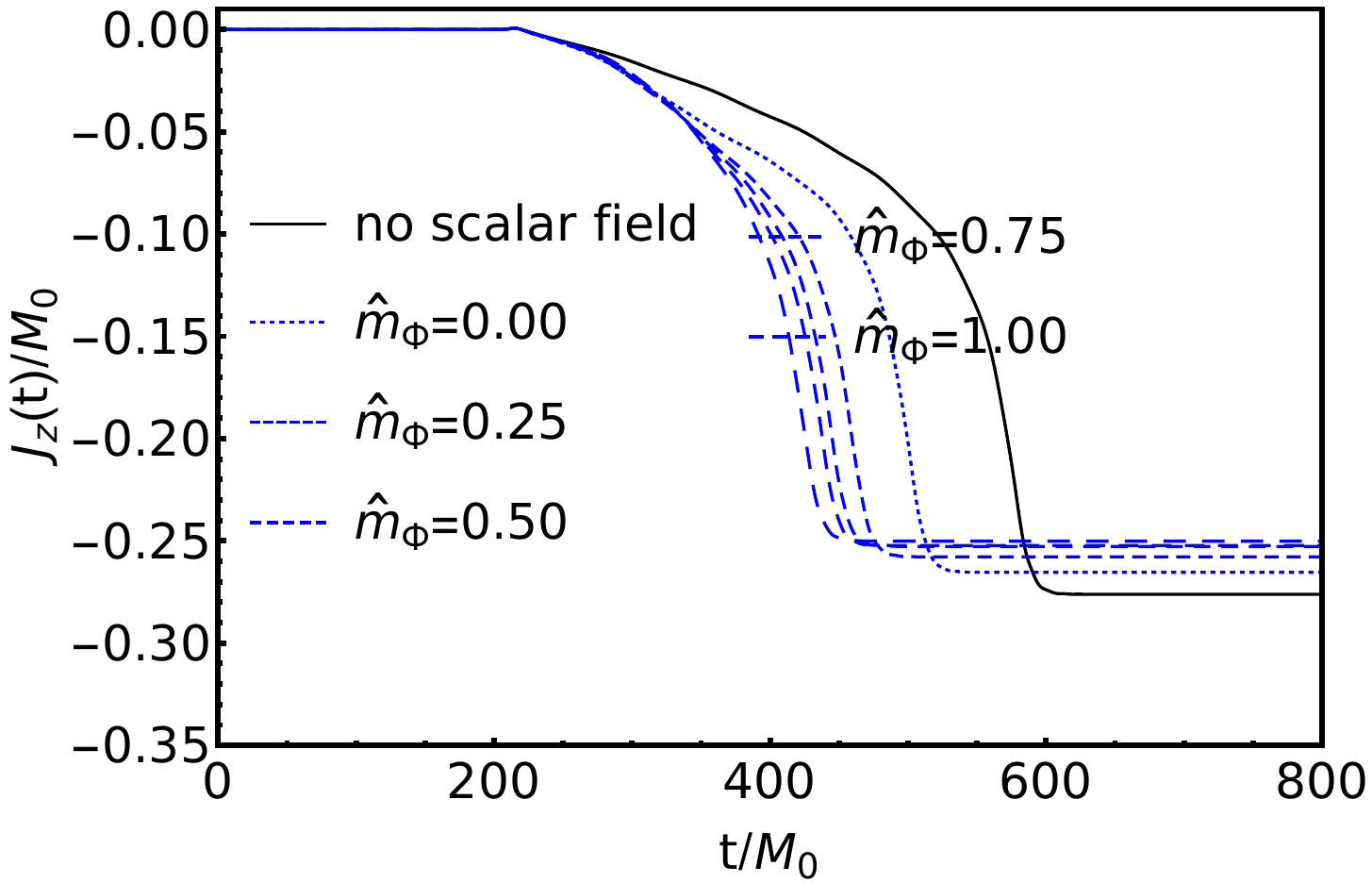}
    \includegraphics[width=0.24\linewidth]{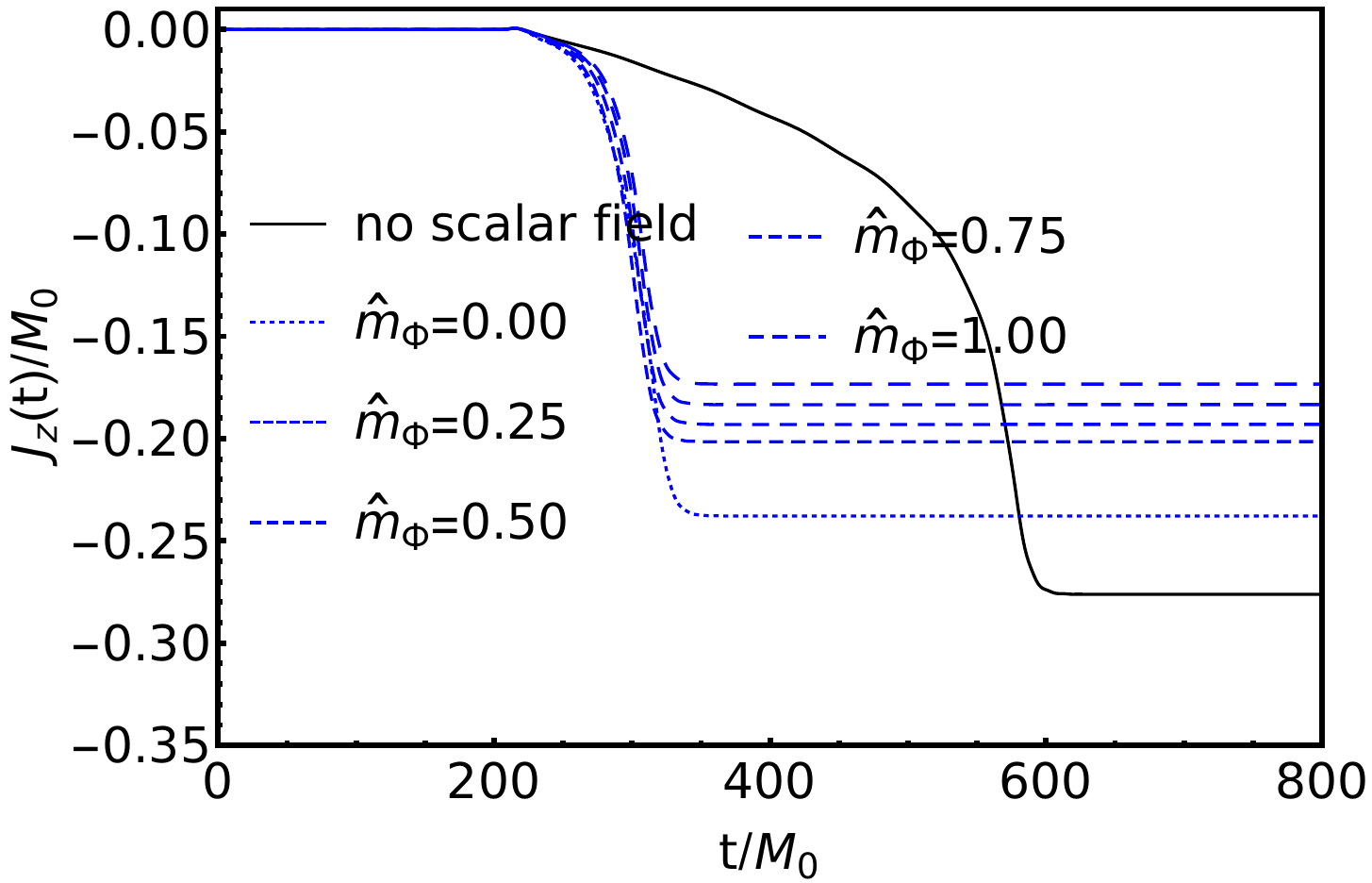}
    \includegraphics[width=0.24\linewidth]{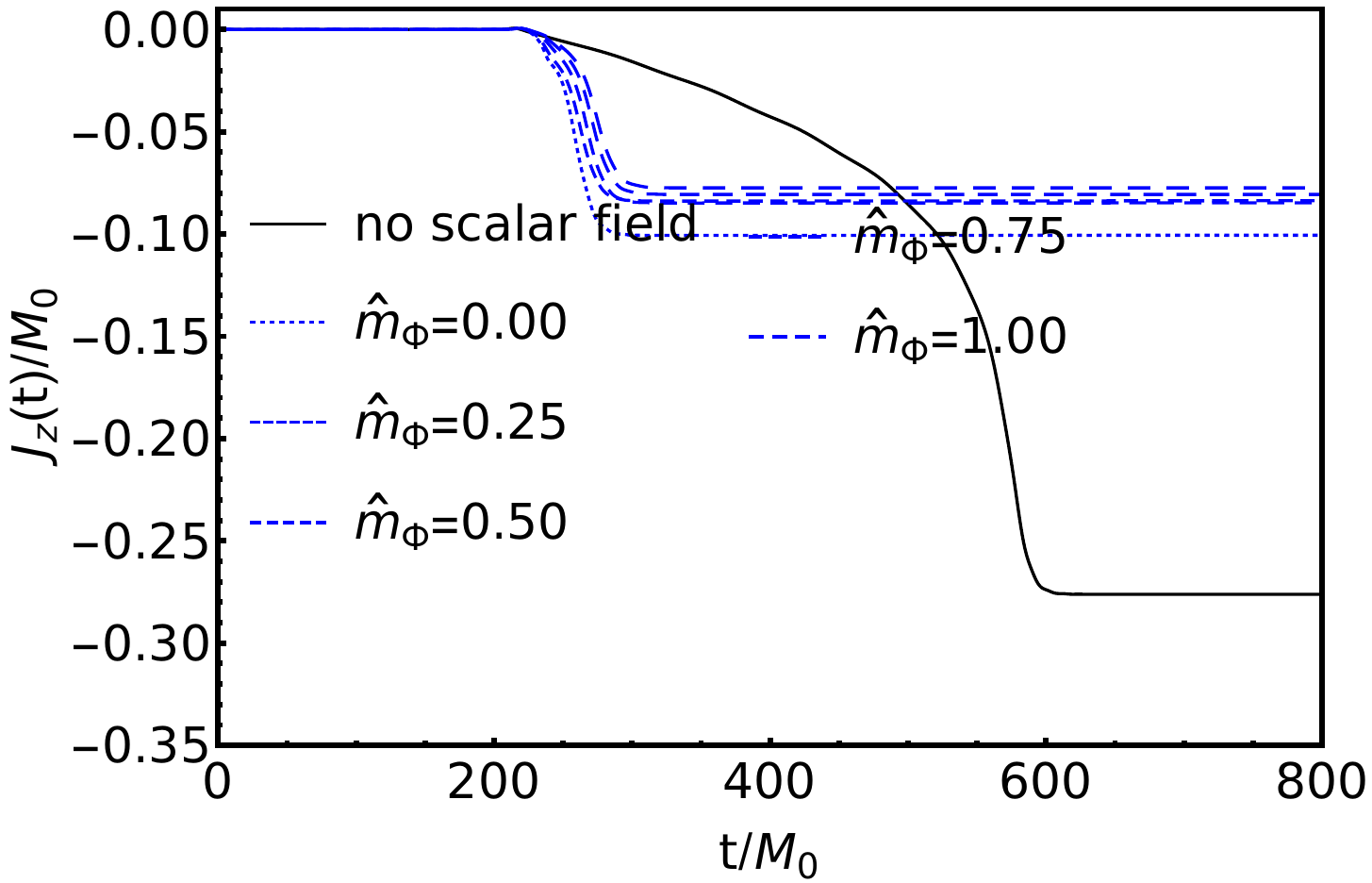}
    \includegraphics[width=0.24\linewidth]{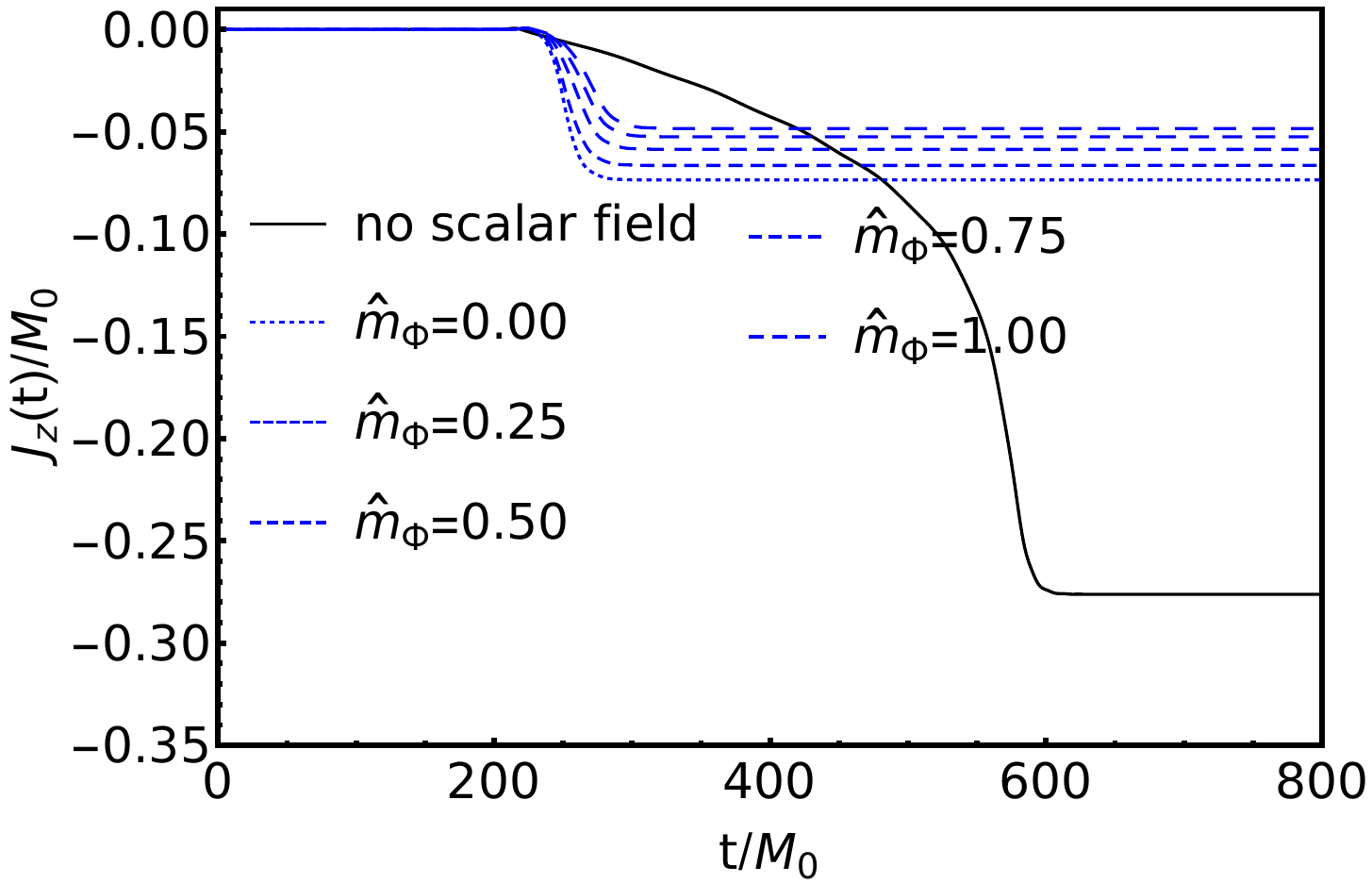}
    \caption{
    Accumulated gravitational-wave angular momentum
    $J_z^{\rm GW}(t)/M_0^2$ for equal-mass BBH mergers, $q=1$.
    From left to right, the panels correspond to
    $\Pi_0=0.0025$, $0.0050$, $0.0100$, and $0.0150$,
    respectively. Different line styles denote
    $\hat m_{\Phi}=0$, $0.25$, $0.50$, $0.75$, and $1.00$.
    The corresponding vacuum result is shown in black.
    All quantities are evaluated at the fixed extraction radius
    $r=200\,M_0$.
    }
    \label{fig:angular_time}
\end{figure*}

We now turn to the angular-momentum budget. Since the binaries
considered here are nonspinning and initially orbit in the
$x$--$y$ plane, we focus on the $z$ component of the angular momentum.
The accumulated gravitational-wave angular momentum is
\begin{equation}
J_z^{\rm GW}(t)
=
\int_{t_0}^{t}
\frac{dJ_z^{\rm GW}}{dt'}\,dt' .
\label{eq:JGW_acc}
\end{equation}
Figure~\ref{fig:angular_time} shows $J_z^{\rm GW}(t)$ for the same
$q=1$ configurations used in the energy analysis.

The angular-momentum evolution follows a trend similar to that of the
gravitational-wave energy. In the presence of the scalar field, the
main angular-momentum loss occurs significantly earlier than in the
vacuum case, consistent with the accelerated merger. More importantly,
the magnitude of the final gravitational-wave angular-momentum loss
decreases strongly as the initial scalar field amplitude is increased.
For $\Pi_0=0.0025$, the scalar field configurations remain relatively
close to the vacuum result, whereas for $\Pi_0=0.0100$ and $0.0150$
the gravitational-wave angular-momentum loss is substantially
suppressed.

The scalar field mass introduces an additional dependence. At low and
intermediate densities, different values of $\hat m_{\Phi}$ lead to
clearly distinguishable final values and emission times. At the
highest scalar field densities, however, the overall suppression
associated with the dense scalar environment becomes dominant, and the
relative differences among the massive-field configurations become
less pronounced.

\begin{figure*}
    \centering
    \includegraphics[width=0.24\linewidth]{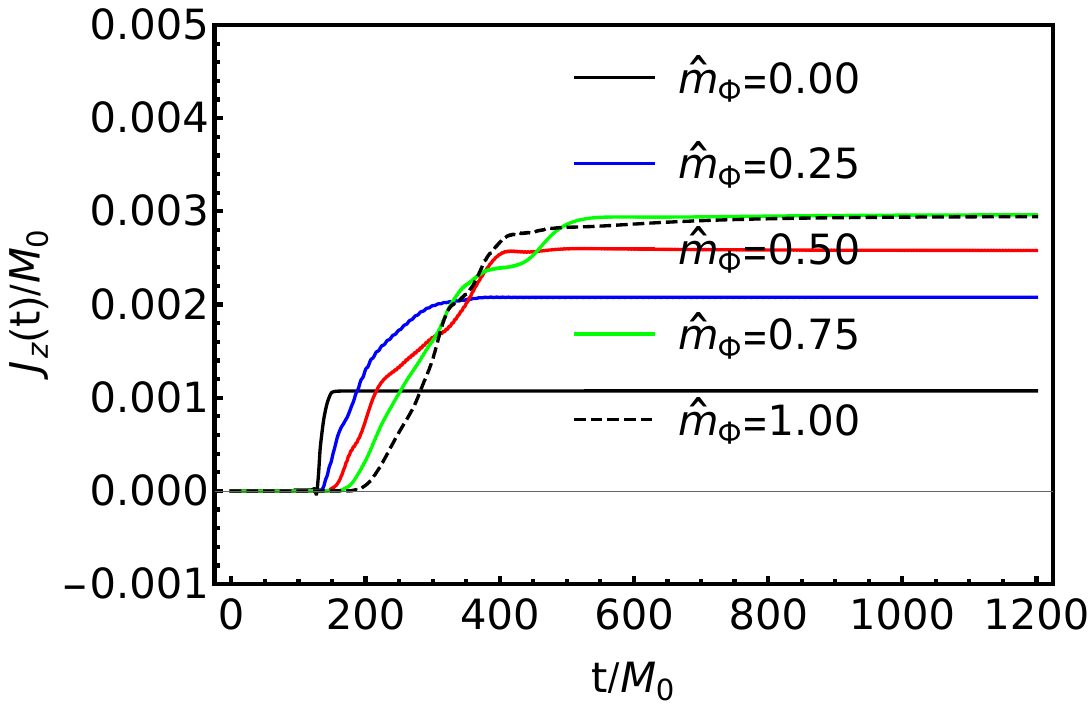}
    \includegraphics[width=0.24\linewidth]{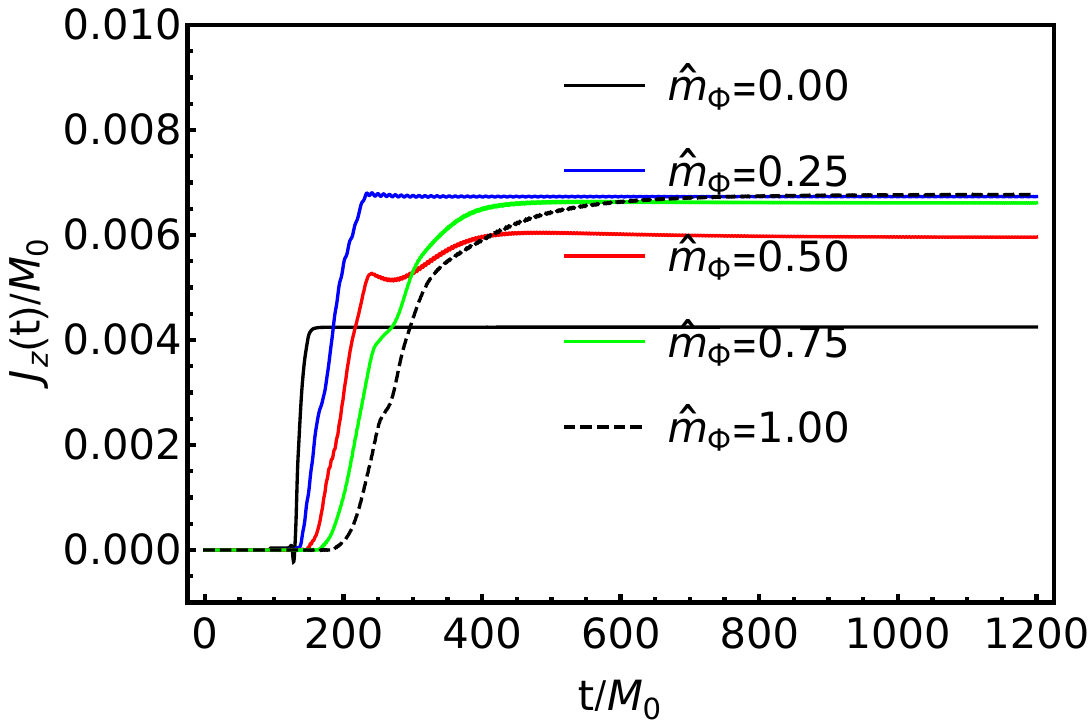}
    \includegraphics[width=0.24\linewidth]{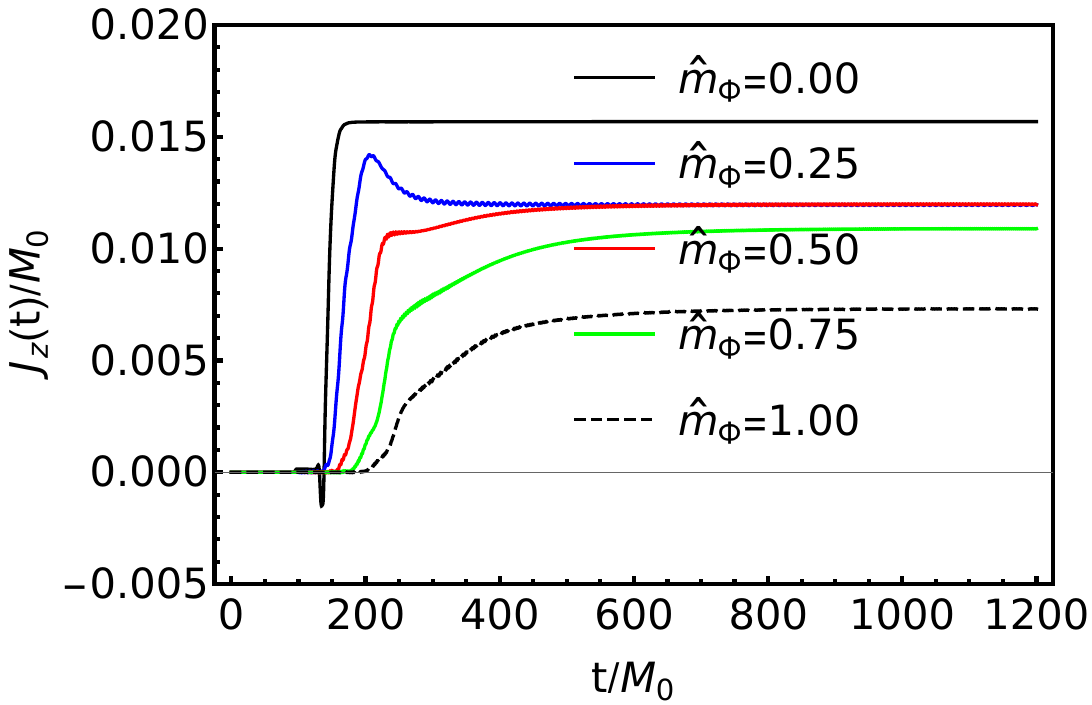}
    \includegraphics[width=0.24\linewidth]{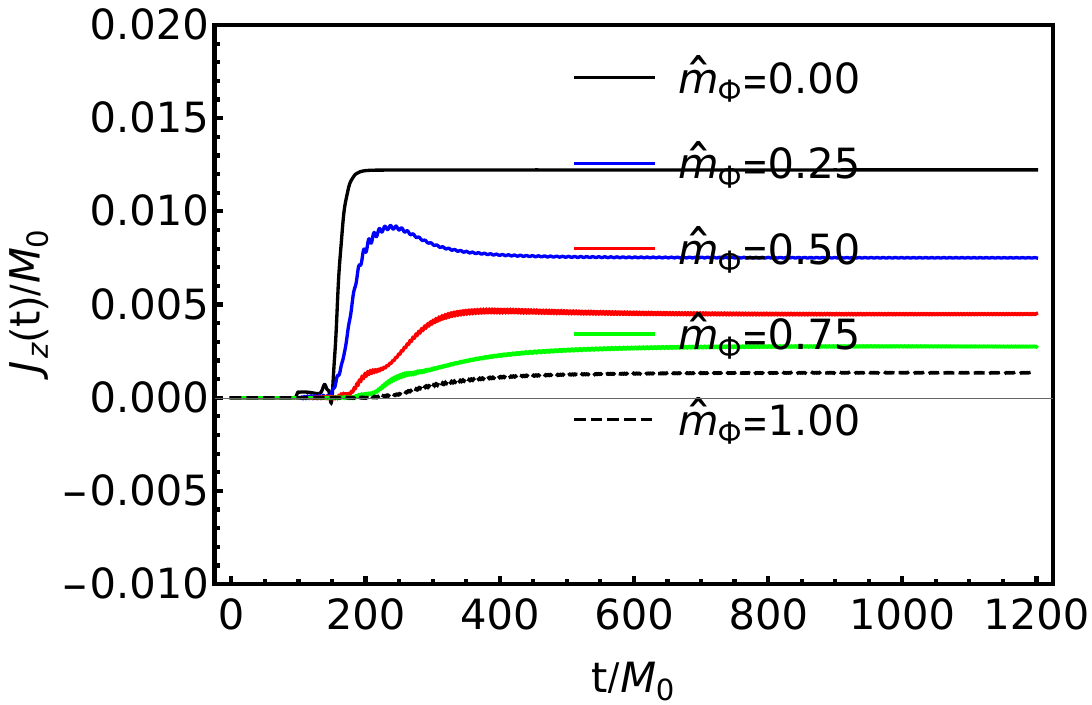}
    \includegraphics[width=0.24\linewidth]{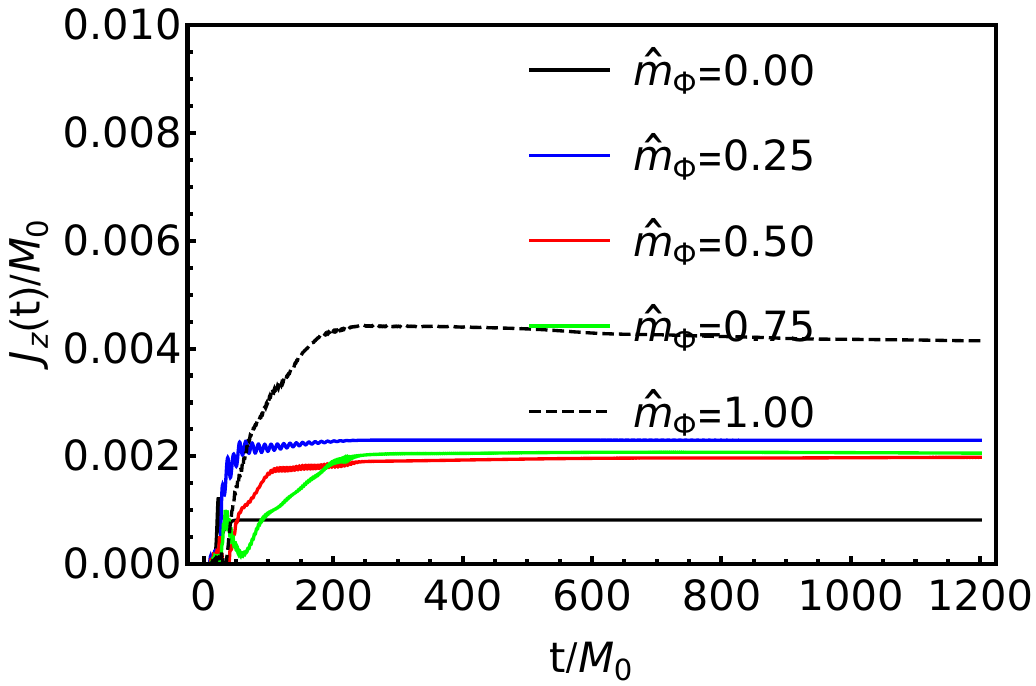}
    \includegraphics[width=0.24\linewidth]{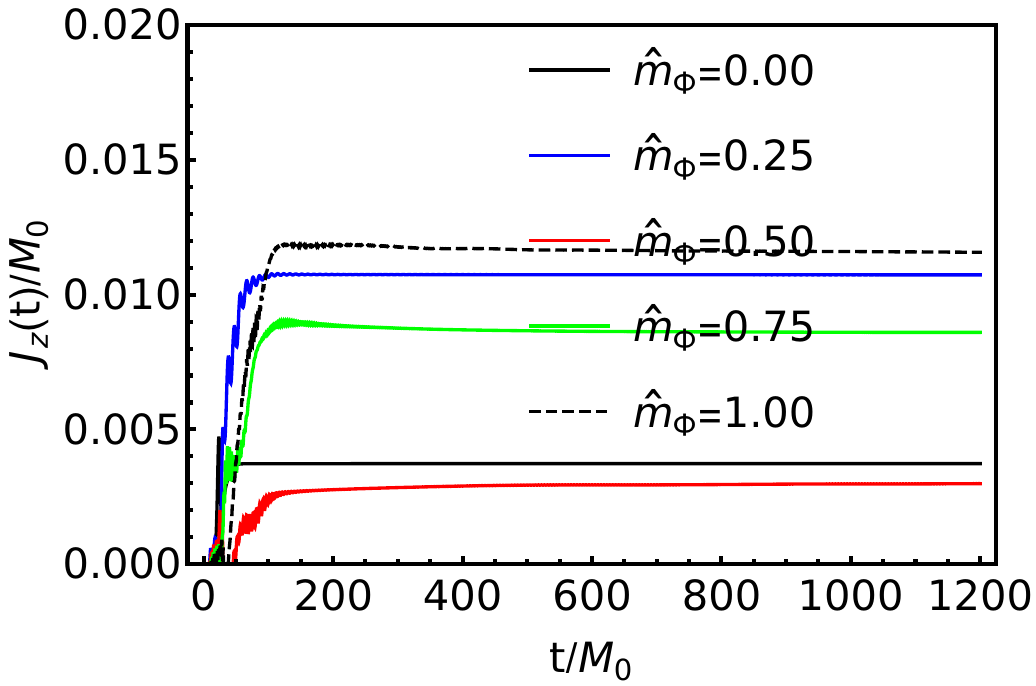}
    \includegraphics[width=0.24\linewidth]{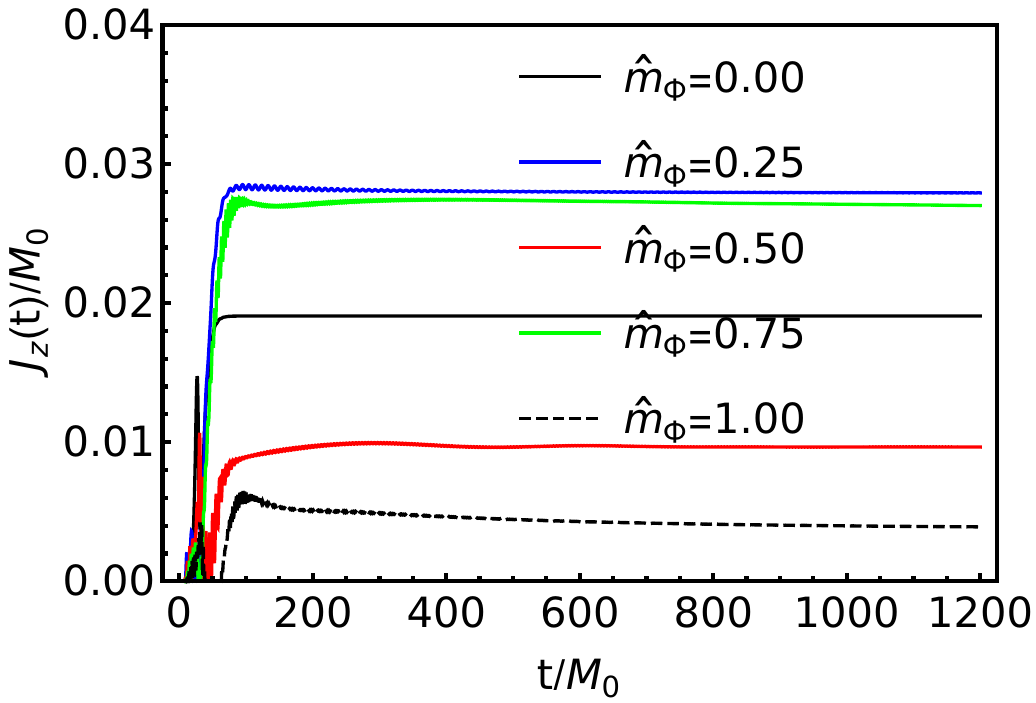}
    \includegraphics[width=0.24\linewidth]{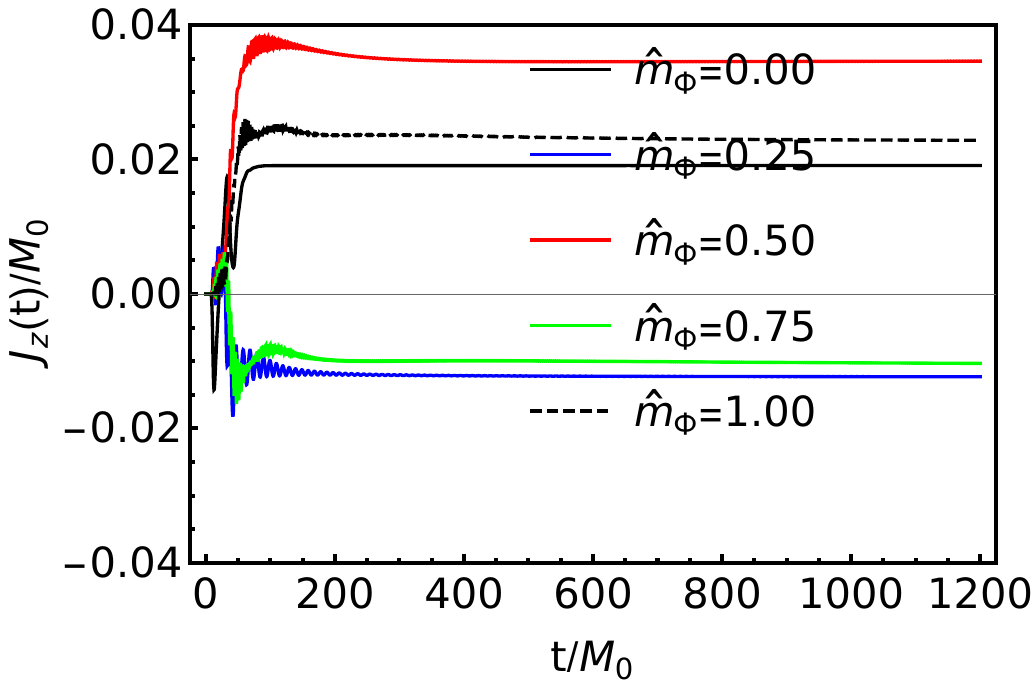}
    \caption{
    Accumulated scalar field angular momentum
    $J_z^{\Phi}(t)/M_0^2$ for equal-mass BBH mergers, $q=1$,
    evaluated at two representative radii.
    The upper and lower rows show the quantities evaluated at
    $r=100\,M_0$ and $r=5\,M_0$, respectively.
    From left to right, the columns correspond to
    $\Pi_0=0.0025$, $0.0050$, $0.0100$, and $0.0150$.
    Different curves denote
    $\hat m_{\Phi}=0$, $0.25$, $0.50$, $0.75$, and $1.00$.
    The outer-radius results characterize scalar angular momentum
    transported away from the binary, while the inner-radius results
    probe angular-momentum exchange in the strong-field region.
    }
    \label{fig:sf_Jz_time}
\end{figure*}

We further analyze the scalar field angular momentum through
\begin{equation}
J_z^{\Phi}(t;r)
=
\int_{t_0}^{t}
\frac{dJ_z^{\Phi}(r,t')}{dt'}\,dt' .
\label{eq:Jphi_acc}
\end{equation}
Figure~\ref{fig:sf_Jz_time} shows the accumulated scalar field angular
momentum at $r=100\,M_0$ in the upper row and at $r=5\,M_0$ in the
lower row, with the same parameter ordering as in
Fig.~\ref{fig:sf_energy_time}.

At the outer radius, the scalar field carries a non-negligible amount
of angular momentum away from the binary. The magnitude of this
contribution increases substantially with the initial scalar field
amplitude. Its dependence on $\hat m_{\Phi}$ is, however, more complex
than that found for the scalar energy. At the lowest density,
$\Pi_0=0.0025$, the outward scalar angular momentum generally
increases with the scalar field mass. At intermediate density the
dependence becomes nonmonotonic, while for $\Pi_0=0.0100$ and
$0.0150$, larger values of $\hat m_{\Phi}$ tend to suppress the
angular momentum transported to large radii. This change of behavior
shows that the effects of scalar field mass and density on
angular-momentum transport are strongly coupled.

The inner-radius results reveal an even richer behavior. At
$r=5\,M_0$, the accumulated scalar angular momentum depends
nonmonotonically on both $\Pi_0$ and $\hat m_{\Phi}$. Unlike the
scalar-energy flux, which shows a robust inward direction at this
radius, the scalar angular-momentum flux does not maintain a single
sign over the entire parameter space. In particular, for the
highest-density configurations, different scalar field masses can lead
to opposite signs of the late-time angular-momentum transport across
the inner surface.

This sign change indicates that the direction of angular-momentum
exchange in the strong-field region is itself sensitive to the
scalar field parameters. The scalar cloud therefore does not simply
remove a fixed fraction of the orbital angular momentum. Instead, it
can exchange angular momentum with the binary in qualitatively
different ways depending on the characteristic scalar field scale and
the amount of scalar matter present. This behavior is naturally more
complicated than the energy transport because angular momentum depends
not only on the amount of scalar matter but also on its spatial and
azimuthal structure around the orbiting black holes.

\subsubsection{Combined energy and angular-momentum budget}

Taken together, the gravitational and scalar channels provide a
consistent picture of the binary evolution. Increasing the
scalar field amplitude strengthens the interaction between the scalar
environment and the BBH, drives the binary to merger earlier, and at
sufficiently high densities strongly suppresses the energy and angular
momentum radiated through gravitational waves. At the same time, the
scalar field provides additional channels for transporting these
quantities both outward and inward.

The scalar field mass determines how this redistribution takes place.
For the energy budget, the trend is relatively clear: smaller
$\hat m_{\Phi}$ favors outward scalar radiation, whereas larger
$\hat m_{\Phi}$ favors localization and inward energy transport. The
angular-momentum budget is more intricate. Its dependence on
$m_{\Phi}$ changes with the scalar field density and can even involve
a reversal of the angular-momentum transport in the strong-field
region. Thus, the scalar field mass controls not only the efficiency
but also the qualitative character of the energy and angular-momentum
exchange between the scalar cloud and the binary.

These results also clarify why the changes in gravitational-wave
energy and angular momentum cannot be understood solely from the
shortened merger time. The scalar environment introduces additional
transport channels that compete with gravitational radiation and
redistribute the conserved quantities of the system. Their relative
importance depends jointly on $\Pi_0$ and $\hat m_{\Phi}$, providing a
natural explanation for the nontrivial mass--density dependence
observed in the gravitational-wave signals.

In the next subsection, we examine the corresponding linear-momentum
budget. Since asymmetric gravitational and scalar radiation directly
determine the recoil of the merger remnant, the combined dependence on
the scalar field mass and density becomes particularly important for
the final black hole kick.

\subsection{Linear momentum and gravitational recoil}
\label{sec:recoil}

We finally turn to the linear-momentum balance and the recoil of the
merger remnant. The energy and angular-momentum analyses presented
above show that the scalar environment introduces additional channels
for redistributing the conserved quantities of the binary. The same
applies to the linear momentum. For asymmetric systems, both
gravitational and scalar radiation can carry a net linear momentum
away from the binary, and the recoil of the final black hole is
determined by their combined contribution.

For each simulation, we calculate the linear momentum carried by
gravitational waves,
\begin{equation}
    \mathbf{P}_{\rm GW}
    =
    \left(
    P_x^{\rm GW},
    P_y^{\rm GW},
    P_z^{\rm GW}
    \right),
\end{equation}
and by the scalar field,
\begin{equation}
    \mathbf{P}_{\Phi}
    =
    \left(
    P_x^{\Phi},
    P_y^{\Phi},
    P_z^{\Phi}
    \right).
\end{equation}
Each Cartesian component used in the final recoil calculation is
evaluated at several finite extraction radii and extrapolated
independently to $r\rightarrow\infty$, following the procedure
described in Sec.~\ref{sec:diagnostics}. The total radiated linear
momentum is then
\begin{equation}
    \mathbf{P}_{\rm rad}
    =
    \mathbf{P}_{\rm GW}
    +
    \mathbf{P}_{\Phi}.
\label{eq:total_momentum_result}
\end{equation}

\subsubsection{Equal-mass binaries}

We first consider the equal-mass binaries, $q=1$, which provide a
useful consistency check for the linear-momentum calculation. Since
the two black holes are nonspinning and the initial scalar field
configuration preserves the symmetry of the binary, neither the
gravitational-wave nor the scalar field channel is expected to carry
a net linear momentum.

Consistent with this expectation, the residual in-plane momentum
components in our $q=1$ simulations are only of order
\begin{equation}
    |P_x|,\ |P_y|\sim10^{-8},
\end{equation}
while the component perpendicular to the orbital plane is even
smaller,
\begin{equation}
    |P_z|\sim10^{-18},
\end{equation}
in our dimensionless units. These values are consistent with zero at
the numerical level.

\subsubsection{Unequal-mass binaries}

The situation changes qualitatively for $q=2$, $3$, and $4$. The mass
asymmetry breaks the symmetry of the equal-mass system and leads to
anisotropic radiation. A nonzero net linear momentum can therefore be
carried away, producing a recoil of the merger remnant.

The scalar environment modifies this momentum balance in two ways.
First, scalar field accretion and backreaction change the orbital
dynamics and merger time. More specifically, the horizon-mass evolution
shown in Fig.~\ref{fig:massratiochange} demonstrates that accretion is
different in unequal-mass binaries: the horizon-mass ratio generally
increases during the inspiral and can depart substantially from its
initial value in dense scalar environments. The scalar field therefore
dynamically changes the binary asymmetry before merger. Since the
gravitational-wave linear momentum is highly sensitive to this asymmetry,
this evolving mass ratio provides an indirect mechanism for modifying
$\mathbf{P}_{\rm GW}$.

Second, the scalar field itself carries linear momentum,
$\mathbf{P}_{\Phi}$. The final recoil therefore depends on the vector
sum of the gravitational and scalar contributions rather than on
either channel separately. In particular, their relative orientation
is important: the two channels may reinforce or partially cancel each
other.

\begin{figure*}
    \centering
    \includegraphics[width=0.24\linewidth]{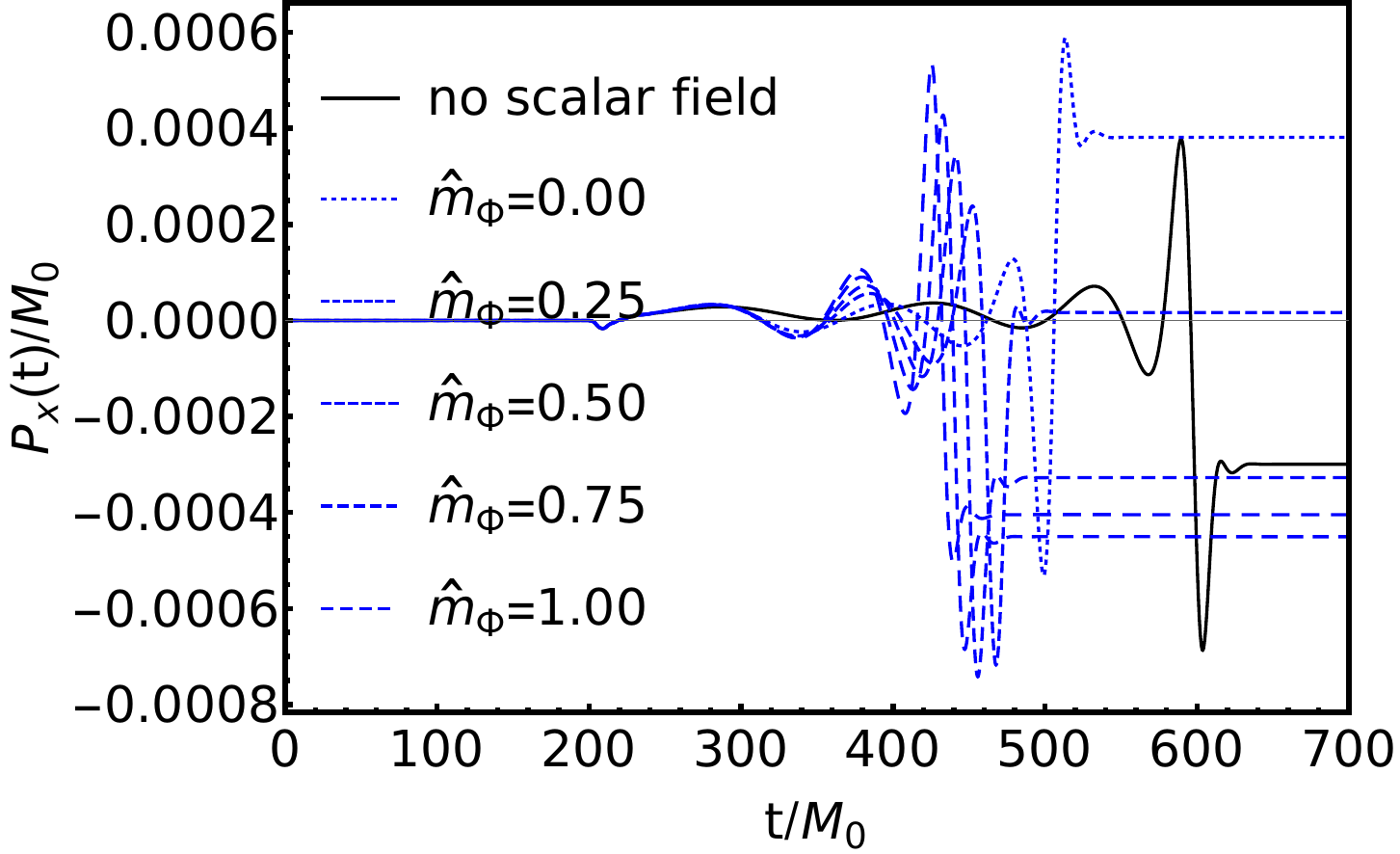}
    \includegraphics[width=0.24\linewidth]{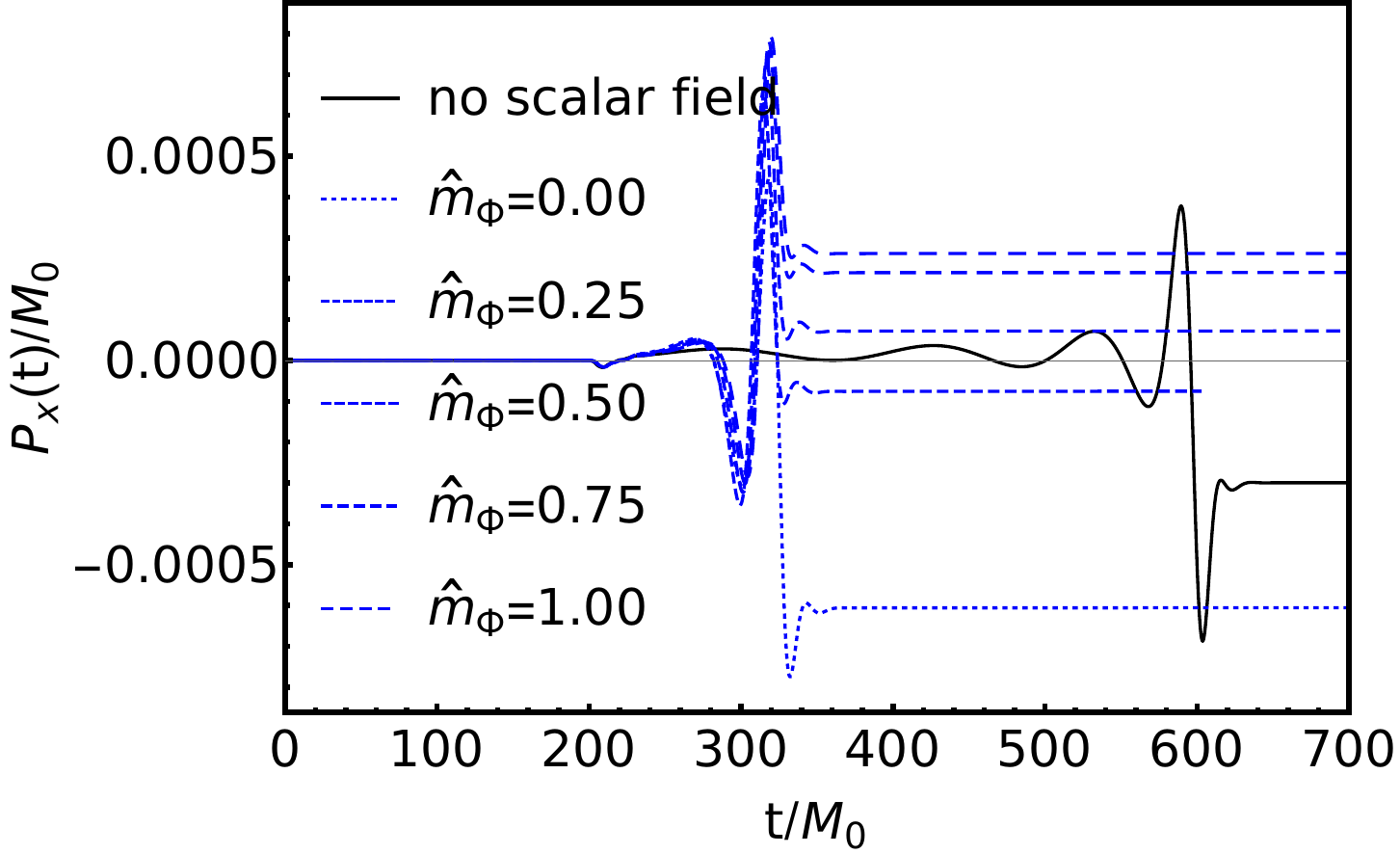}
    \includegraphics[width=0.24\linewidth]{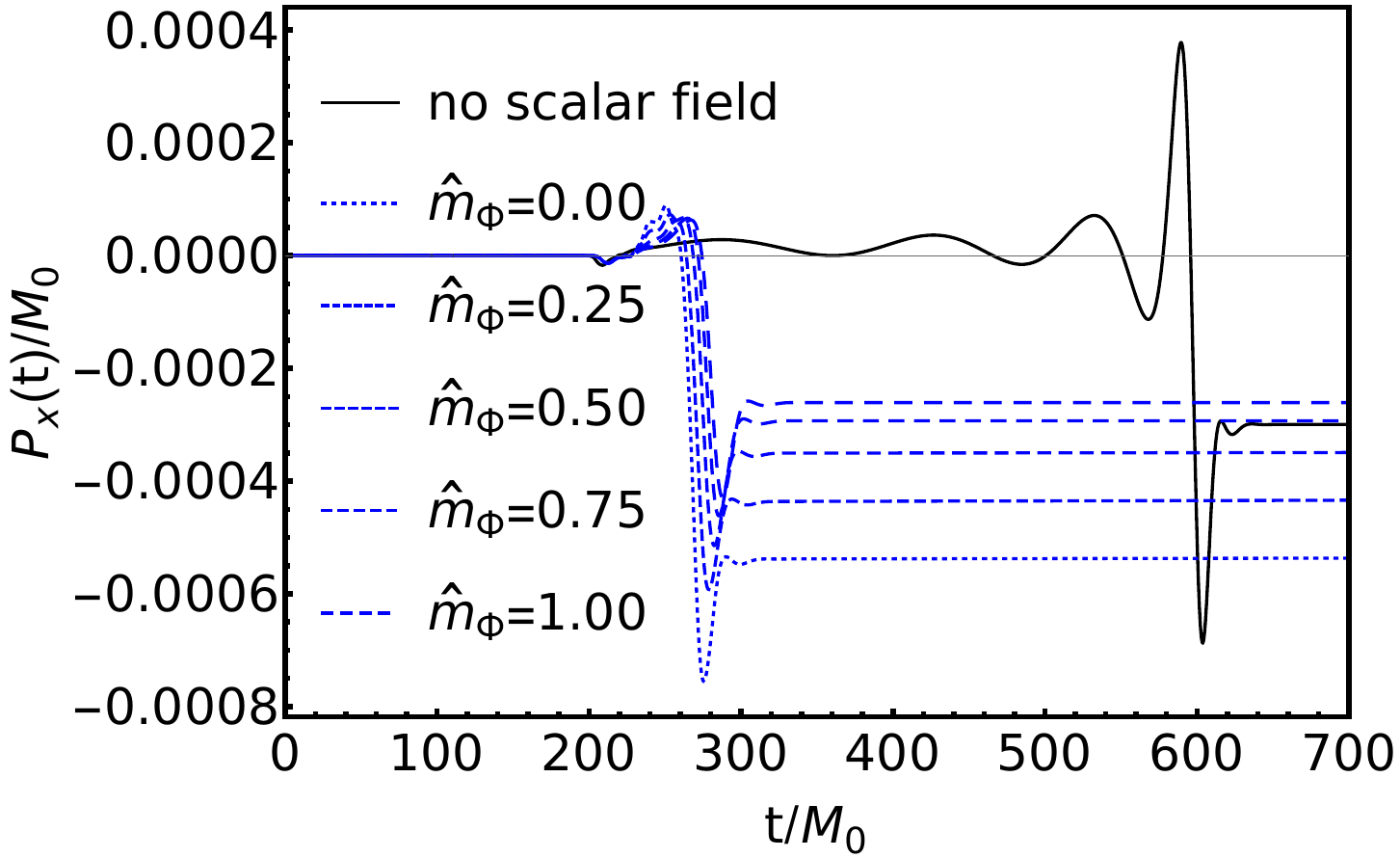}
    \includegraphics[width=0.24\linewidth]{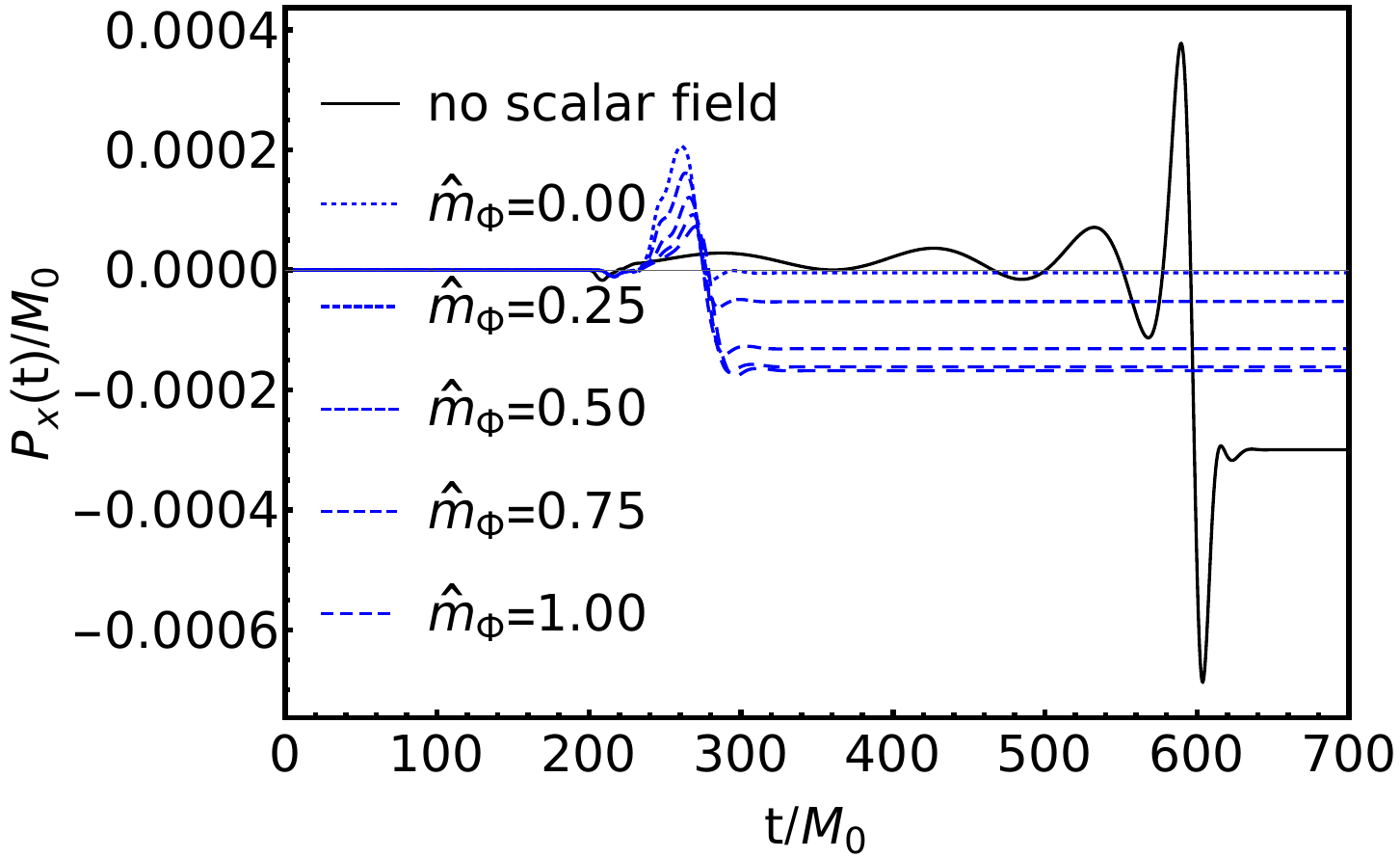}
    \includegraphics[width=0.24\linewidth]{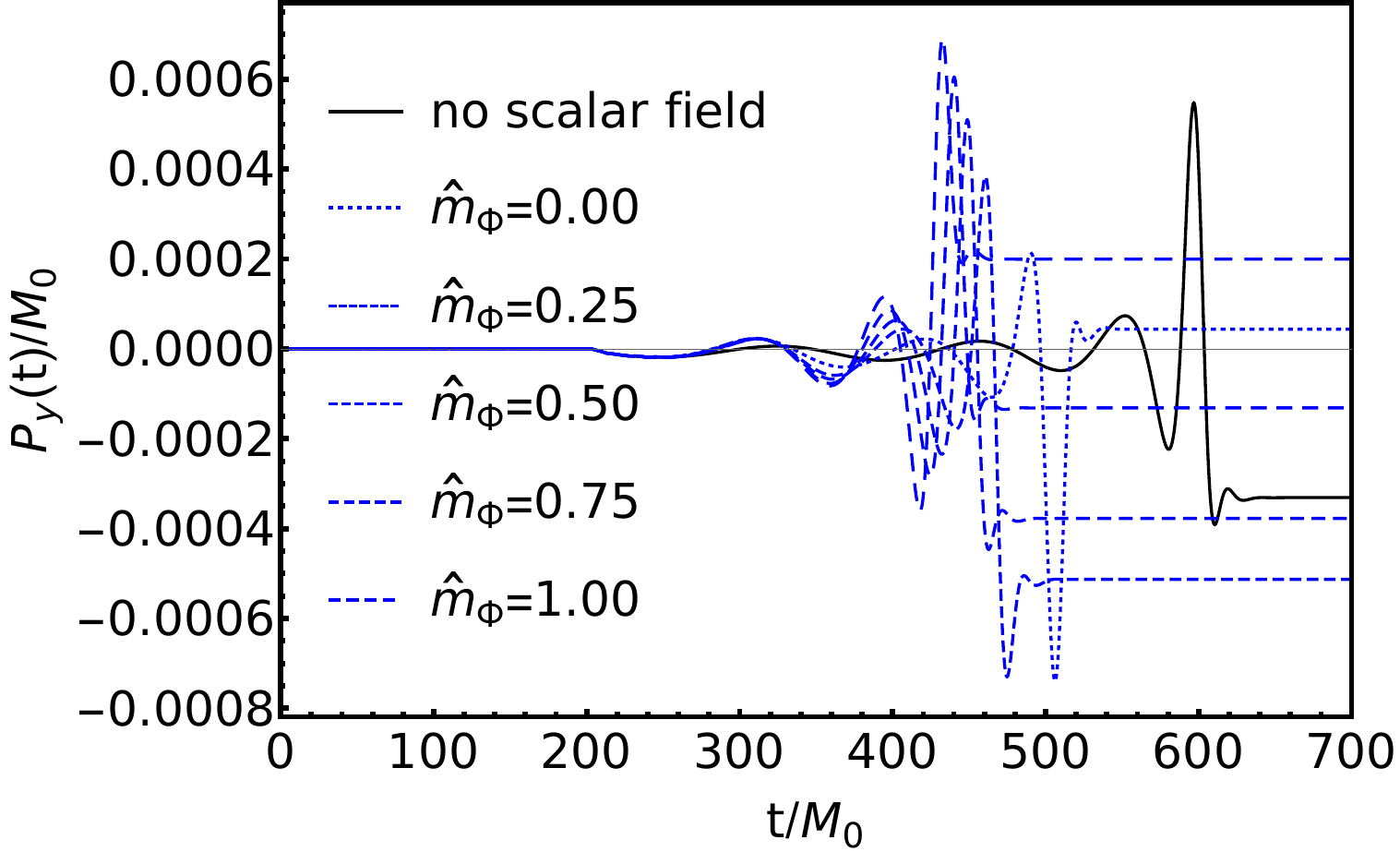}
    \includegraphics[width=0.24\linewidth]{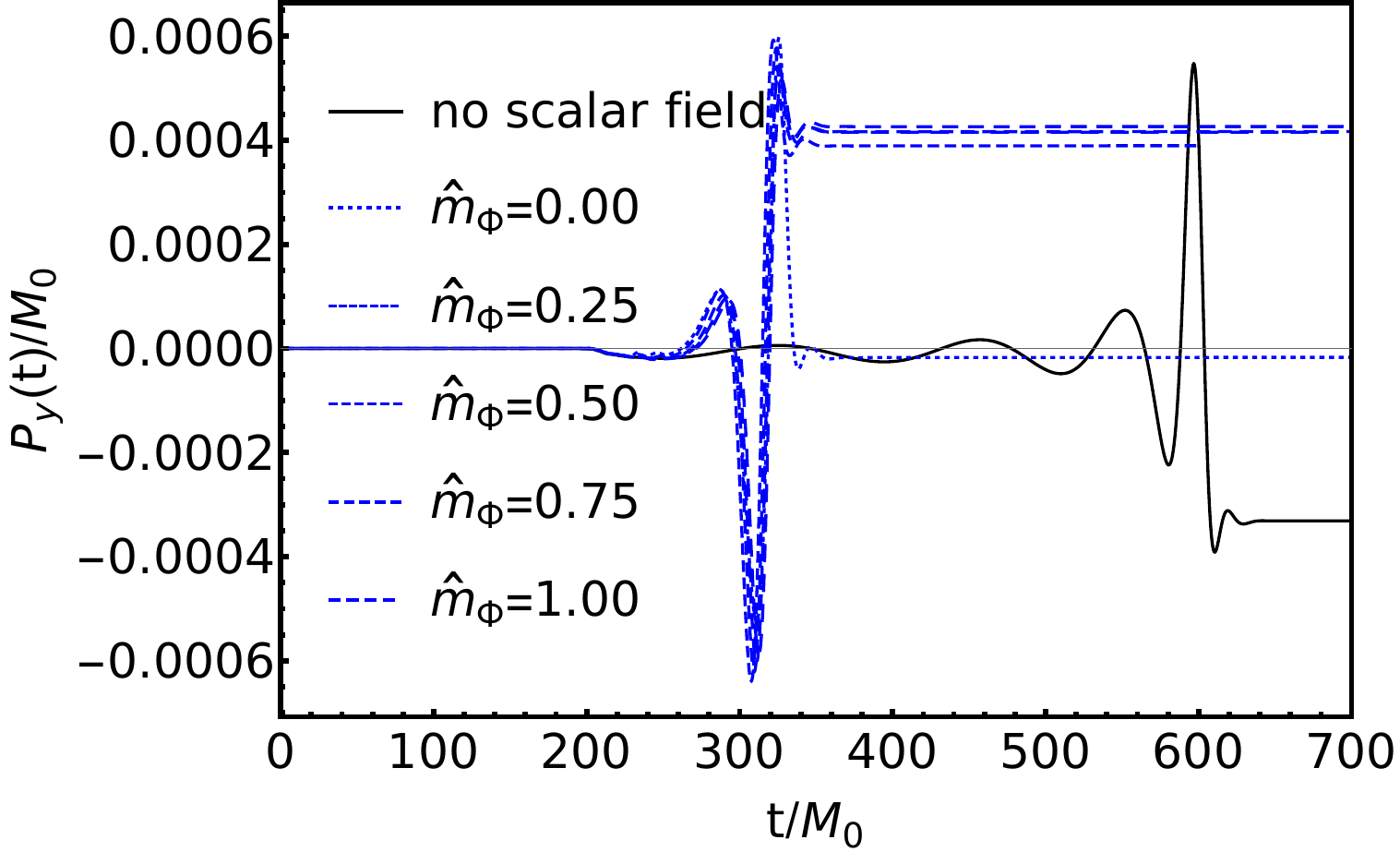}
    \includegraphics[width=0.24\linewidth]{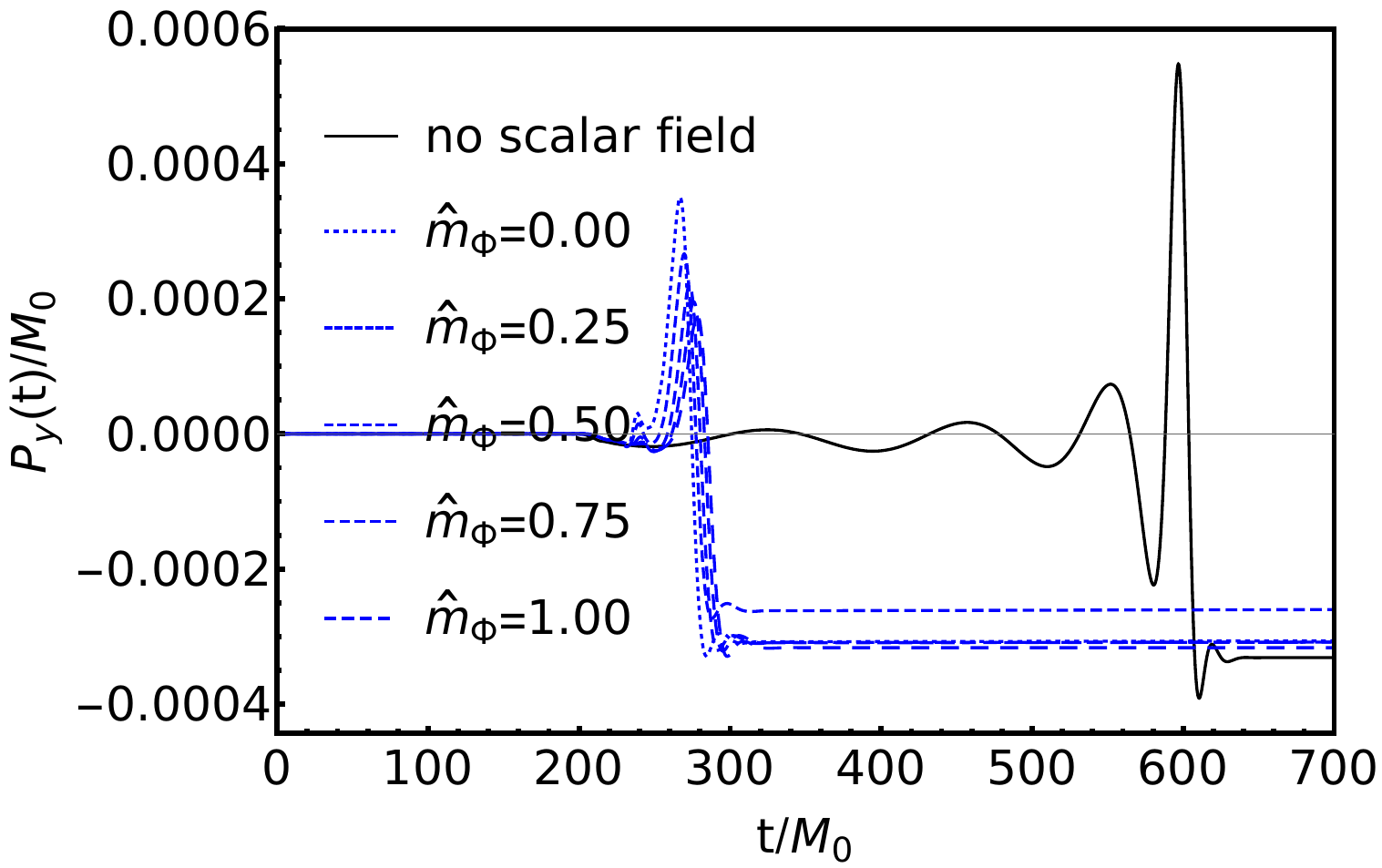}
    \includegraphics[width=0.24\linewidth]{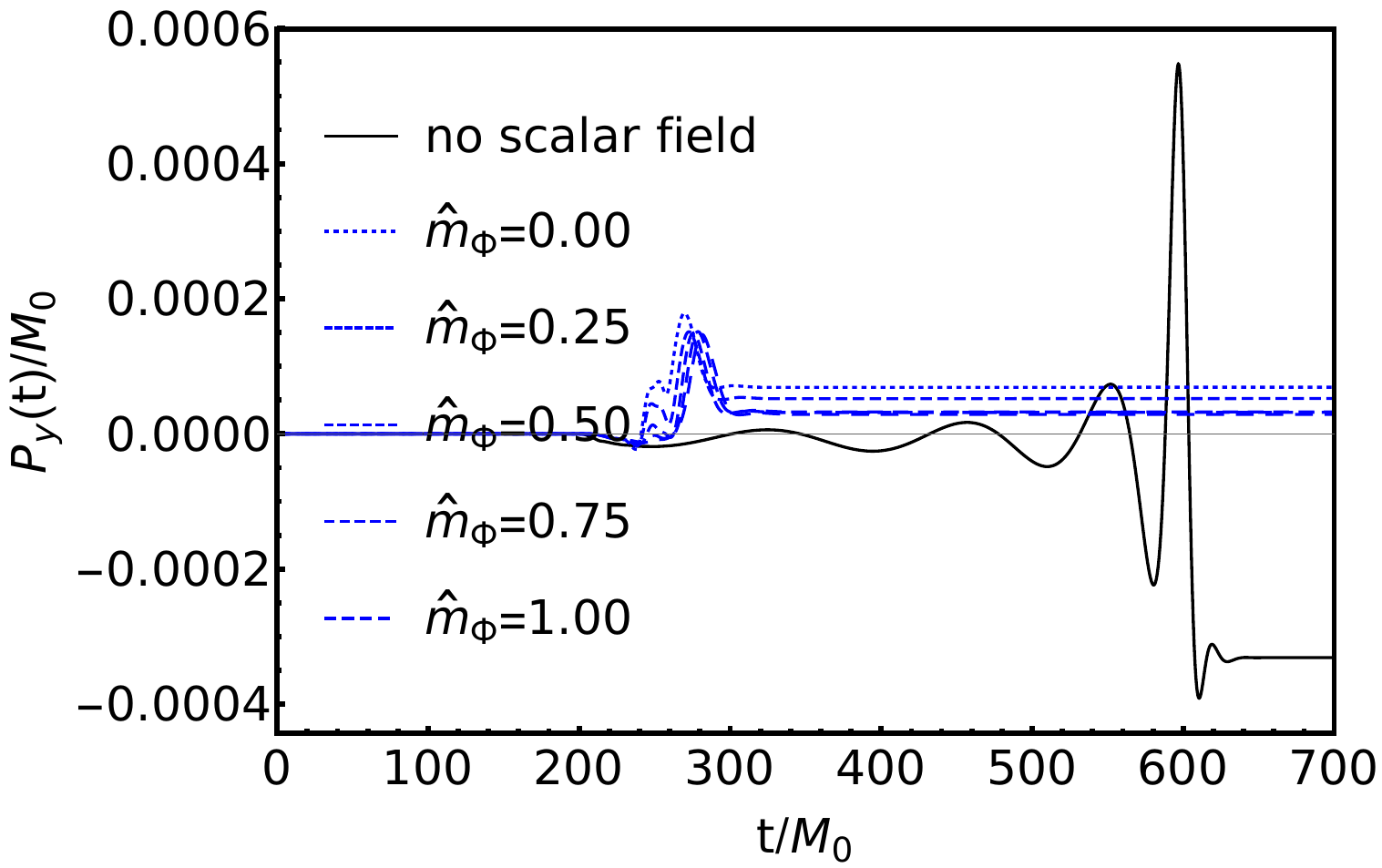}
    \caption{
    Accumulated linear momentum carried by gravitational waves for
    unequal-mass BBH mergers with $q=2$.
    The upper and lower rows show $P_x^{\rm GW}(t)$ and
    $P_y^{\rm GW}(t)$, respectively.
    From left to right, the columns correspond to
    $\Pi_0=0.0025$, $0.0050$, $0.0100$, and $0.0150$.
    Different curves denote
    $\hat m_{\Phi}=0$, $0.25$, $0.50$, $0.75$, and $1.00$,
    while the corresponding vacuum result is shown in black.
    }
    \label{fig:gw_momentum_q2}
\end{figure*}

\begin{figure*}
    \centering
    \includegraphics[width=0.24\linewidth]{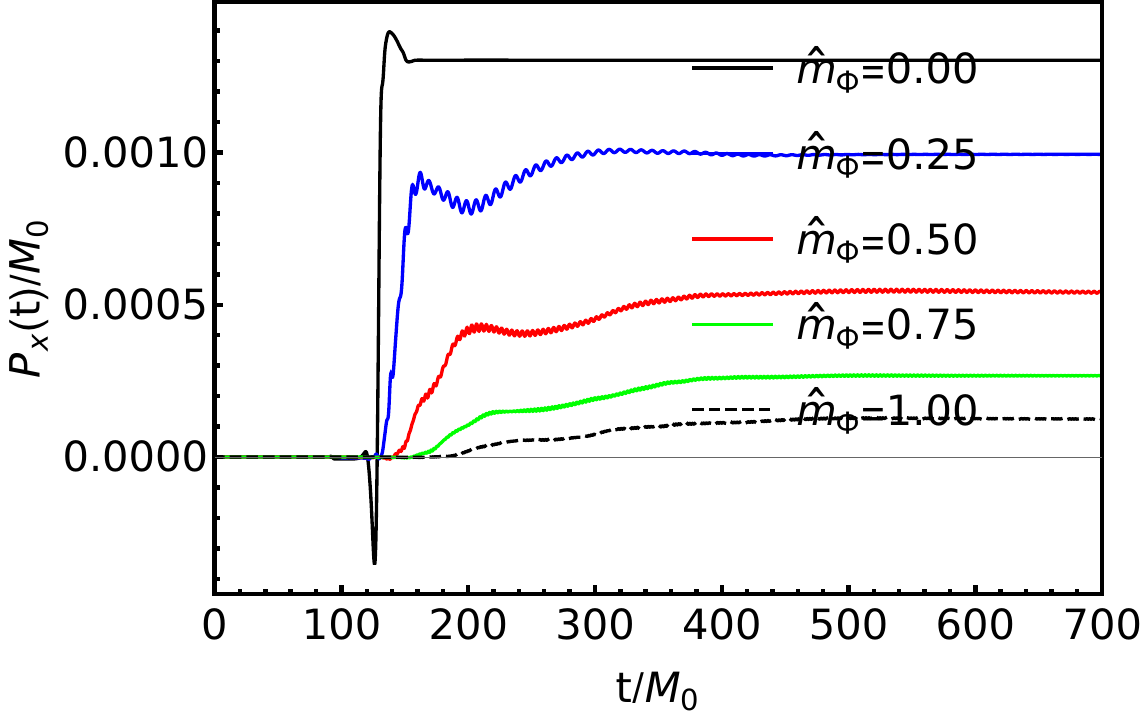}
    \includegraphics[width=0.24\linewidth]{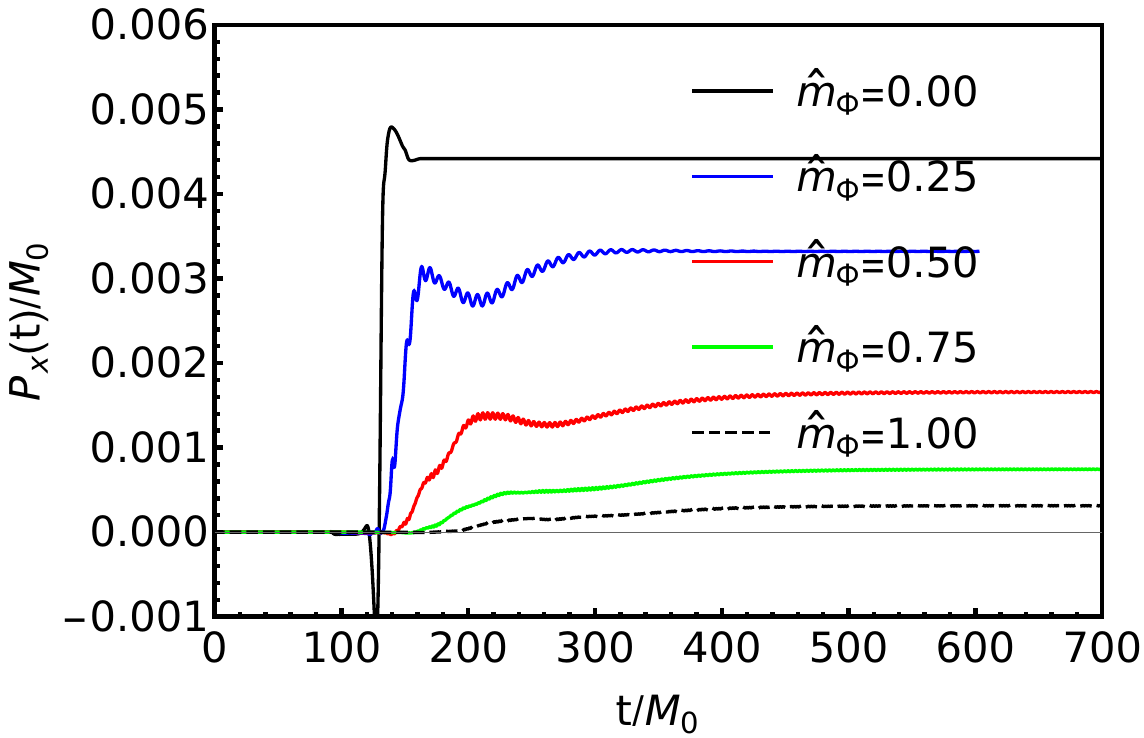}
    \includegraphics[width=0.24\linewidth]{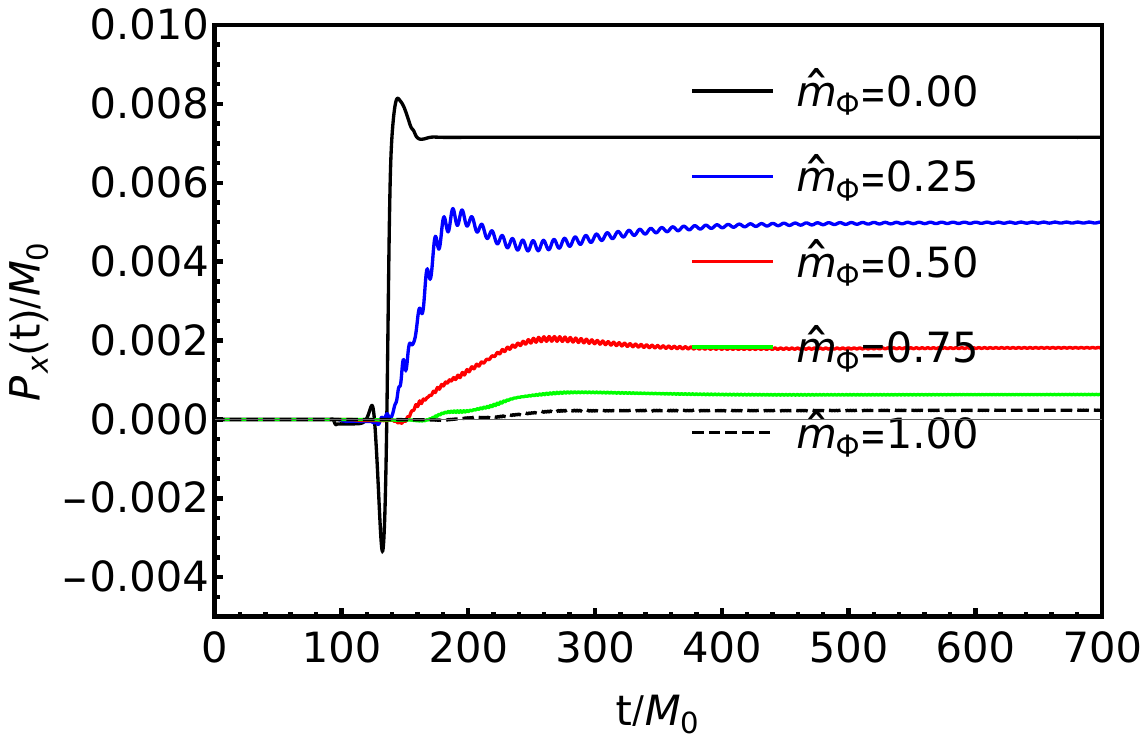}
    \includegraphics[width=0.24\linewidth]{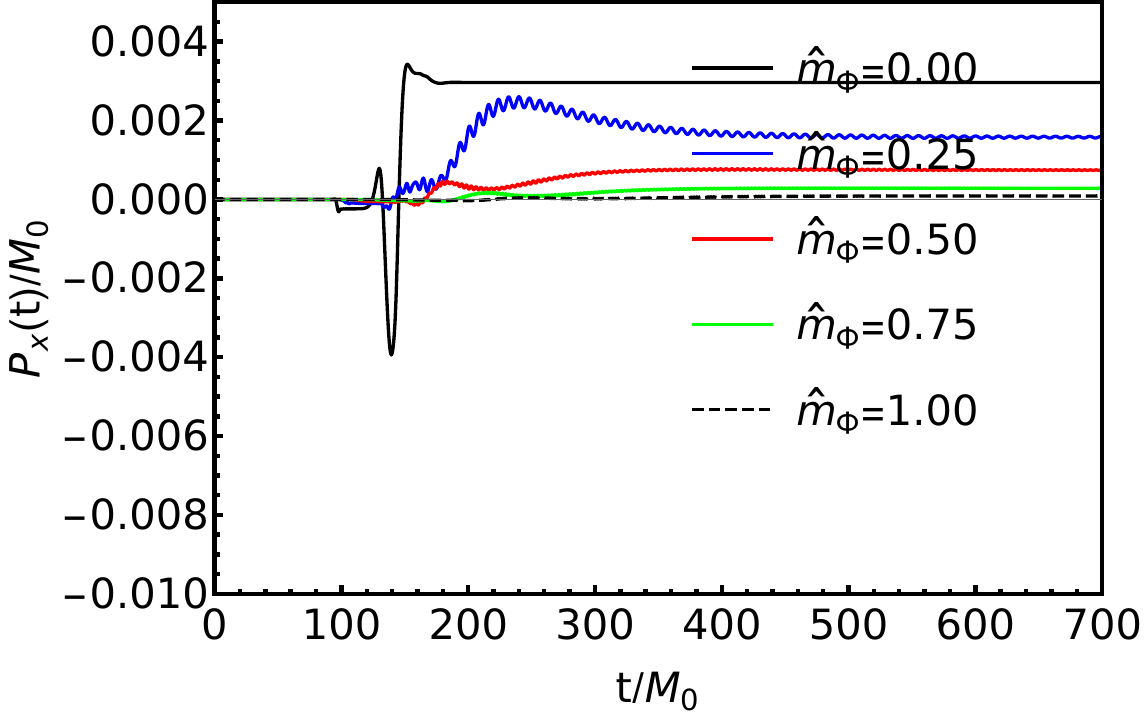}
    \includegraphics[width=0.24\linewidth]{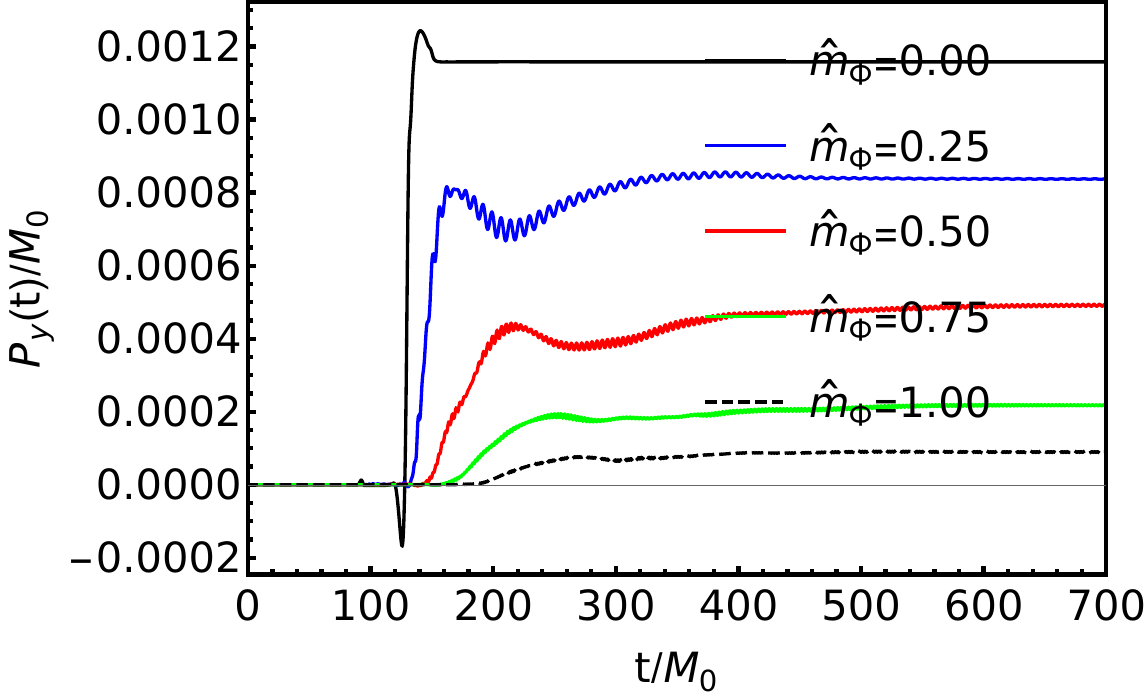}
    \includegraphics[width=0.24\linewidth]{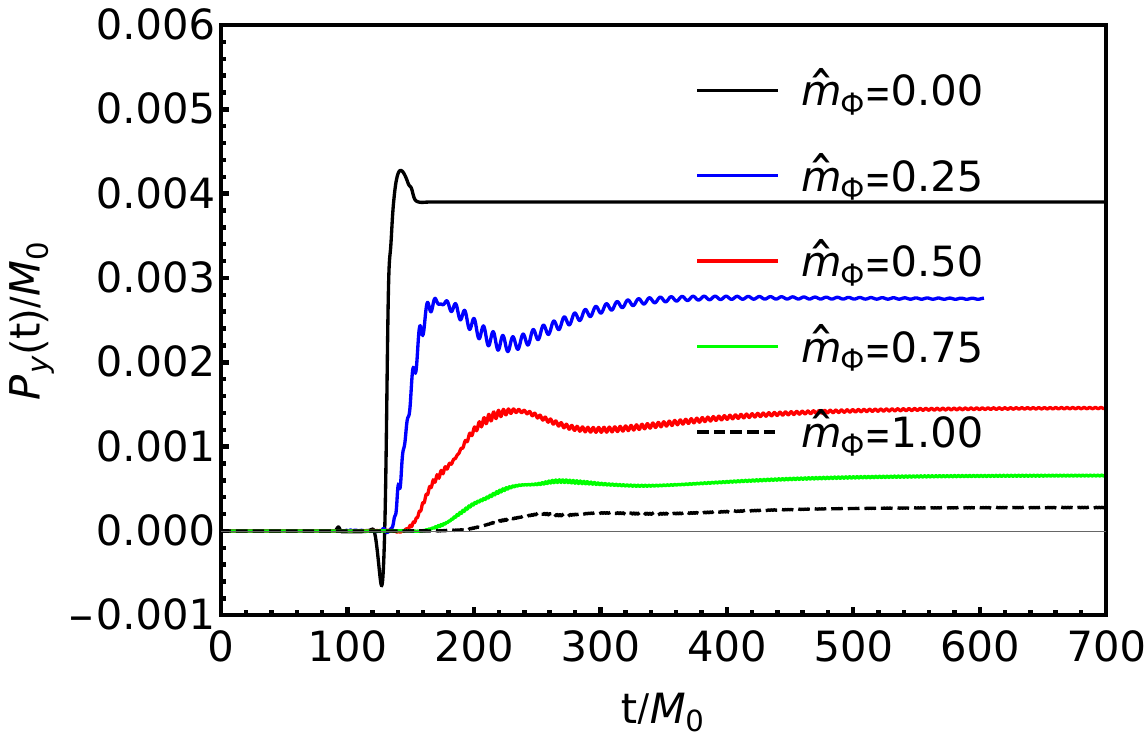}
    \includegraphics[width=0.24\linewidth]{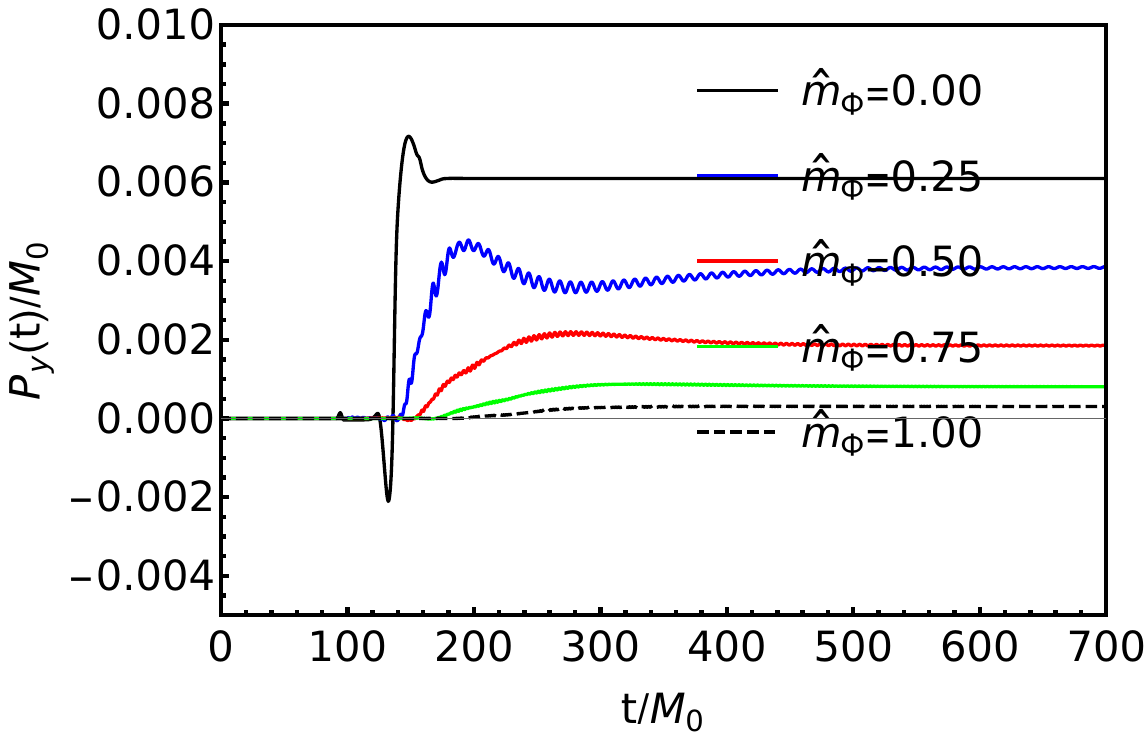}
    \includegraphics[width=0.24\linewidth]{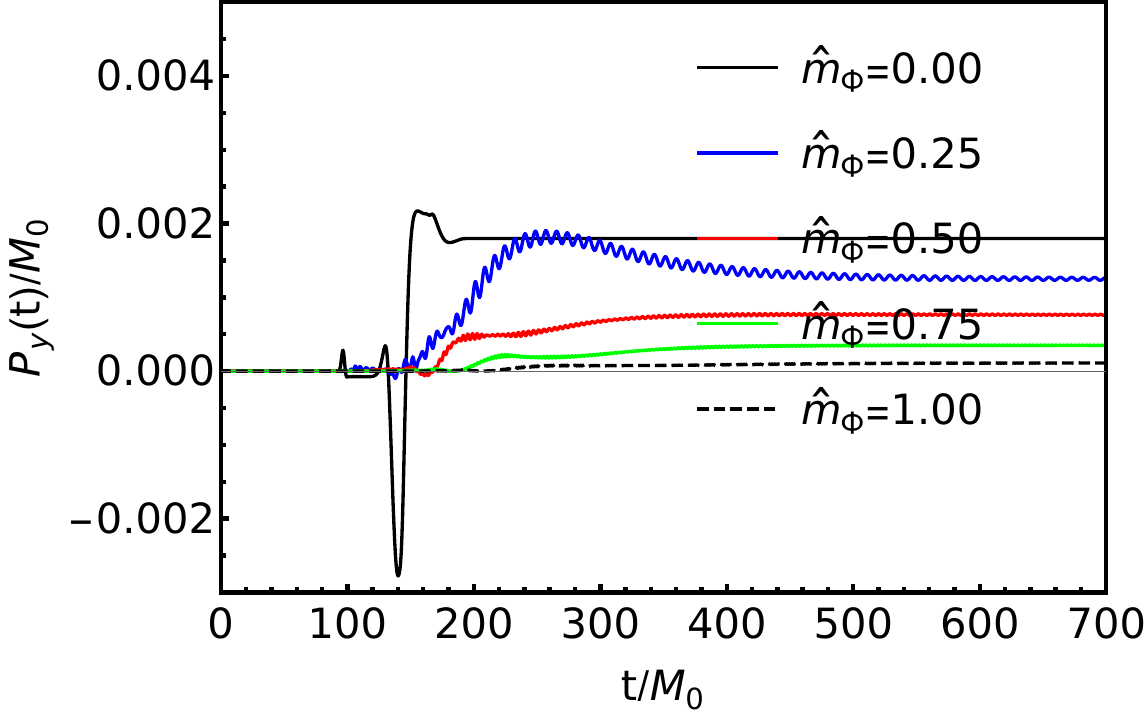}
    \caption{
    Accumulated linear momentum carried by the scalar field for
    unequal-mass BBH mergers with $q=2$.
    The upper and lower rows show $P_x^{\Phi}(t)$ and
    $P_y^{\Phi}(t)$, respectively.
    From left to right, the columns correspond to
    $\Pi_0=0.0025$, $0.0050$, $0.0100$, and $0.0150$.
    Different curves denote
    $\hat m_{\Phi}=0$, $0.25$, $0.50$, $0.75$, and $1.00$.
    }
    \label{fig:sf_momentum_q2}
\end{figure*}

To illustrate this behavior, we first consider the $q=2$ binaries.
Because reflection symmetry with respect to the orbital plane is
preserved, the radiated momentum is dominated by its $x$- and
$y$-components. Figures~\ref{fig:gw_momentum_q2} and
\ref{fig:sf_momentum_q2} show the accumulated linear momentum carried
by gravitational waves and the scalar field, respectively.

The two momentum channels exhibit qualitatively different behaviors.
For the gravitational-wave contribution,
Fig.~\ref{fig:gw_momentum_q2}, most of the net momentum is accumulated
during the late inspiral and merger. The rapid change occurs
systematically earlier in the scalar field configurations than in the
vacuum binary, in agreement with the earlier merger found from the
orbital dynamics, waveform, energy, and angular-momentum analyses.

Both the magnitude and direction of the final gravitational-wave
momentum depend sensitively on $\Pi_0$ and $\hat m_{\Phi}$. In
particular, changing the scalar field parameters can reverse the signs
of $P_x^{\rm GW}$ and/or $P_y^{\rm GW}$. The scalar environment
therefore modifies not only the amount of linear momentum carried by
gravitational waves but also its direction in the orbital plane.

The scalar contribution in Fig.~\ref{fig:sf_momentum_q2} shows a
different parameter dependence. Once the equal-mass symmetry is
broken, a substantial net momentum can be carried by the scalar field.
For fixed $\Pi_0$, the scalar momentum is generally larger for the
massless and low-mass scalar configurations and decreases as
$\hat m_{\Phi}$ increases. This behavior is consistent with the
scalar-energy results discussed above: lighter scalar fields propagate
more efficiently to large distances, whereas heavier fields are more
strongly localized in the central region.

The dependence on $\Pi_0$ is, however, strongly nonmonotonic.
Increasing the scalar field amplitude does not necessarily increase
the net scalar momentum. This reflects the fact that linear momentum
depends on the anisotropy of the radiation rather than simply on the
total amount of scalar energy emitted.

More importantly, a direct comparison of
Figs.~\ref{fig:gw_momentum_q2} and
\ref{fig:sf_momentum_q2} shows that
$\mathbf{P}_{\rm GW}$ and $\mathbf{P}_{\Phi}$ need not point in the
same direction. To quantify this vector nature of the momentum
balance, we define
\begin{equation}
    P_{\rm GW}=|\mathbf{P}_{\rm GW}|,
    \qquad
    P_{\Phi}=|\mathbf{P}_{\Phi}|,
\end{equation}
and introduce their relative angle,
\begin{equation}
    \cos\theta_P
    =
    \frac{
    \mathbf{P}_{\rm GW}\cdot\mathbf{P}_{\Phi}
    }{
    P_{\rm GW}P_{\Phi}
    }.
\label{eq:momentum_angle}
\end{equation}
Thus, $\cos\theta_P\simeq1$ indicates nearly aligned momentum vectors,
whereas $\cos\theta_P\simeq-1$ corresponds to nearly opposite
directions.

For vanishing initial total linear momentum, momentum conservation
gives the recoil velocity
\begin{equation}
    \mathbf{v}_{\rm kick}
    =
    -
    \frac{
    \mathbf{P}_{\rm GW}+\mathbf{P}_{\Phi}
    }{M_f},
\label{eq:vkick_result}
\end{equation}
where $M_f$ is the mass of the remnant black hole. We perform this
analysis for the full set of unequal-mass binaries with
$q=2$, $3$, and $4$.

\begin{figure*}
    \centering
    \includegraphics[width=0.32\linewidth]{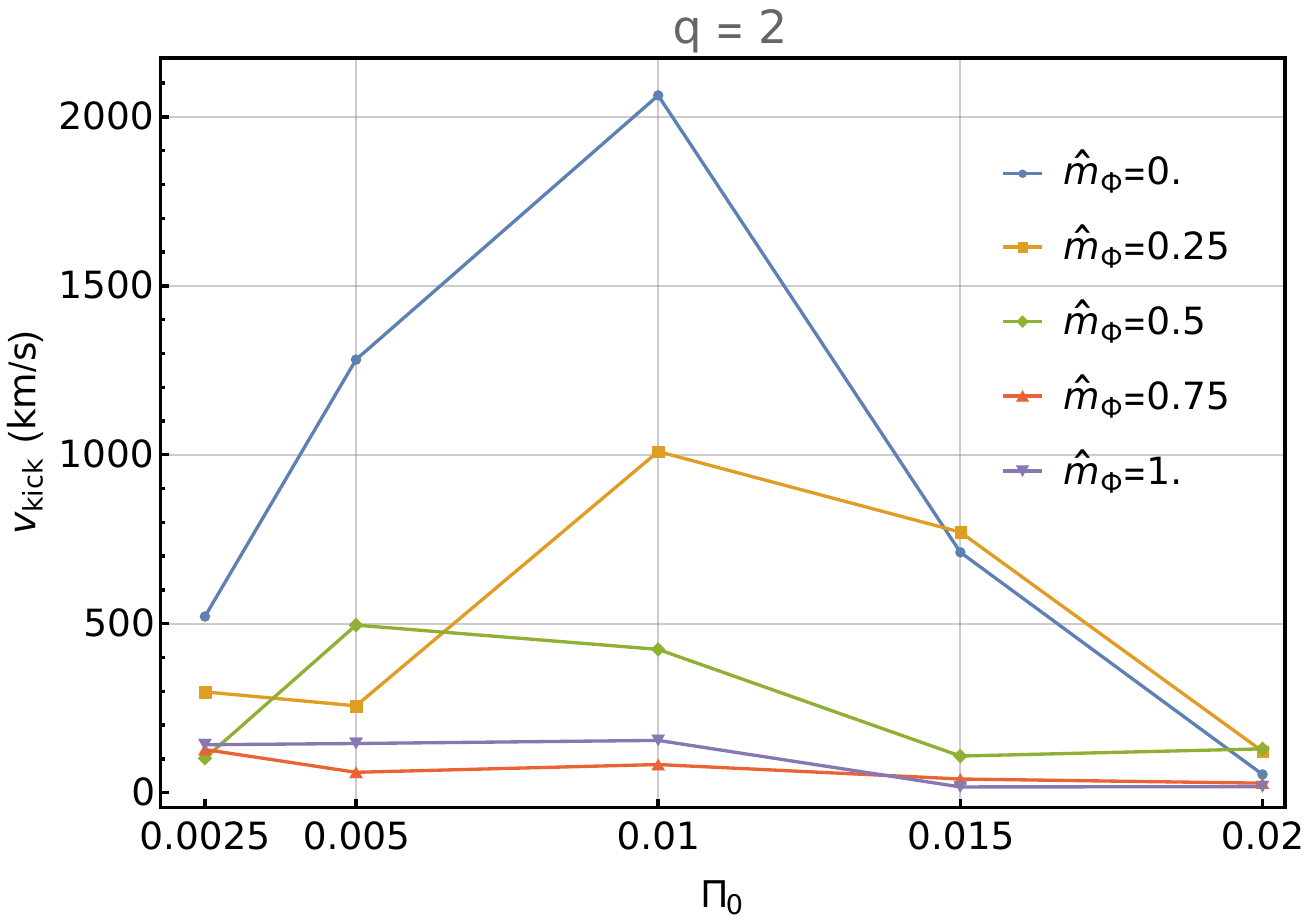}
    \includegraphics[width=0.32\linewidth]{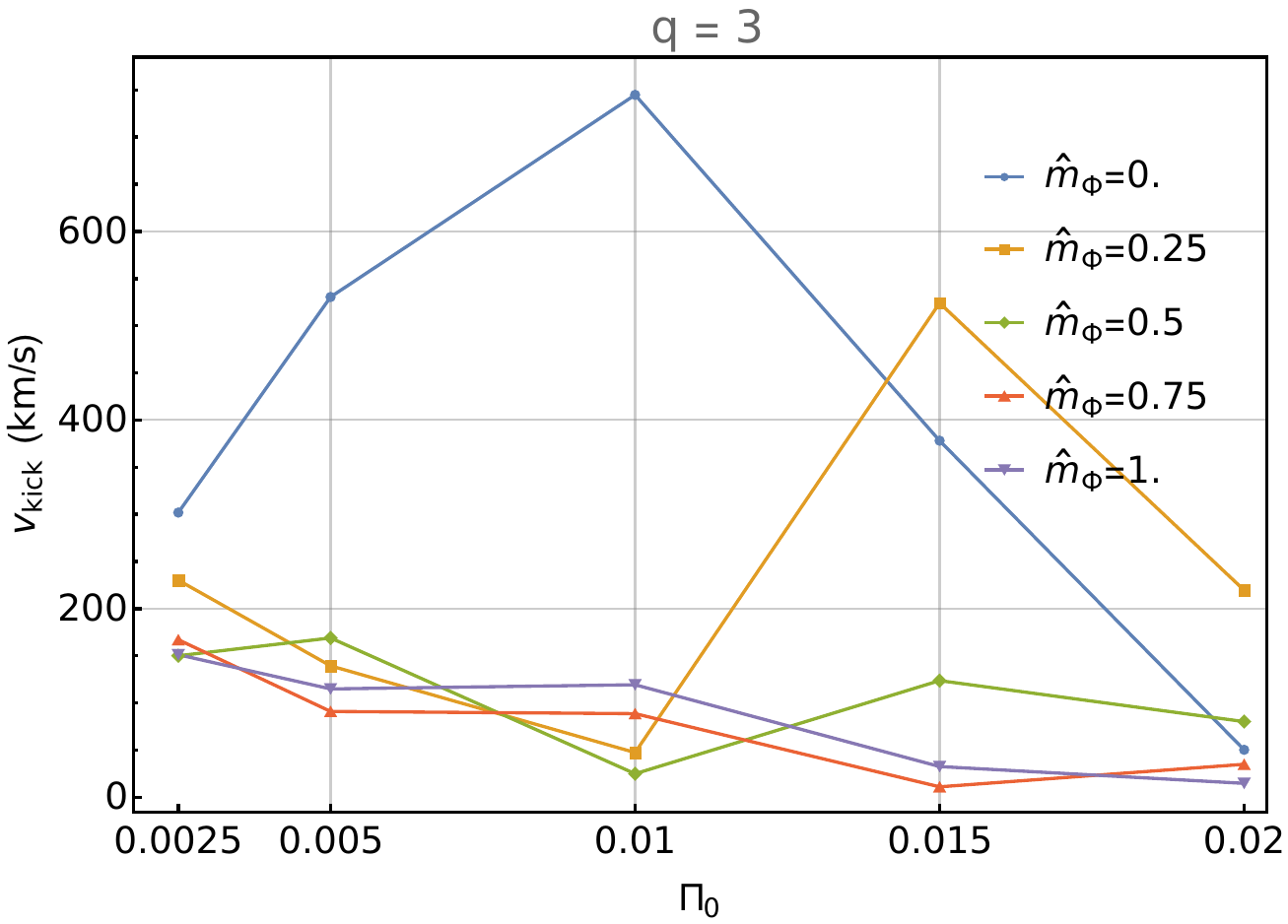}
    \includegraphics[width=0.32\linewidth]{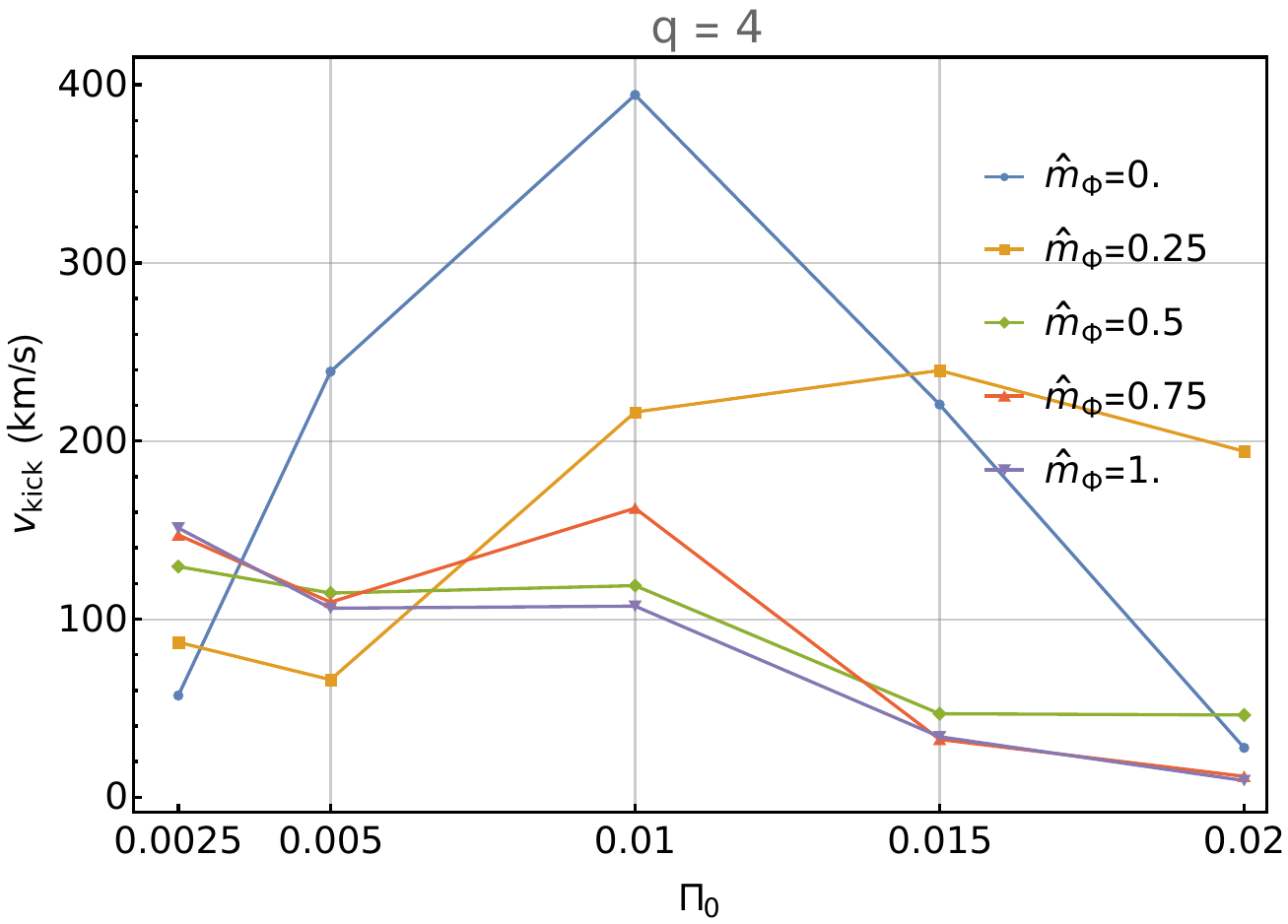}
    \caption{
    Recoil velocity $v_{\rm kick}$ of the remnant black hole as a
    function of the initial scalar field amplitude $\Pi_0$.
    From left to right, the panels correspond to
    $q=2$, $3$, and $4$, respectively.
    Different curves denote
    $\hat m_{\Phi}=0$, $0.25$, $0.50$, $0.75$, and $1.00$.
    }
    \label{fig:kick}
\end{figure*}

Figure~\ref{fig:kick} shows that the recoil depends strongly and
nonmonotonically on both the scalar field mass and density. The effect
is particularly pronounced for $q=2$. In the massless case, the recoil
increases rapidly with $\Pi_0$, reaches
\begin{equation}
    v_{\rm kick}=2063.6~{\rm km\,s^{-1}}
\end{equation}
at $\Pi_0=0.0100$, and subsequently decreases as the scalar field
amplitude is increased further. This value is more than an order of
magnitude larger than the corresponding vacuum recoil,
$v_{\rm kick}^{\rm vac}=137.8~{\rm km\,s^{-1}}$.

A similar, although weaker, enhancement is found for $q=3$. The
largest recoil for this mass ratio is
$744.8~{\rm km\,s^{-1}}$, obtained for
$\hat m_{\Phi}=0$ and $\Pi_0=0.0100$, compared with the vacuum value
$162.9~{\rm km\,s^{-1}}$. For $q=4$, the maximum recoil decreases
further to $394.4~{\rm km\,s^{-1}}$, again occurring for the massless
field at $\Pi_0=0.0100$, while the corresponding vacuum value is
$139.0~{\rm km\,s^{-1}}$.

The massless scalar field therefore produces the largest recoil found
in our parameter survey for all three mass ratios. This is consistent
with the efficient outward propagation of the massless field found in
the scalar-energy analysis, which allows a substantial amount of
anisotropic scalar momentum to escape from the system. Nevertheless,
the dependence on $\hat m_{\Phi}$ is not monotonic in general.
For example, configurations with $\hat m_{\Phi}=0.25$ can also produce
large kicks, reaching $1009.6~{\rm km\,s^{-1}}$ for $q=2$, while
heavier scalar fields can either enhance or suppress the recoil
depending on $\Pi_0$ and $q$.

Importantly, increasing the scalar field density does not
systematically increase the recoil. In all three mass-ratio families,
the kick can increase by a large factor over part of the parameter
space and then decrease again at larger $\Pi_0$. In several
configurations the recoil even falls below the corresponding vacuum
value. The environmental effect on the recoil is therefore governed
by the detailed momentum balance rather than simply by the amount of
scalar matter surrounding the binary.

\begin{figure*}
    \centering
    \includegraphics[width=0.32\linewidth]{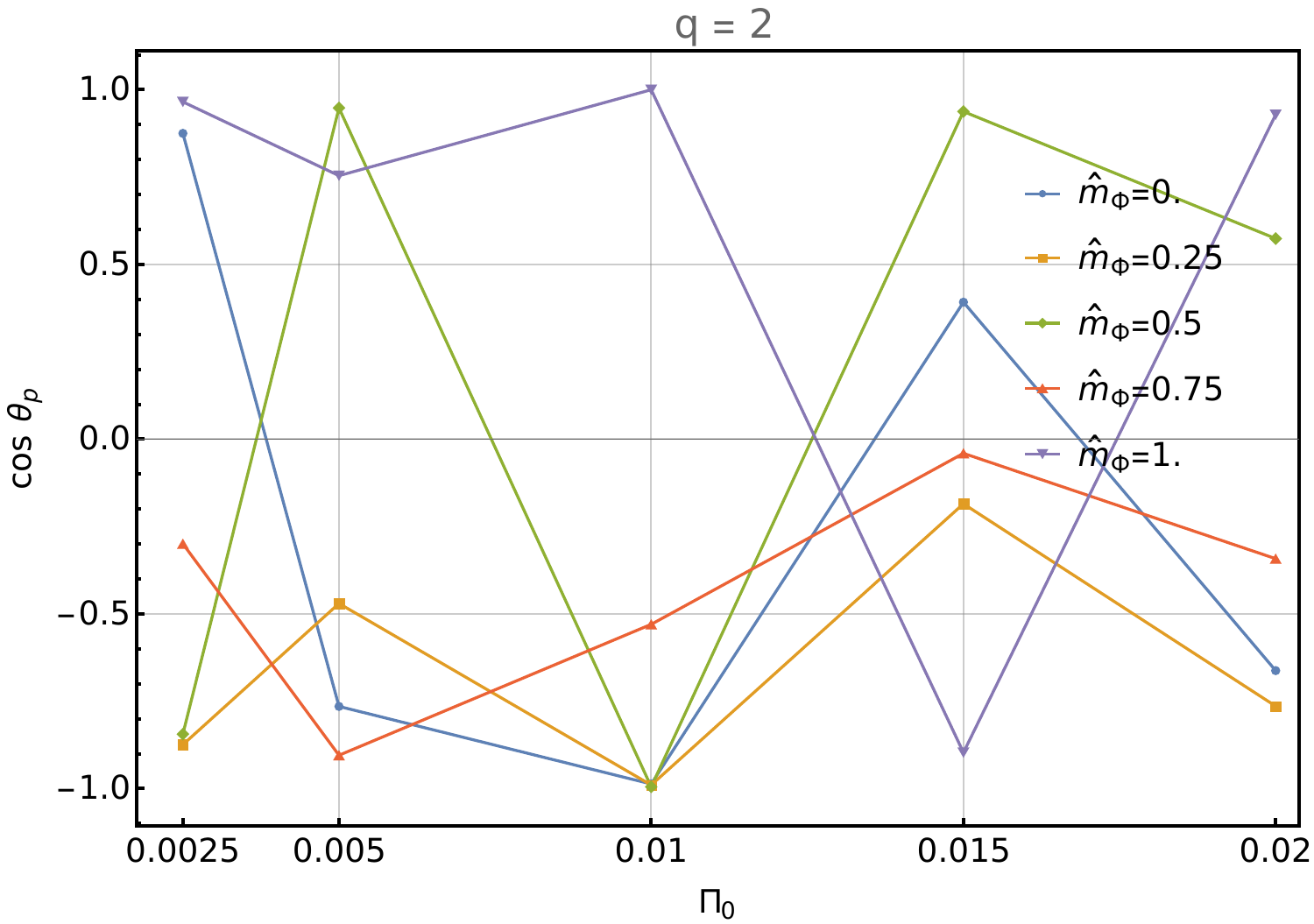}
    \includegraphics[width=0.32\linewidth]{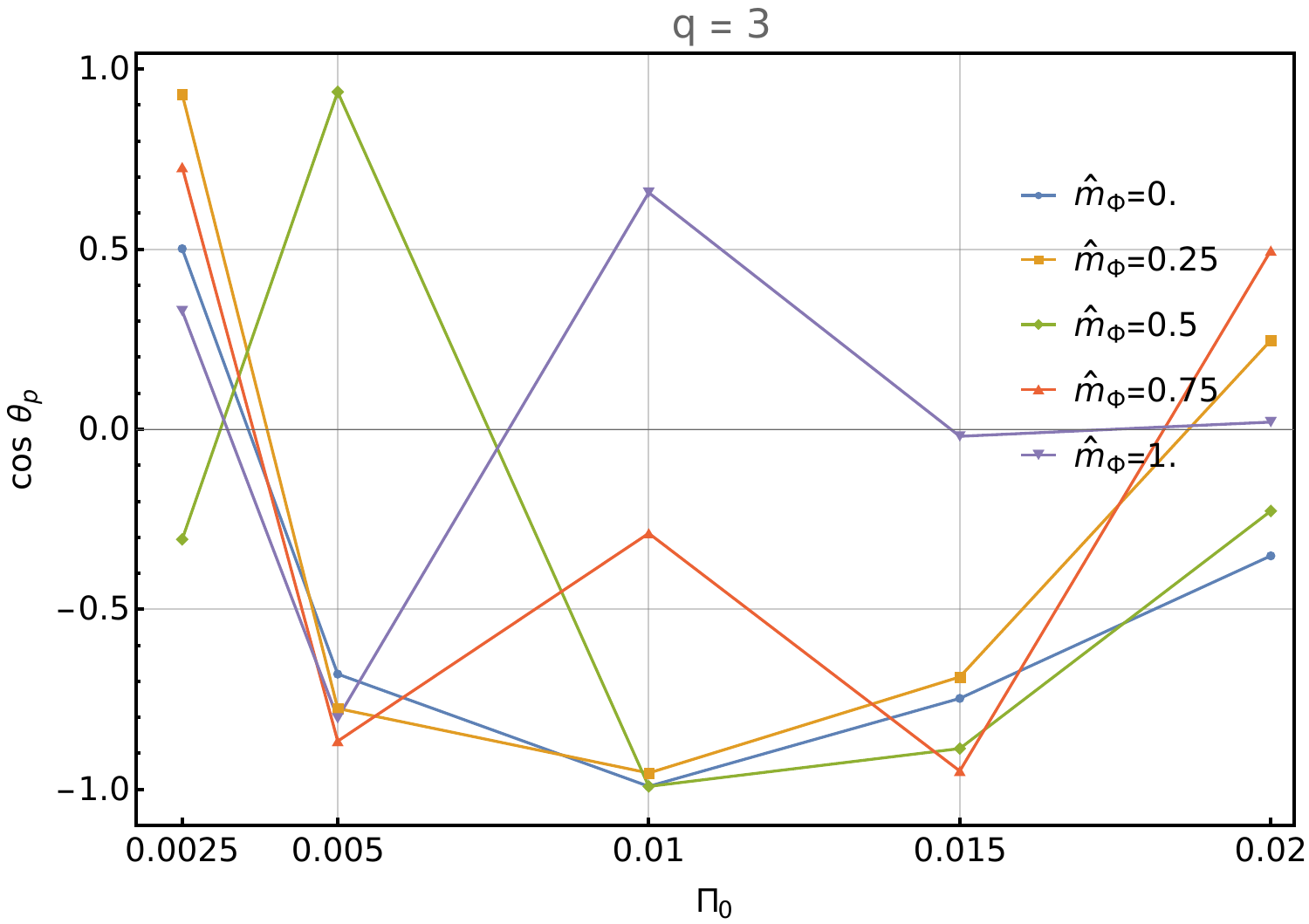}
    \includegraphics[width=0.32\linewidth]{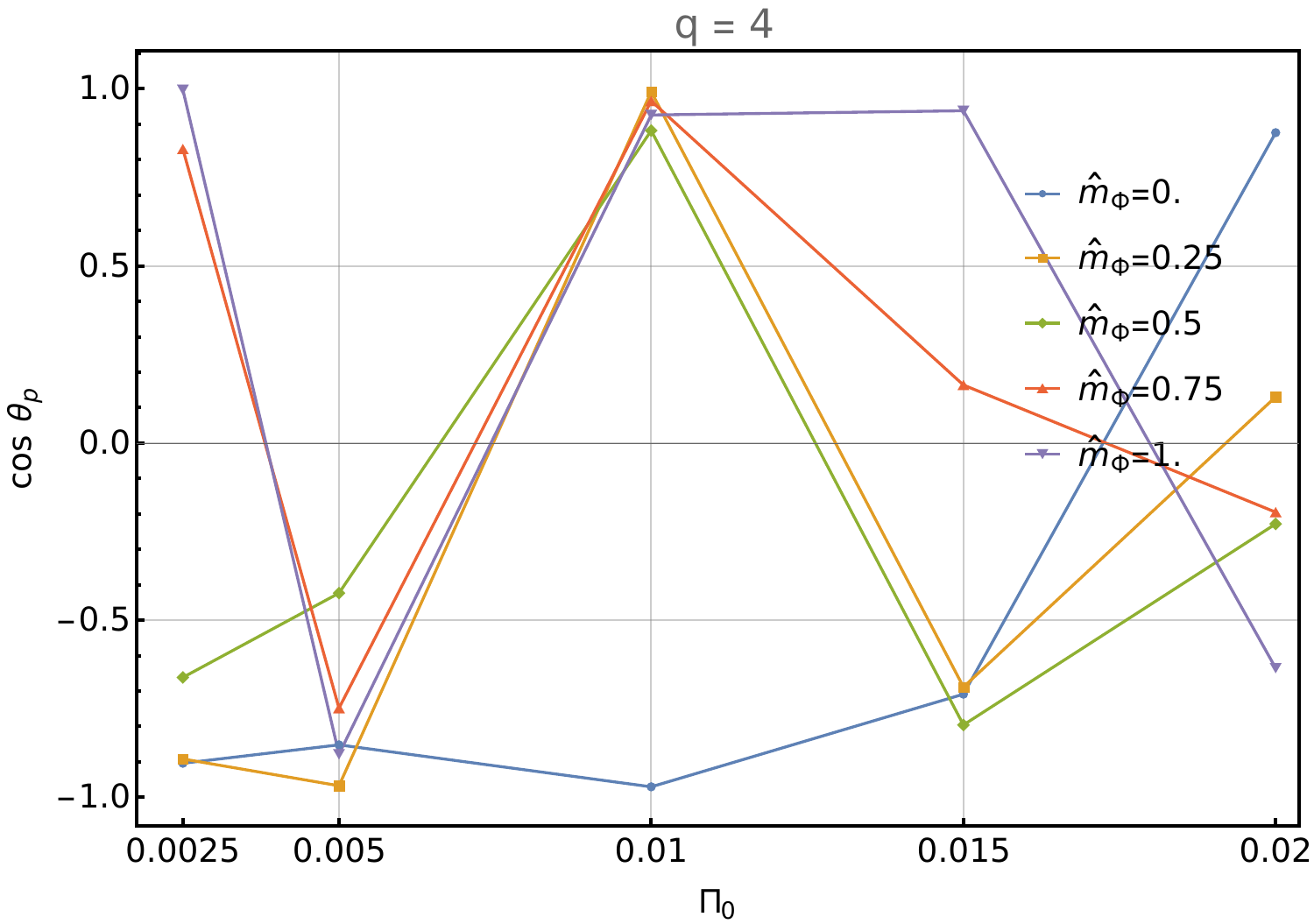}
    \caption{
    Relative orientation of the gravitational-wave and scalar field
    linear momenta, characterized by $\cos\theta_P$, as a function of
    the initial scalar field amplitude $\Pi_0$.
    From left to right, the panels correspond to
    $q=2$, $3$, and $4$, respectively.
    Different curves denote
    $\hat m_{\Phi}=0$, $0.25$, $0.50$, $0.75$, and $1.00$.
    Values $\cos\theta_P\simeq1$ and $-1$ correspond to nearly aligned
    and anti-aligned momentum vectors, respectively.
    }
    \label{fig:kickthetap}
\end{figure*}

The complex parameter dependence of the recoil becomes clearer when
the relative direction of the two momentum channels is examined.
Figure~\ref{fig:kickthetap} shows that $\cos\theta_P$ spans almost the
entire range from $-1$ to $1$ throughout the parameter space. Thus,
the gravitational-wave and scalar field momenta can be nearly aligned,
nearly anti-aligned, or oriented at an intermediate angle depending on
$q$, $\Pi_0$, and $\hat m_{\Phi}$.

When $\cos\theta_P>0$, the two radiated momentum vectors tend to add
constructively, whereas for $\cos\theta_P<0$ they partially cancel.
However, the recoil magnitude cannot be inferred from
$\cos\theta_P$ alone. Even nearly anti-aligned momentum vectors can
produce a large kick when one channel is substantially larger than the
other. Conversely, two approximately aligned contributions need not
produce a large recoil if both momentum magnitudes are small.
The final kick is therefore controlled jointly by
$P_{\rm GW}$, $P_{\Phi}$, and $\theta_P$.

This explains why the recoil exhibits a much stronger and more
irregular dependence on the scalar field parameters than the total
radiated energy. Energy is a scalar quantity and simply adds between
different radiation channels, whereas the linear-momentum balance is
intrinsically vectorial. Changes in the scalar field mass and density
modify both the magnitudes and directions of the two momentum
contributions, naturally producing the strong nonmonotonic behavior
seen in Fig.~\ref{fig:kick}.

\begin{table}
\centering
\caption{
Recoil velocities $v_{\rm kick}$ (in ${\rm km\,s^{-1}}$) of the
remnant black holes for unequal-mass binaries with
$q=2$, $3$, and $4$.
The scalar field mass is denoted by $\hat m_{\Phi}$.
For each mass ratio, the vacuum result is listed separately, while the
remaining columns correspond to different initial scalar field
amplitudes $\Pi_0$.
}
\label{tab:kick_velocity}
\begin{tabular}{c|c|c|c|c|c|c|c}
\hline\hline
\multirow{2}{*}{~$q$~} &
\multirow{2}{*}{~$\hat m_{\Phi}$~} &
\multirow{2}{*}{Vacuum} &
\multicolumn{5}{c}{$\Pi_0$}
\\
\cline{4-8}
& & &
$0.0025$ &
$0.0050$ &
$0.0100$ &
$0.0150$ &
$0.0200$
\\
\hline

2 & 0.00 & 137.8 & 521.3 & 1281.6 & 2063.6 & 711.4 & 53.9  \\
2 & 0.25 & --    & 298.2 & 257.0  & 1009.6 & 771.1 & 122.5 \\
2 & 0.50 & --    & 101.8 & 495.8  & 424.0  & 108.3 & 129.6 \\
2 & 0.75 & --    & 127.6 & 60.4   & 83.2   & 40.4  & 28.1  \\
2 & 1.00 & --    & 141.8 & 145.5  & 154.7  & 16.9  & 17.6  \\
\hline

3 & 0.00 & 162.9 & 301.9 & 530.5  & 744.8  & 378.1 & 50.3  \\
3 & 0.25 & --    & 230.2 & 139.7  & 47.4   & 524.1 & 219.6 \\
3 & 0.50 & --    & 150.1 & 168.9  & 24.9   & 123.7 & 80.2  \\
3 & 0.75 & --    & 167.1 & 91.0   & 88.7   & 11.1  & 35.0  \\
3 & 1.00 & --    & 151.2 & 114.9  & 119.3  & 32.6  & 14.8  \\
\hline

4 & 0.00 & 139.0 & 57.1  & 239.0  & 394.4  & 220.5 & 27.8  \\
4 & 0.25 & --    & 87.0  & 66.0   & 216.3  & 239.6 & 194.3 \\
4 & 0.50 & --    & 129.5 & 114.7  & 118.8  & 46.9  & 46.2  \\
4 & 0.75 & --    & 147.3 & 109.5  & 162.3  & 32.5  & 11.7  \\
4 & 1.00 & --    & 151.2 & 106.2  & 107.3  & 34.0  & 9.4   \\

\hline\hline
\end{tabular}
\end{table}

Table~\ref{tab:kick_velocity} summarizes the numerical recoil
velocities and further demonstrates the strong nonmonotonic dependence
on the scalar field mass and density. The largest recoil in the entire
parameter survey is obtained for
$q=2$, $\hat m_{\Phi}=0$, and $\Pi_0=0.0100$, with
$v_{\rm kick}=2063.6~{\rm km\,s^{-1}}$.

These results extend our previous study, in which the scalar field
mass was fixed and only the initial scalar density was varied. The
present parameter survey demonstrates that the recoil enhancement or
suppression is not determined by the scalar field density alone.
Instead, it results from the combined dependence of the binary
dynamics, gravitational-wave momentum, scalar field momentum, and
their relative orientation on both $\Pi_0$ and $\hat m_{\Phi}$.

\section{Conclusions}
\label{sec:conclusions}

We have performed a systematic numerical-relativity study of
nonspinning BBH mergers in complex scalar field environments, varying
the binary mass ratio, scalar field mass, and initial scalar density.
The simulations cover $q=1,2,3,$ and $4$,
$\hat m_{\Phi}=0,0.25,0.50,0.75,$ and $1.00$, and a broad range of
$\Pi_0$. Because the initial scalar configuration satisfies
$\Phi|_{t=0}=0$, simulations with the same $q$ and $\Pi_0$ but different
$\hat m_{\Phi}$ start from the same binary geometry and scalar
energy-density profile. The subsequent differences therefore provide a
controlled measure of the dynamical role of the scalar field mass.

The orbital and waveform results show that the scalar environment can
substantially shorten the inspiral. For the lowest densities,
increasing $\hat m_{\Phi}$ produces a clear additional acceleration of
the merger, whereas at larger $\Pi_0$ the density-driven effect becomes
dominant and the mass dependence becomes weaker or nonmonotonic. For
unequal-mass binaries, the scalar cloud also drives differential accretion
onto the two components. The horizon-mass ratio $q(t)$ generally
increases during the inspiral, demonstrating that the larger black hole
gains mass more efficiently in fractional terms. This effect is modest at
low density but becomes pronounced for dense clouds, where the mass ratio
can change by order unity before merger. The scalar environment therefore
modifies not only the inspiral timescale but also the intrinsic binary
asymmetry. These changes are directly imprinted on the dominant
gravitational-wave mode through shifts in merger time and modifications of
the waveform amplitude and phase.

The energy and angular-momentum budgets reveal how the scalar
environment redistributes the conserved quantities of the system.
Dense scalar configurations strongly suppress both the energy and
angular momentum carried by gravitational waves. At the same time, the
scalar field provides additional transport channels. The scalar-energy
results show a clear dependence on $\hat m_{\Phi}$: lighter fields
propagate efficiently to large radii, whereas heavier fields remain
more localized and produce stronger inward energy transport through the
inner strong-field region. Scalar angular-momentum transport is more
complex. Its dependence on both $\Pi_0$ and $\hat m_{\Phi}$ is
nonmonotonic, and the direction of the transport across the inner
surface can even reverse for different scalar masses.

The strongest environmental response is found in the linear-momentum
balance. For $q=1$, symmetry suppresses the net momentum to the
numerical level, with $|P_x|$ and $|P_y|$ of order $10^{-8}$ and
$|P_z|$ of order $10^{-18}$. For $q=2,3,$ and $4$, both gravitational
and scalar radiation carry net linear momentum. Their magnitudes and
directions depend sensitively on the scalar parameters, and the two
momentum vectors can be nearly aligned, anti-aligned, or oriented at an
intermediate angle. Consequently, the scalar field can either enhance
or suppress the final recoil.

The recoil exhibits a particularly strong and nonmonotonic dependence
on the scalar mass--density parameter space. The largest kick in our
survey is $v_{\rm kick}=2063.6~{\rm km\,s^{-1}}$ for
$q=2$, $\hat m_{\Phi}=0$, and $\Pi_0=0.0100$, compared with
$137.8~{\rm km\,s^{-1}}$ in vacuum. For $q=3$ and $q=4$, the largest
values are $744.8~{\rm km\,s^{-1}}$ and
$394.4~{\rm km\,s^{-1}}$, respectively, again for the massless
reference sequence at $\Pi_0=0.0100$. The corresponding vacuum kicks
are $162.9~{\rm km\,s^{-1}}$ and $139.0~{\rm km\,s^{-1}}$. Other
regions of parameter space instead yield recoils below the vacuum
value. The kick is therefore governed by the full vector momentum
balance, not simply by the total amount of scalar matter.

Overall, the initial scalar density sets the amount of matter available
to interact with the binary, whereas the scalar field mass controls
the characteristic propagation and dynamical scales. These two effects
are strongly coupled and jointly determine differential accretion, the
evolving binary mass ratio, the merger timescale, gravitational-wave
signal, energy and angular-momentum transport, and remnant recoil. The present results extend our previous fixed-mass
study to a two-dimensional scalar mass--density parameter space and
show that conclusions obtained at a single scalar mass need not persist
across the broader parameter space. Future work should include spinning
binaries, alternative scalar field profiles, and scalar
self-interactions, and should quantify the detectability of these
environmental signatures in current and future gravitational-wave
observations.

\section{Acknowledgments}

This work was supported in part by the National Key Research and Development Program of China (Grant No. 2023YFC2206701), the National Natural Science Foundation of China (Grants No. 12575055, 12475056, 12475055, and 12247101), the Fundamental Research Funds for the Central Universities (Grant No. lzujbky-2025-jdzx07), the Natural Science Foundation of Gansu Province (No. 22JR5RA389, No.25JRRA799), and the `111 Center' under Grant No. B20063, the Gansu Province's Top Leading Talent Support Plane. Yu-Peng Zhang was supported by ``Talent Scientific Fund of Lanzhou University".

\section*{References}
\bibliographystyle{unsrt}

\end{document}